\documentclass[a4paper]{article}
\usepackage{braket}
\usepackage{physics}
\usepackage{geometry}
\usepackage{authblk}
\usepackage{subfigure}
\usepackage{caption}
\usepackage{verbatim}
\usepackage[table,xcdraw]{xcolor}
\usepackage{enumitem, kantlipsum}
\usepackage{fancyvrb}
\usepackage{longtable}
\usepackage{lscape}
\usepackage{amssymb,amsmath}
\usepackage{graphicx}
\usepackage{float}
\usepackage{multirow}
\usepackage{tabularx}
\usepackage{marginnote}
\usepackage{amsmath}
\usepackage{flexisym}
\usepackage{amsmath}
\usepackage[utf8]{inputenc}
\usepackage{xparse}
\usepackage{listings}
\usepackage{tikz}
\NewDocumentCommand{\tens}{t_}
 {%
  \IfBooleanTF{#1}
   {\tensop}
   {\otimes}%
 }
\NewDocumentCommand{\tensop}{m}
 {%
  \mathbin{\mathop{\otimes}\displaylimits_{#1}}%
 }

\newcommand{\projname}{Quantum Computing }

\newcommand{\QPU}{Quantum Accelerator } 
 
\newcommand{\NN}{nearest-neighbour constraint }
\newcommand{\NNN}{nearest-neighbour constraint, }
\newcommand{\QX}{QBeeSim }

\providecommand{\keywords}[1]
{
  \small	
  \textbf{\textit{Keywords---}} #1
}
\title{Quantum Accelerator Stack: A Research Roadmap}
\author[1]{K.~Bertels}
\author[1,3]{A.~Sarkar}
\author[3]{A.~Krol}
\author[1]{R.~Budhrani}
\author[1]{J.~Samadi}
\author[4]{E.~Geoffroy}
\author[2]{J.~Matos}
\author[2]{R.~Abreu}
\author[5]{G.~Gielen}
\author[1,6]{I.~Ashraf}
\affil[1]{QBee.eu}
\affil[2]{University of Porto, Portugal}
\affil[3]{Delft University of Technology, Netherlands}
\affil[4]{ENSTA Brest, France}
\affil[5]{University of Leuven, Belgium}
\affil[6]{HITEC University, Pakistan}

\begin{document}
%\begin{titlepage}
\maketitle
%\end{titlepage}

\begin{abstract}
This paper presents the definition and implementation of a quantum computer architecture to enable creating a new computational device - a quantum computer as an accelerator\footnote{This research is done without any financial support. }, with clear consequences for almost all scientific fields.
A key question addressed is what such a quantum computer is and how it relates to the classical processor that controls the entire execution process.
In this paper, we present explicitly the idea of a quantum accelerator which contains the full stack of the layers of an accelerator. Such a stack starts at the highest level describing the target application of the accelerator. Important to realise is that qubits are defined as perfect qubits, implying they do not decohere and perform good quantum gate operations.
The next layer abstracts the quantum logic outlining the algorithm that is to be executed on the quantum accelerator.
In our case, the logic is expressed in the universal quantum-classical hybrid computation language developed in the group, called OpenQL, which visualises the quantum processor as a computational accelerator. We also have to start thinking about how to verify, validate and test the quantum software such that the compiler generates a correct version of the quantum circuit.
The OpenQL compiler translates the program to a common assembly language, called cQASM, which can be executed on a quantum simulator.
The cQASM represents the instruction set that can be executed by the micro-architecture implemented in the quantum accelerator.  In this context, we think about the necessity to develop a quantum operating system that manages all the hardware of the micro-architecture and also makes sure the qubits are in the right place.  The layer below the micro-architecture is responsible of the mapping and routing of the qubits on the topology that allows the qubits to be placed such that two-qubit gates can be easily applied on the qubits, even when the \NN-constraint needs to be respected.
At any moment in the future when we are capable of generating multiple good qubits, the compiler can convert the cQASM to generate the eQASM, which is executable on a particular experimental device incorporating the platform-specific parameters.
This way, we are able to distinguish clearly the experimental research towards better qubits, and the industrial and societal applications that need to be developed and executed on a quantum device. We introduce two quantum applications. The first is already quite advanced and deals with the DNA-research that modern medicine is using for all its analyses.  The second topic has  started a couple of years ago and looks how quantum concepts can be used on financial challenges.  For each, we emphasise the use of perfect qubits and supercomputers to compute the result.

\end{abstract}
\section{Introduction}

\begin{figure}[hbt]
\centering
\includegraphics[width=\textwidth]{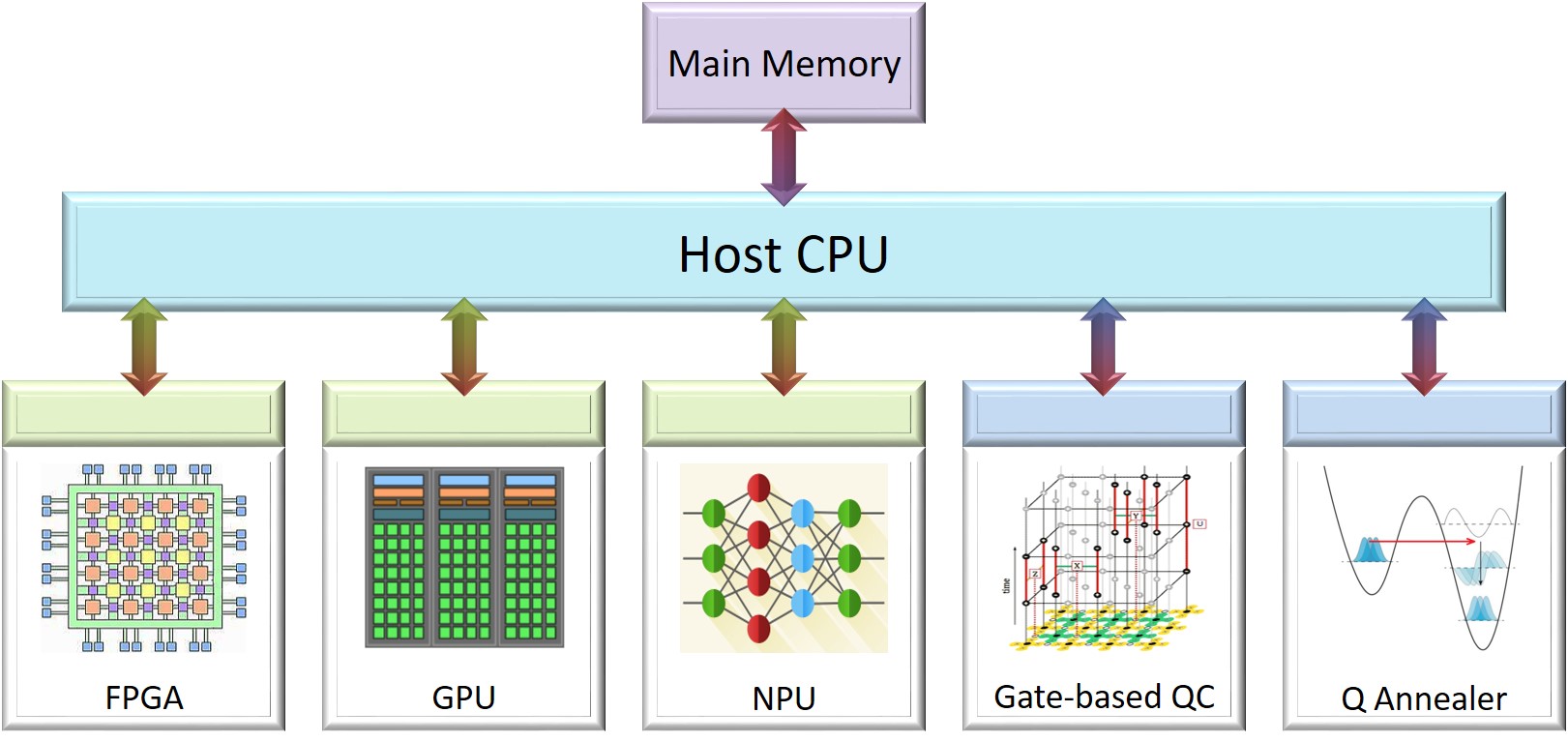}
\caption{System architecture with heterogeneous accelerators}
\label{fig:accel}
\end{figure}

The history of computer architecture dates back several decades and is marked by evolutionary spurts associated with innovations in core technologies, process and  industrial paradigm shifts and co-technologies such as computer languages and operating systems.
An important extension has been the emergence of accelerators~\cite{vassiliadis2004} as specialised processing units to which the host processor offloads computational tasks with the objectives of improving performance, capacity and the physical envelope of systems such as energy consumption and thermal characteristics.
With a view to the prospect of paradigm breaking computational capacity, of late, computer architecture research has looked to quantum computing as a means of eschewing the real-estate and clock frequency based battle with Moore's law. Despite providing fertile ground and sustaining a complex global ecosystem there is a shared intuition of the transformational possibilities that quantum devices could afford to science, industry and society.
In the next 5 to 10 years of quantum computer development, it does not makes sense to talk about quantum computing in the sense of a universal, Turing computer that can be applied within all application domains.  
Given the recent insights leading to e.g. Noisy Intermediate-Scale Quantum~(NISQ) technology as expressed in~\cite{preskill2018quantum}, % as well as randomised compiler techniques as described in~\cite{wallman2016noise}, 
we are much more inclined to believe that the first industry-based and societal relevant application will be a hybrid combination of a classical computer and a quantum accelerator.
This view is based on the observation that many end-applications contain multiple computational kernels and the properties of these parts are better executed by a particular accelerator which can be, as shown in Figure~\ref{fig:accel}, either field-programmable gate arrays~(FPGA), graphics-processing units~(GPU), neural processing units (NPU) like Google's tensor processing unit (TPU), etc. 
The formal definition of an accelerator is, indeed, a co-processor linked to the central processor that is capable of accelerating the execution of specific computational intensive kernels, as to speed up the overall execution according to Amdahl's law.
We now propose to add two classes of quantum accelerator as additional co-processors.
The first is based on quantum gates, the second on quantum annealing.
A classical host processor keeps the control over the total system and delegates the execution of certain parts to the available accelerators.
  
Computer architectures have evolved dramatically over the last couple of decades.
The first computers that were built did not have a clear separation between compute logic and memory.
It was only with von Neumann's idea to separate and develop these distinctly that the famous von Neumann architecture was born.
This architecture had for a long time a single processor and was driven forward by the ever increasing number of transistors on the chip, which doubled every 18 months. This supported a pervasive, renaissance-like surge of activity that rippled through academia, industry and society.
At the beginning of the 21st century, having spawned an appetite for computing power and fed this for more than three decades, single cores became too complex and started to falter as a means of providing more than incremental processing improvement.
This spurred the development and proliferation of multiple cores and associated architectures which had formerly been the province of high-end computing.
In turn, the homogeneous multi-core processor dominated the processor development for the better part of a decade but companies such as IBM and Intel, with deep insight not only into computer architecture but also a wide gamut of applications, industries, algorithms and data were able to see that heterogeneous architectures offered a different equation and economics in terms of improving, tailoring and packaging computational power.
GPUs and FPGAs provided natural building blocks and extensions for computer architecture. It is in this light that we suggest a quantum accelerator is the next emerging step and from which the lessons learned with the previous generation of hardware and toolsets should be incorporated into research roadmaps. As is evidenced by government, industrial and societal interest, these lessons go beyond purely point solutions and a uni-dimensional TRL-level. They encompass the evolution of a  whole quantum technology, co-technologies and ecosystem stack. What is very important to emphasise and realise is that the introduction of the quantum computational paradigm will have a significat impact on all scientific fields. The main reason for this is that a mere compilation to the newest processor will not provide any of the anticipated benefits. A complete redesign, implementation and testing of quantum algorithms is required as the logic, concepts and operations  need to be substantially invented and integrated in any quantum language.  The long-term benefit is to have substantial better results for almost all problems the earth has to solve and the short-term benefit is to build inside any organisation a team of quantum competent people capable to start looking at the problems one has to solve.

In the technical quantum computing world that this paper describes, there exist two important challenges.  
The \textbf{first} is to have enough numbers of good quality qubits in the experimental quantum processor.
The current competing qubit technologies include Ion traps, Majoranas, semi-conducting and superconducting qubits, NV-centers and even Graphene.  Since one of the first papers about quantum computing by R.\ Feynman in 1982 \cite{feynman1982simulating}, research on quantum computing has focused on the development of low-level quantum hardware components like superconducting qubits, ion trap qubits or spin-qubits.  The design of proof-of-concept quantum algorithms and their analysis with respect to their theoretical complexity improvements over classical algorithms has also received some attention.  A true quantum killer application that demonstrates the exponential performance increase of quantum over conventional computers \emph{in practice} is, however, still missing but is urgently needed to convince quantum sceptics about the usefulness of quantum computing and to make it a mainstream technology within the coming decade. Recent world-news coming from, for instance, Google, should be interpreted in this context. The errors that these devices still produce are so big that it is very difficult to make any valid statement about their operational quality. 

\begin{figure}[hbt]
    \centering
    \includegraphics[width=60mm]{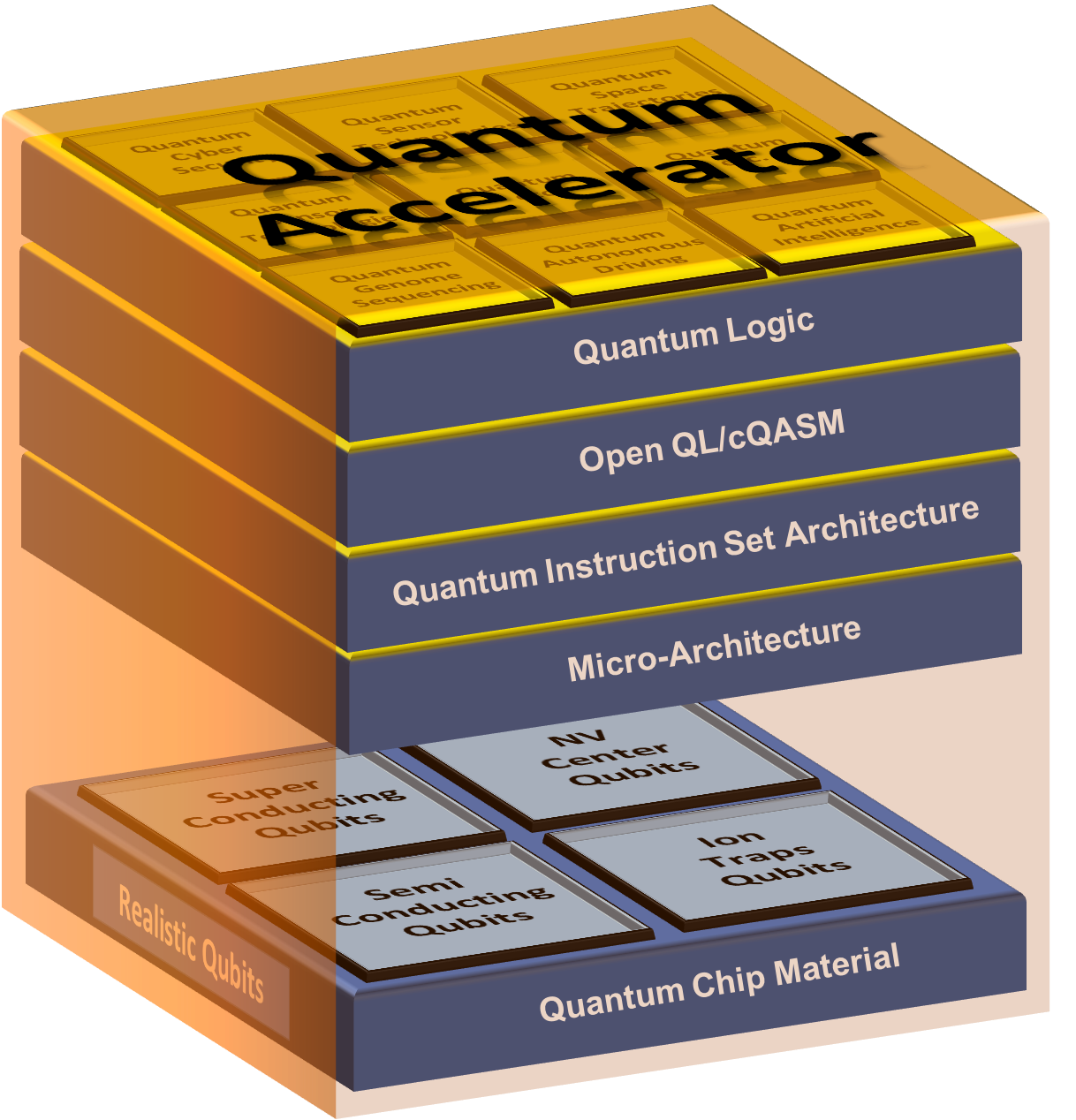}
    \caption{ Experimental full-stack with real qubits (old schema) }
    \label{fig:QAC}
\end{figure}

Improving the overall status of the qubits is challenging as these suffer from decoherence that introduces errors when performing  quantum gate operations. It is only when the quantum physical researchers overcome those challenges that the quantum accelerator will be a widespread adopted solution. It is important to understand that any quantum accelerator that will be developed in the next 5 to 10 years will have the layers as shown in  Figure~\ref{fig:QAC}. 
The \textbf{second} challenge is to formulate at a high level the quantum logic that companies and other organisations need to be able to use high-performance accelerators for certain computations that can only run on the quantum device.
This requires a long-term investment in terms of people and technical know-how from companies that want to pursue this direction and reap the benefits.
As will be explained in this paper, we will implement and test the  industrial or societal applications for which  the required quantum logic  can be executed using the full-stack, evaluated and tested on a quantum simulator, running on a supercomputer. It is important to emphasise that the qubits are called perfect qubits that do not decohere or have any other kind of errors generated by them. With the emergence of huge amounts of data, commonly called big data, it is understood that this paradigm is not scalable to super-large data sets.  
The key factor is the huge amount of data that needs to be processed by multiple computing cores which is a very tough problem to solve.
The data communication between the cores is a very difficult programming problem and the data management problem is substantially slowing down the overall performance.

%However, when building a computer, it is important to base any design decision on the understanding of a couple of decades of experience on how to build it. In the quantum computing context, it is therefore important to do the same especially as the computer industry is facing a serious problem with respect to improving the performance of the compute power. Even though the accelerator idea is very interesting, also other paths are intensively studied. One of those paths is the in-memory computing architecture.  The current supercomputers have a very large number of processors and terabytes of memory.  The von-Neumann architecture always sends the data to the processor and writes back the result.  
Based on our group's research since 2004~\cite{vassiliadis2004} and as shown in Figure~\ref{fig:QAC}, an important concept that we have been implementing in the quantum computing world is the implementation of a full stack for a quantum accelerator as will be described later in this paper.
The basic philosophy of any accelerator is that a full stack needs to be defined and implemented.  
The last 10 to 15 years have shown a large number of accelerators that were developed as part of any modern computer architecture.  
It always consists of the same following layers: it starts at the highest level describing the logic that needs to be mapped on the accelerator.  
Examples are video processing, security, matrix computation, etc.
These application-specific algorithms can be defined in various languages such as C++ or Fortran.
In the case of FPGAs, these algorithms are translated into VHDL or Verilog.
In the case of GPUs, the language is often formulated using mathematics or other libraries and translated by the compiler to an assembly language that can be mapped on the GPU-architecture.
Especially in the case of FPGAs, there is no standard micro-architecture on which the VHDL or Verilog can be executed. 
Such an architecture needs to be developed for every application that needs to be accelerated.
The final layer is a chip based implementation of the micro-architecture combined with the hardware accelerator blocks that are needed.

 We use the term 'accelerator' as it is very unknown how to define a quantum computer. The terms 'Perfect Qubits' and 'Virtual Qubits' refer to the need to currently overcome  the challenge that the current, experimentally produced qubits have, meaning  very high error-rates and very short coherence-times. \textbf{Supercomputers} will be used to test any execution of the quantum algorithm and the \QX simulator will be capable of addressing all the computational features that modern supercomputers can have, which will primarily consist of the memory resources that are available to represent the qubit states.

%\begin{figure}[htb] 
%    \centering
%    \captionsetup{justification=centering}
%    \subfigure[Experimental full-stack with real qubits]
%    { 
%        \centering
%        \includegraphics[width=.42\textwidth]{fig/QAC.png}
%        \label{fig:QAX} 
%    } 
%    \subfigure[Simulated full-stack with perfect qubits]
%    { 
%        \centering
%        \includegraphics[width=.42\textwidth]{fig/QAS.png}
%        \label{fig:QAS}
%    }
%    \caption{Two current full-stack quantum accelerators} 
%    \label{fig:qcube}
%\end{figure}

\subsection*{Background}

One of the first proposals on quantum computing was written by R.\ Feynman in 1982 \cite{feynman1982simulating} which launched a world-wide research on quantum computing focusing on important low-level challenges leading to the development of  superconducting qubits, ion trap qubits or spin-qubits.  
He formulated the use of quantum computers as an important scientific instrument to allow us to understand the quantum phenomena that quantum physics tries to understand.
The design of proof-of-concept quantum algorithms and their analysis with respect to their theoretical complexity improvements over classical algorithms has also received some attention.  
However, we still need substantial progress in either of those domains. 
Qubits with a sufficiently long coherence time combined with a true quantum killer application are still crucial achievements on which the community is working.  
These are vital to demonstrate the exponential performance increase of quantum over conventional computers \emph{in practice} and are urgently needed to convince quantum sceptics about the usefulness of quantum computing such that it can become a mainstream technology within the coming 10 to 15 years.  
However, as we will describe in this paper, we need much more before any kind of computational device can been developed, which ultimately connects the algorithmic level with the physical chip.
What is needed involves a compiler, run-time support and more importantly a micro-architecture that executes a well-defined set of quantum instructions.

An interesting and quite high-level kind of description was published in 2013~\cite{van2013blueprint}. 
The authors describe their understanding of the blueprint of a quantum computer. 
They correctly emphasised the need to look at computer engineering to better understand what the similarities and differences are between quantum and classical computing.  
As mentioned before, the most important difference is the substantially higher error rate that qubits and quantum gates ($10^{-3}$) have compared to CMOS-technology ($10^{-15}$).
Guaranteeing fault-tolerant computation can easily consume more than 90\% of the actual computational activity.  
The second difference focuses on the nearest-neighbour constraint which imposes that two-qubit gates can only be applied if the qubits reside next to each other. 
The no-cloning theorem prohibits copying quantum states. 
The way that two-qubit gates are applied requires the two qubits to be sufficiently close to each other.  
They also describe a hierarchical layered structure but rather than defining these layers in terms of more computer engineering concepts, the schema is more expressed in terms of the different, relevant fields and research domains. 
Examples are Quantum Error Correction (QEC) theory, programming languages, fault-tolerant (FT) implementation and so on. 
There are also other mechanisms with undefined time costs that are necessary to make FT-quantum computing (hopefully) efficient and performing. 
Examples are state distillation for ancilla factories and the emergence of a wide variety of defects and errors, which all impose an additional burden on the micro-architecture and the corresponding run-time management.

An older but conceptually quite similar paper was published by DiVincenzo in 2000~\cite{divincenzo2000physical}.  
This article outlines 5 criteria needed to build a quantum computer: i) a scalable physical system with well characterised qubits, ii) the ability to initialise the state of the qubits to simple fiducial state, iii) long relevant coherence times, iv) a universal set of quantum gates and v) a qubit specific measurement capability. Two additional criteria needed for quantum communication are, the ability to inter-convert stationary and flying qubits and the ability to transmit flying qubits between specified locations.
Considering currently available quantum processors, we could say that they already comply to DiVincenzo's criteria and thus we already have a quantum computer.
However, an important and missing criterion is the number of qubits that we need for any kind of reasonable application. 
Depending on the application domain, the estimates of the number of qubits goes from relatively low, such as a couple of hundreds, to several billions.  
Being less critical, we could say that the first criterion explicitly formulates the size of the system, which is still a very considerable challenge to compute in a reliable way.

% The rest of the paper is structured as follows and as shown in Figure~\ref{fig:qcube}.
% We first describe the quantum algorithm layer and present the programming language OpenQL and the quantum assembly language cQASM to which the OpenQL compiler translates.  
% We then introduce the micro-architecture, including the mapping of quantum circuits to the quantum chip, and conclude the paper with a detailed discussion of two particular examples of an accelerator that we are currently developing.

The rest of the paper is structured as follows.
First, we describe the various layers like, application, algorithmic logic, software testing, programming language and OpenQL compiler, micro-architecture, mapping and simulator and the \QX. Each layer is developed in terms of  perfect qubits.  We also introduce the basic Tensor mathematics that is needed for performing any quantum operation. We then present the two quantum accelerator applications on which we are working.  Our longer term vision is to develop a full system stack, that goes from whatever experimental quantum chip up to quantum accelerator applications.

\section{State of Qubits }

The overarching objective of \projname is to provide, world-wide, the  industry and institutions with the necessary infrastructure to move into the quantum-computing era and lift them on a competitive basis against other countries or organisations. The developed infrastructure will cover the full  spectrum from key application kernels of leading  companies to algorithmic definitions,  compilers, all the way down to mapping into the qubits topology.    
The key definition for  \projname is that more people will have to work on various  application domains and develop the relevant quantum logic for each of them. That quantum logic will ultimately be written in any quantum programming language.  We developed the OpenQL as flexible programming language. Two aspects are very important to highlight and underline. The first is to start reasoning in perfect qubits and the second is to develop all that is needed to have a quantum accelerator.

An important extension that is introduced for our line of research is the use of three kinds of qubits, namely real, realistic and perfect qubits. Before we explain the different kinds of qubits, it is important to highlight one important distinction between real and perfect qubits. Irrespective of the quantum technology used to produce qubits such as superconducting or photonics qubits, there is only one way of getting the final result of the quantum circuit, which is through measurement of the qubit. When measuring any real qubit multiple times when executing the quantum circuit, we will get the probability  which defines the likelihood that this qubit contains the good result of the algorithm. This is also the reason why we have to execute multiple times the same algorithm on the same qubits, to have an overview of the final and good result of the quantum computation. When we generate perfect qubits on the classical (super)computer, we will have access to both the amplitude and derive the probability from that amplitude by the known $|\alpha|^2 +|\beta|^2 $ equation where $\alpha$ and $\beta$ are the amplitudes from which the probability is derived.  The creation of qubits and the diversity in the underlying technology is similar to the pre-transistor period we had when building the classical computer. In this section, we define them in detail and how these relate to each other.

\subsection{Real qubits} 
The first qubit type is the experimental qubit, called the real qubit, which refers to experimentally realised system with challenges such as decoherence and error-rates. These features need to be substantially improved for any commercially available quantum device. The real qubits are investigated by the experimental quantum physicists community.  The goal is to improve the quality of the real qubits such that these become more easy to scale to large numbers and allow for a pragmatic micro-architectural control. This implies that there is a need to study how long the qubits can stay in a particular state and maintain their fidelity, called the coherence time. Most of the real qubits go to the ground state in a very short time (ranging from micro to milliseconds) after these are created in a particular state. Adding to that, all the quantum gates that need to be applied to the qubits generate errors. In quantum gate operations the errors and the error-rates need to be better than the current $10^{-2}$ rates. \footnote{We will limit ourselves now to quantum gates but will introduce later the quantum annealing approach.} %Without going into a detailed discussion, we should not forget that all the qubit and quantum phenomena such as superposition and entanglement are analogue phenomena and thus subject to various changes caused by contextual influences.  
There are currently many quantum technologies experimenting to produce good quality qubits for reasonable quantum computation.  The use of real qubits is very important as the physicists need to understand the dynamic and static behaviour of the qubits under different circumstances. Many large companies implement physical system for quantum computing such as IBM, Google, Rigetti, D-Wave Systems, IonQ, etc. However, the quality as well as the number of these qubits is very limited and the decoherence and error-rates as mentioned before are currently problematic for application development as these tend to influence the overall result that the quantum device is computing.  An important use of physical qubits is to combine many of those qubits in one logical qubit. An important approach for many years was the Surface Code where up to 49 physical qubits can be combined in one logical qubit.  It was early 2018 that the quantum physicist John Preskill wrote an article in which he urgently argued to postpone the use of Surface Code as we are loosing too many ancilla qubits to make one logical qubit. \cite{preskill2018}
A final comment we can make is to refer to the metric launched by IBM and which is called Quantum Volume. It measures the computational capacity of near-term quantum computers with a relatively low number of qubits and will be presented later in this paper.\cite{Cross_2019}

% a single-number metric, quantum volume, that can be measured using a concrete protocol on near-term quantum computers of modest size (n <∼ 50), and measure it on several state- of-the-art transmon devices, finding values as high as 16
%This would mean that the qubits have only perfect behaviour. In certain quantum gate operations there will be some errors and the error-rates will be better than the current $10^{-2/-3}$ and only when decoherence will take longer than the microseconds that we see appearing in current experimental devices.
\subsection{Realistic qubits} 
Realistic qubits represent what improvements we are trying to achieve on the real qubits. The real qubits have error rates around $10^{-2/-3}$ and what does it take if we reach $10^{-4/-5}$ as the error rates.
Any computer architecture needs functionality to continuously monitor the quantum system to detect and recover possible errors, as we describe here. For quite a long period, the focus has been mostly on planar surface codes as it was considered one of the most promising QEC codes for short-term implementations and for scalability concerns in the FT-era and manufacturing.  %As shown in Figure \ref{fig:logical_qubits},  
Qubits are generally manufactured in a regular 2-D lattice connectivity with only nearest-neighbor (NN) interactions. The array comprises of two kinds of qubits, namely the data and ancilla qubits. Data qubits are used to store the quantum information for the computation, whereas ancilla qubits are helper qubits which are used to detect bit-flip and phase-flip errors by performing error syndrome measurements (ESM). This implies that after every sequence of quantum gates, the system needs to measure out its state and interpret those measurements to see if an error has been produced. Given the constraints of the coherent qubit lifetime, it implies that a very large graph needs to be processes and interpreted in real-time such that any error can be identified. Measurements themselves can be erroneous and therefore need to be repeated multiple times before a final conclusion is reached. In 2018, Preskill~\cite{preskill2018} introduced a counter-argument to this approach because surface code requires too many ancilla-qubits for logical protection.   This led to the re-initiation of the small-codes which were first defined almost 20 years ago. The impact on the system architectural and compiler level is yet unclear but this is currently the focus of a lot of research. 
%Companies such as Fujitsu, offer simulation platforms based on  realistic qubits, implementing a quantum-inspired annealing approach.

\subsection{Perfect qubits}
Companies, governments and other organisations interested in building a quantum accelerator need to evaluate the availability of quantum computing resources in terms of quantum algorithms and have a way to test the correctness of the quantum logic. To serve these needs, we use perfect qubits, such that any of the erroneous behaviour arising due to qubit quality can be avoided during application development phase. These qubits modelled in the simulator do not decohere and stay in ideal state required for the algorithm. Using these perfect qubits guarantees that the end-users can verify and check the algorithm that they are working on and test if the computed results have a meaning that can be easily interpreted. We are not the only ones who use this but it is a very clear concept that separates the two directions that we are investigating in the Quantum Computer Architecture lab.  What is very encouraging to notice is that companies such as IBM have a version in their Qiskit platform that can compile assuming the qubits are perfect.\cite{ibmqiskit} As explained above, we introduce in OpenQL a datatype which represents the perfect qubit which has a more stable behaviour than the realistic qubits. Whether or not the nearest-neighbour constraint applies, is a discretion of the designer. The compiler may or may not compute a route for the qubits. These decisions are based on the requirement and maturity of the application development stage before translating to realistic experimental testing. In the rest of this paper, we will in most cases assume that we have perfect qubits. It will be a challenge to other field to improve the quality of the qubits but we, as computer engineers, make abstraction of all the errors that qubits will have and assume the perfect coherent behaviour and the correction computational quantum gate results.
 
Perfect qubits can be also named virtual qubits as they refer to the non-mapped, all-to-all connected, high-level qubits that define a virtual platform where the logic of quantum algorithms is defined using a quantum programming language. If no topology is specified, this is the default model of qubits considered for the quantum accelerator developments. Logical qubits consider the constraints of the connectivity provided by the interconnection topology of a given chip where the virtual qubits are mapped to a particular topology. 

\section{The Full-Stack of the Quantum Accelerator}
\begin{figure}[ht]
    \centering
    \includegraphics[width=90mm]{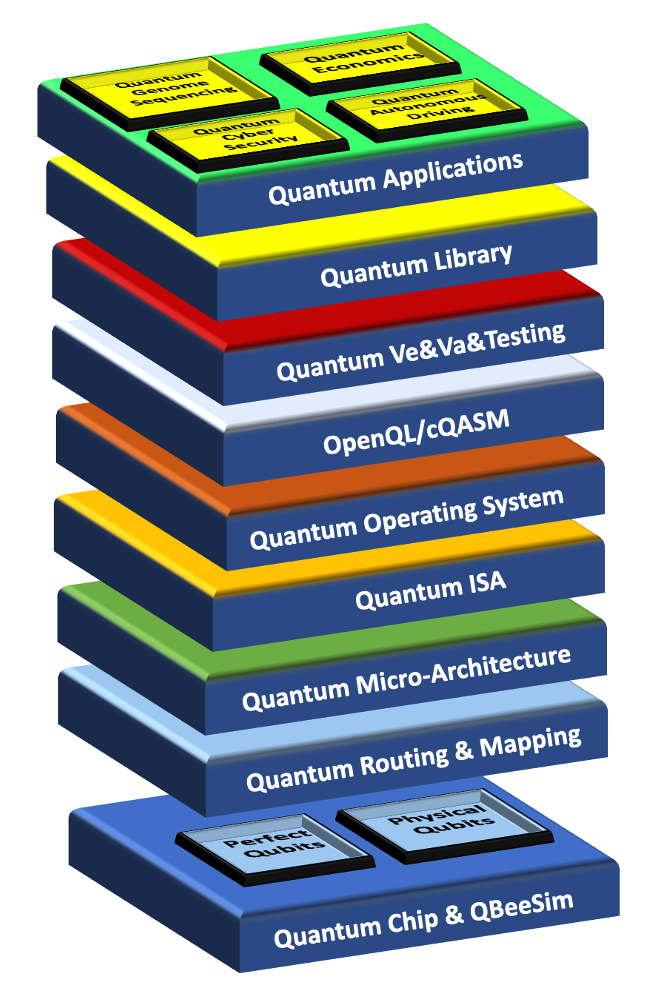}
    \caption{  Full-Stack with Perfect and Physical Qubits }
    \label{fig:FullStackNew}
\end{figure}

In this section and as given in Figure~\ref{fig:FullStackNew}, we describe the different layers the quantum accelerator system stack most likely will have, at least assuming the current state of the classical computers.  What is important to realise is that in the long term, we will have quantum computational devices that have a correct behavior and result but for now we have to use classical and very  powerful computers to realise the result of any quantum application. It will be explained in detail later in the paper.

We do not yet have a full implementation of the micro-architecture for logic expressed in terms of the perfect qubits. Our current understanding of a generic micro-architecture is given in Figure~\ref{fig:microarchitecture}. That architecture is being used for a first prototype of a quantum accelerator and will be described later in this paper. 
It is important to define the QISA needed for any quantum accelerator application and fine-tune the corresponding micro-architectural blocks needed to execute the quantum instructions on the \QX  simulator.

Every classical computer  consists of different components that are very integrated with each other, ranging from a classical chip where the classical circuits are implemented, a system memory and a bus interconnect that always to bring data to the processor and write back the result, an operating system, a programming language and an application layer. We transpose those layers to the quantum world and make the changes that we think are needed for any application that will use the quantum logic. A very important feature in the quantum domain is in-memory computing implying that the quantum logic is brought to the qubit rather than the qubit brought to the quantum processor. As described earlier, the following concepts need to be understood to comprehend the full stack:

\begin{figure}[bt]
\centering
\includegraphics[width=0.6\textwidth]{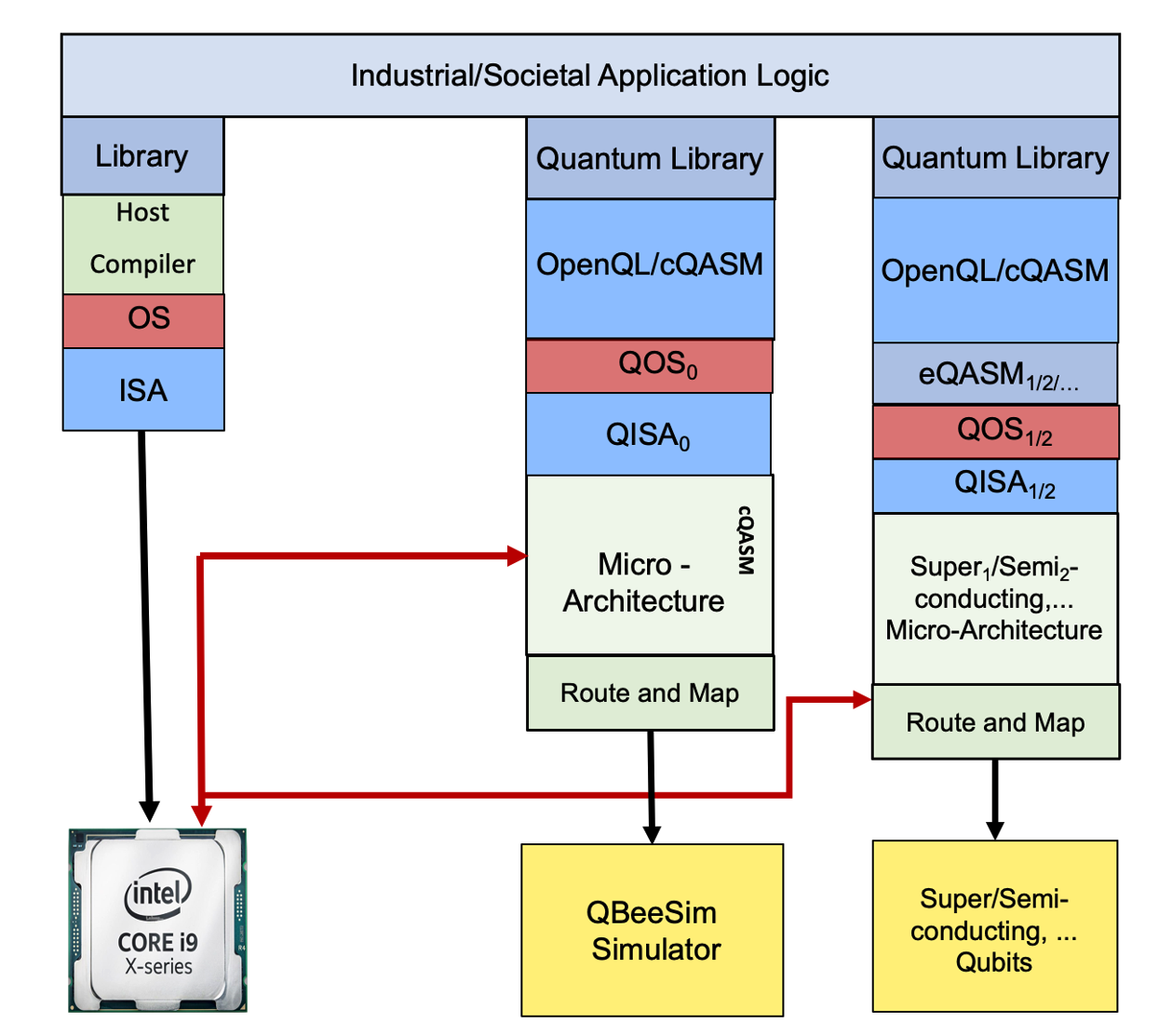}
\caption{Full-stack execution}
\label{fig:layers}
\end{figure}

\begin{itemize}
    \item \textbf{Software stack and the different layers: } The cube as shown in Figure \ref{fig:FullStackNew} is based on a full system stack and more information can be found in \cite{bertelsAccel91}.   A full software stack will be designed and developed that integrates all scientific and technological results from different fields, including algorithm development and quantum programming language extensions. This will allow users to easily express quantum algorithms using an appropriate programming language and execute it on quantum simulators and physical quantum devices. It is important to emphasise that we are capable of controlling super- and semiconducting qubits so our approach is independent to whatever quantum technology will dominate the market in the future.
   
    \item \textbf{Quantum accelerator application:}  The most important step towards promoting a wider interest and adoption of quantum technologies is to showcase their commercial viability through application in concrete domains which are key to European economy. Indeed, one of the strongest aspects of \projname is that we focus on some real industrial domains that have the potential to show Quantum Advantage, where a quantum device will outperform any classical processor mechanism. Examples are  bioinformatics, finance, autonomous cars, aerospace industry, material sciences and chemistry. \cite{Bauer_2020}
    
    \item \textbf{Quantum Library:} Quantum Library describes the overall structure of the applications in terms of quantum operations that are needed. Quantum Library only looks at the algorithmic and operational requirements that a particular application needs. It is not looking at lower-level constraints, which are also important but that tend to obfuscate the higher level application needs. The application logic will be translated in some high-level pseudo-code and will be refined and ultimately expressed in OpenQL, which is the programming language that was originally developed. But it can be any pseudo-language that allows to express the main logic in a non-executable and human-friendly way.
    
    \item \textbf{Quantum Software Validation, Verification and Testing:} An application from any domain and written in any programming language needs to be tested to guarantee it behaves and computes in an expected and reliable way. This is why an essential layer in our full-stack relates to quantum software testing.  The main challenges that need to be analysed can be easily separated in two directions. The first is the quantum software that will be executed on a classical computer. We have to test whether the functional behaviour is correct and that the software is computing the results we are expecting. That means that we need to have an expectation of the range of results we are looking for and what ranges we expect to change in the future. A second constraint related to this challenge is that we need to be able to verify if the parallel version of the software is correctly generated by the developer. That is a very complex problem as the parallelisms will rapidly grow if the number of superpositioned or fully-entangled grows.  There are clearly implications for the quantum application and the programming language that is being used.

     \item \textbf{Programming language and compiler:} Any application that needs to run on a computational platform needs to be expressed in a programming language.  The programming language that will be used in the research is OpenQL that results in an executable assembler, common QASM (cQASM). QASM is a term that was defined by Nielsen and Chuang for the Quantum Computation and Quantum Information book \cite{nielsen2010quantum}. We use the same term but prefix it with the letter 'c' that represents the assembly components of the QASM which can be executed on the \QX simulator. Those components  can also serve as the input for another back-end compiler, that we call the eQASM back-end compiler passes, that translates them into an executable version that can run on, for instance, a superconducting or semiconducting quantum chip.   %The OpenQL-language was developed by DUT and will be used to translate the Quantum Logic into an executable OpenQL-based version, which will be translated  in a cQASM-version which can be executed on the \QX -simulator.
     
     \item \textbf{Quantum Operating System (QOS):} When many hardware blocks need to collaborate in a very precise way in terms of data and time, a quantum operating system is needed. There are several examples that have been developed so far but what is important in any QOS is that there is a part of it that takes care of the variety of instructions that are being executed by the micro-architecture and those instructions and hardware blocks that require an extremely precise timing.  Any quantum application will be a combination of classical programming structures such as loops and branches. But once the micro-architecture in combination with the QOS is sending quantum instructions to the quantum processor or the simulator, the timing becomes very deterministic and strict.
    
    \item \textbf{Tensor Mathematics:} Any quantum circuit will be expressing the logic and the quantum gates that will be applied on the qubits. The mathematics used to express the quantum gates is based on a known mathematical field, called Tensor mathematics. In this context, a tensor is an algebraic object that describes a relationship between different  algebraic objects that form a vector space. We describe the elementary operations that also explain why the matrices describing the quantum states grows exponentially with the number of qubits and we therefore need very large memories to execute a quantum application.
    \item \textbf{Quantum micro-Architecture:} A quantum chip will behave differently than any classical chip as the overall behavior of a quantum chip is completely different as the focus in on bringing the quantum logic to the qubits. We will describe our current understanding of the quantum computer architecture as far as the classical architecture is involved. We just have to be realising that a quantum accelerator will always have a quantum part and a classical part. 
    
    \item  \textbf{Quantum Routing and Mapping:} When we are using a large number of (perfect or physical) qubits, we need to take care of the communication overhead and routing of qubits. The main factor playing in this context is the nearest-neighbour constraint that exists to execute quantum gates that involves at least two or more qubits. Currently, there is the constraint that these qubits need to be close to each other, which involves the mapping and routing of the qubits in a way that is consistent with the physical space that is assigned to the qubits. Given that the quantum gates will be different and evolving over the quantum execution of the algorithm, there will be a mapping and routing objective at any moment in time.  Depending on the quality state of the quantum technology, the qubits will be arranged in specific topologies with more or less constraints. Evidently, we can reason in terms of perfect qubits or physical qubits. The scheduling, placement, and routing of the quantum circuits is part of the software stack and might at some point be executed by an operation system and no longer by the application logic.
    
    \item \textbf{\QX:}  The lowest layer is the quantum simulator platform on which the tests will take place of any accelerator logic that will be developed in this project. The lowest layer contains the simulator that DUT developed and that can execute the cQASM generated by the OpenQL-compiler.  There is no hardware that needs to be developed but the \QX  can execute both realistic as well as perfect qubits.
    
\end{itemize}

In the context of quantum accelerator development, the same full-stack approach is adopted for  perfect qubits. 
The execution can be either on an experimental quantum chip or on the \QX  simulator. 
The highest level starts at the end-user application for which a part of that application is developed in a quantum language, such as OpenQL.  The quantum part of any industrial or societal application can be executed on any kind of available quantum prototype.
For any quantum logic that is specified, a specific and target-related micro-architecture needs to be defined and used.  We present the considerations for the various layers in this section.
Besides gate-based quantum computing approach, we also include the quantum annealer based system/simulator in Figure~\ref{fig:layers} as we currently investigate the components of all types of architectures currently in the market.

\section{Quantum Full-Stack Explained}
In this section and as shown in Figure~\ref{fig:FullStackNew}, we present the different layers that we have already developed and on which we base our future research activities. When combined together, they present a potential and executable version of the quantum full-stack that can be executed on any powerful computer platform.  We start at the highest level and introduce the necessary concepts and explanations needed to have a good starting point.

%\begin{figure}[h]
%    \centering
%    \includegraphics[width=170mm]{fig/fullStackNew.png}
%    \caption{Full stack in detail }
%    \label{fig:full_stack_in_detail}
%\end{figure}

\subsection{Quantum accelerator application}

The highest layer in the full-stack focuses on the application that needs to be developed for any organisation. On current, modern architectures, there are a large number of initiatives developed that run on either the Field Programmable Gate Array (FPGA), the Graphical Processing Unit (GPU) or the Tensor Processing Unit (TPU) as the accelerator platform. When envisioning the quantum accelerator idea, many similar topics are well suited, such as security, artificial intelligence, autonomous driving, genome sequencing, sensors and trajectories for aeroplanes and rockets. For the three application examples that we are currently developing, we assume the use of perfect qubits such that the focus can be completely given to the algorithm logic and the micro-architecture design.

Quantum algorithms \cite{stephen2011quantum} are described by high-level quantum programming languages \cite{abhari2012scaffold, wecker2014liqui}. %\cite{green2013introduction} 
Such algorithm description is agnostic to the faulty quantum hardware and assumes that both qubits and quantum operations are perfect. In the compilation layer, quantum algorithms are expressed in terms of what the community tends to call the perfect qubits. Such qubits do not have any decoherence nor any error rate when executing a quantum gate on one or more qubits.  Approaches by some of the partners in the past \cite{fu2016heterogeneous} focus on the micro-architectural control of any quantum device but in this project, we just want to focus on the software aspects.

In any modern computer, different  accelerators are needed to define diverse models of computation and provide the appropriate consistency in order to support a software ecosystem.   As we will expand in this paper, the combination of the classical with the quantum logic is needed when investigating and building a full quantum accelerator where classical logic is combined with quantum logic. The  \QPU can be compared to accelerator processors such as FPGAs, GPUs or TPUs, which we find in mostly all computers today.  As we said before, the quantum accelerators are in the pre-transistor phase as it is not clear what quantum technology will dominate.

\textbf{Applicability and scalability -} All quantum software developments  focus on applicability and scalability.  \footnote{Performance is already mentioned several times but is more difficult to asses at this moment.}  To ensure the industrial relevance and applicability of the software solutions developed in different application domains will develop, implement, and test quantum algorithms relevant to their field. We briefly discuss  several application domains such as bioinformatics, aerospace, finance, the automotive industry and material science.  

\begin{itemize}
    
    \item \textbf{Bioinformatics -} For modern medicine, all medical decisions will be based on a thorough DNA analysis of the patient. The fields of genomics and proteonomics generate immense volumes of data that need to be processed for semantic understanding down the application pipeline. Quantum algorithms in the field of bioinformatics has been one of the major focus areas for near-term applicability via optimisation and learning approaches.
    
    \item \textbf{Aerospace -} The aerospace industry has already embraced quantum technology. Airbus has initiated quantum-oriented aerospace research and issued the Airbus Quantum Computing Challenge in 2019. Companies such as Airbus and Boeing are already looking at \projname to investigate what the impact can be of this technology on the development of new airplanes.
    
    \item \textbf{Finance - } An increasingly important service to any population world-wide is based on the money they have earned in their life and would like to invest in financial products, anywhere in the world.  One of those services is portfolio management that decides what financial products to buy or sell. These financial products are in most of the cases shares of companies on a stock market. With the fast economic development of countries like China and India, the financial markets will get bigger and accessible for anybody in the world. Examples can be found in \cite{BAAQUIE_2003}.
    
    \item \textbf{Automotive -} The automotive industry is facing major technological challenges, one of which being autonomous driving. Large amounts of data need to be used to train vehicles to take the right decisions in known and unknown scenarios. The emergence of new cars such as Tesla make the classical car manufacturers increasingly aware that the electrical and autonomous cars have a real future.  Quantum computing will aid the training, as well as the prediction on the edge. Another very interesting line of research is the development of new batteries, for cars but also for many other fields.
    
    \item \textbf{Chemistry and Material Science -} Quantum computing holds great promise for (ab-initio) chemistry and material-science applications. Many newly started companies are looking at quantum-enhanced computational-chemistry solutions, which it aims to bring to market once (near) fault-tolerant quantum-processors of reasonable size (100+ qubits) are commercially available. %However, that may still be about 5 to years away.
    
\end{itemize}

Many more examples can be given, very similar to how many accelerator applications have already been developed for the mainstream classical hardware.%\textbf{QX-simulator and experimental quantum devices - } 
An important message that we want to bring with this paper is the fact that we will be testing the quantum applications on very powerful, classical computers.  In order to assess the performance of any quantum accelerated algorithms, it is essential to have as many qubits as possible. Not only does this allow the computation of relevant problem sizes, but also the scaling of the algorithms can be assessed. It should not be forgotten that when combining 3 qubits, we have a superposition and new qubit of $2^3$ number of states. If we combine 20 qubits, the superpositioned qubit will have $2^{20}$ states. For each state, we need to represent the amplitude which consists of a floating point number and an imaginary  number combined with a floating point. This is the main explanation of why we need a supercomputer as these machines have very large internal memory such that we can represent the state vectors for all the qubits that we use in the application.
Therefore most developments will be implemented on a quantum simulator with perfect qubits. A quantum simulator is a reliable and robust test platform, contrary to existing quantum devices.  The quantum simulator used within this paper is the \QX simulator that the QCA-team developed. The assumption of perfect qubits increases the number of usable qubits.

Following Figure \ref{fig:FullStackNew}, we will describe in the following sections the different abstraction layers on which this project is focusing its research and contributions, starting at the application layer and going down to execution on a simulator platform. Given the potential of quantum acceleration, this top-down approach is necessary to understand how investing in the development of quantum computing has the potential to become a world-wide technology that can be used by every country, organisation or individual. 
%In section \ref{3stack}, we will present the three accelerators in more detail.

\subsection{Quantum Library}

For this section, we always consider perfect qubits.
The highest level is the application layer where a potential end-user of the quantum compute power instructs what exactly needs to be computed.
Quantum computing promises to become a computational game changer, allowing the calculation of various algorithms much faster (in some cases exponentially faster) than their classical counterparts.
Especially, applications requiring manipulation of a large set of data items to produce a statistical answer are very suitable to be processed by quantum computers, which we call in this paper quantum accelerators.
Currently, there is no generally acknowledged or accepted functional domain where quantum technology would be the game changer.
Potential promising domains include physical system simulation, cryptography and machine learning.
Evidently, the cryptography domain is a clear candidate as algorithms such as Shor's factorisation showed that potentially a quantum computer can break any RSA-based encryption, as it leads to finding the prime factors of the public key~\cite{shor1994algorithms} based on which the private key can be easily calculated.
However, the cryptography domain has actively establishing a new research theme, namely the post-quantum cryptography such that the attacks emerging from such a compute power can be rebuffed.

Another potential application area is the biological domain where chemistry, medication and pharmacology belong to.
We focus on one such candidate application, namely genome sequence reconstruction.
For instance, quantum computational power would be imperative if we aim to compute the DNA-profile of every human being in the world, which takes around one week on a large network of very powerful servers for one person's DNA.
With the availability of enough qubit capacity, the entire parallel input data-set can be evolved simultaneously as a superposition of a wave function.\footnote{By our estimate, given the size of the human genome and currently available sequencers, the number of qubits required will be around 150 logical qubits.}
This particular property makes it possible to perform the computation of the entire data-set in parallel.  
This kind of computational acceleration provides a promising approach to address the computational challenges of DNA analysis algorithms. 
The essence of accelerating sequence reconstruction is the ability to run parallel search operations on the short reads obtained from sequencing an individual DNA from a sequencing machine, onto an already available reference of the organism. 
In recent years, GPU, FPGA and cluster computing frameworks like Hadoop and Spark have been used to reduce the total run-time. 
Potentially, quantum computation offers a fundamentally different way to address the enormous volume of data by employing superposition of reads in the search process, thereby reducing the memory requirement maybe even exponentially.
The quantum search primitive (Grover's search) itself is provably optimal~\cite{zalka1999grover} over any other classical or quantum unstructured search algorithm.
The rather modest quadratic speedup in cycles however becomes extremely relevant for industrial application due to the total CPU run-time involved in the big data manipulation (in order of 1000s of CPU hours~\cite{houtgast2018hardware} for a single human DNA sequence reconstruction).

% where a difference is made between character-based as well as sequence-based correlation analysis. Sequence-based correlation analysis is based on performing Fourier transformations on e.g. the DNA-sequences to be analysed. However, quantum computers are known to be able to speed up the computation of Fourier transformations using superposition. The well-known Shor-factoring algorithm uses a Quantum Fourier Transformation  operation, which has a complexity of $\mathcal{O}(N^2)$ as compared to $\mathcal{O}(N2^N))$  for classical implementations, where $N$ is the length of the sequence. This represents a significant reduction in complexity~\cite{lin2014shor}. Going back to genome sequencing, the QFT approach can be used to calculate the similarity between two different DNA sequences by transforming them into the Fourier domain. Depending on the coding system of the sequences, such analysis requires from four to eight Fourier transformation operations, making Fourier transformation a bottleneck in the calculation. But further research is being performed in that domain so that we have experimental validation of this approach.
%Quantum Fourier Transformation would allow to perform such analysis efficiently and in a scalable way.   As will be described in the next section, an important part of any quantum algorithm is the language in which it is described. 

\subsection{Programming language, compiler and run-time support}

The quantum algorithms and applications presented in the previous section can be described using a high-level programming language such as Q\#~\cite{svore2018q}, Scaffold~\cite{ abhari2012scaffold}, QisKit~\cite{ibmqiskit},  Quipper~\cite{ green2013introduction} or OpenQL~\cite{khammassi18} and compiled into a series of instructions that belong to the (quantum) instruction set architecture.  Many of other languages that are available in the world are mostly developed for supporting the quantum physical experiments such as Forest~\cite{rigetti}, CirQ~\cite{google}, Strawberry Fields~\cite{xanadu}, XACC~\cite{open1} and Ocean Software~\cite{dwave}.  Interesting to mention is that Qiskit can compiler for  very gate oriented (Terra) code up to very application oriented (Aqua) implementations. 
% As shown in Figure \ref{fig:compiler}, the compiler infrastructure for such a heterogeneous system will consist of the classical or host compiler combined with a quantum compiler. %that generates the Quantum Assembly language QASM, which is increasingly used in both experimental as well as simulation settings, and that can be executed on the quantum device. It is clear that there will be one binary file produced consisting of, in our case, two distinct phases : the classical with the quantum logic.

\begin{figure}[bt]
\centering
\includegraphics[width=0.75\textwidth]{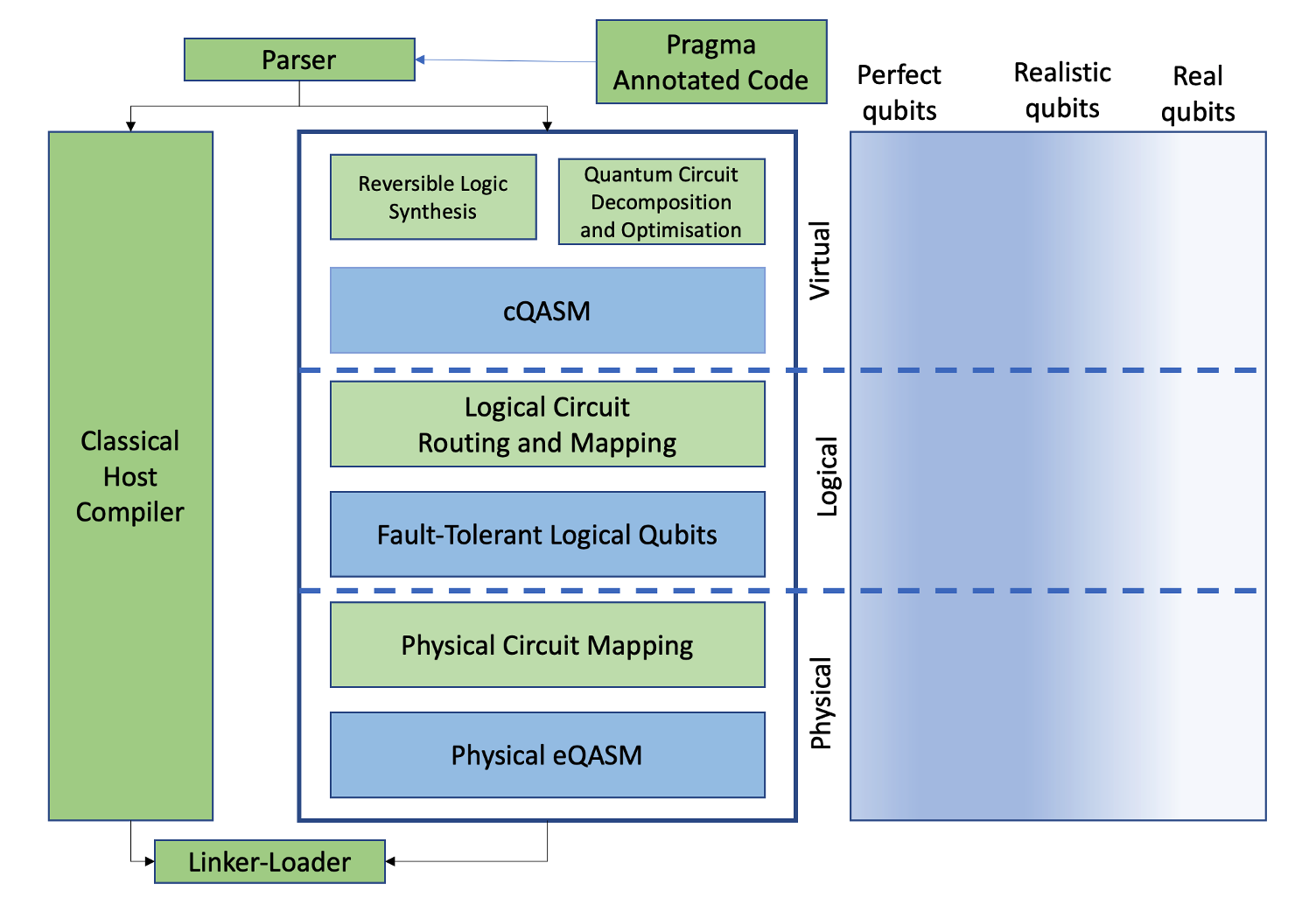}
\caption{Compiler infrastructure}
\label{fig:compiler}
\end{figure}
 
Consistent with our distinction between perfect, realistic and real qubits, the compiler is capable of adapting to the requirements of the end-user.
So there is an option that translates the qubits in perfect, realistic or real manner.
As shown in Figure~\ref{fig:compiler}, the compiler infrastructure for such a heterogeneous system consists of the classical compiler for the host processor combined with the quantum compiler.
It is important to note that the architectural heterogeneity where classical processors are combined with different accelerators such as the quantum accelerator, imposes a specific compiler structure where each compiler part can target the different instruction sets and ultimately generates one binary file which can be executed on different instruction set architectures.
For the computer architecture envisioned in our research, any high-level implementation of the system application will consist of two interleaved types of logic: the classical logic which will be executed by the micro-architecture of the controlling processor and the quantum logic which will be mapped onto the quantum processor.
The quantum logic can be encapsulated by classical language structures such as decision and loop constructs.
The micro-architecture extracts the quantum part and send it to the quantum processor.
 
As we adopt the quantum circuit model as a computational model, the quantum compiler translates the quantum logic into quantum circuits for which reversible circuit design, quantum gate decomposition and circuit mapping are needed.
The output of this compiler is a series of instructions, expressed in a quantum assembly language, such as cQASM, that belongs to the defined instruction set architecture. \footnote{QASM is one candidate for such a language and was originally produced by Nielsen and Chuang to generate the \LaTeX~figures for the quantum circuits for their book.} The definition of a shared quantum assembly language is a key challenge such that there is uniformity in the algorithmic descriptions of different research groups. 

There are two interesting new initiatives. The first is the QisKit software platform developed by IBM and to which our language can be connected. The second  interesting development is that  work is being done on the development of adequate operating systems. A recent introduction was from the UK company Riverlane which announced the DeltaFlow operating system.\footnote{More information can be found on https://www.riverlane.com/products.} It is important that these tools are evaluated and assessed in simulated or real settings.

      \textbf{Perfect qubits:} The compiler can also target the use of perfect qubits. As defined above, that implies that these qubits live as long as they are needed and have principally no error-rates in the quantum gates that are executed. Depending on the state of the execution platform, connectivity constraints can be imposed for mapping and routing. When we generate everything in terms of perfect qubits, that also implies that there is no separation anymore between logical and physical qubits as there is no requirement for error coding.  
%\end{enumerate}

\subsection{Tensor Mathematics}
Wikipedia gives a very good introduction to Tensor Mathematics  as multidimensional array.\footnote{ https://en.wikipedia.org/wiki/Tensor} We give an example taken from D. Desurvire's book on the mathematics behind quantum computing. \cite{desurvire2009}\footnote{This section is heavily based  on chapter 16 of the book of Desurvire. Any error in the text is our responsibility.} We give some examples based on relatively small circuits but useful in many applications.  We will conclude this section by presenting the no-cloning theorem which is an important property and even limitation of quantum qubits and computing.  In the following sections, we will always assume that the quantum states are in a pure state that corresponds to a vector in a Hilbert space. The eigenvalues of the operator correspond to the possible values of the quantum state.

\subsubsection{Tensor Products}
For single-qubit gate operations, we always used tensor products to obtain the result. Let us assume two qubits $|a>$ and $|b>$ that we can multiply with each other. The tensor product is written as follows  $ |a>\tens |b>=|a>|b> $. where $\tens$ stands for the tensor-product operation.  Assuming the base $V={|0>,|1>}$, we can formulate the 4-dimensional extended based for 2-qubits $V\tens V$ as follows:
\[
%V\tens V= (|0>,|1>) \tens (|0>,|1>) 
V\tens V= (\ket{0},\ket{1}) \tens (\ket{0},\ket{1})
\]
\[
= (\ket{0}\ket{0}, \ket{0}\ket{1},\ket{1}\ket{0},\ket{1}\ket{1})
\]
\[
= (\ket{0} \tens \ket{0}, \ket{0} \tens \ket{1},\ket{1} \tens \ket{0},\ket{1} \tens \ket{1})
\]

In this extended base,  $\ket{a} \tens \ket{b}$ is represented by four coordinates $(u_1,u_2,u_3,u_4) $. Let us formulate the qubits as follows: $\ket{a}=a_0\ket{0}+a_1\ket{1}$ and $\ket{v}=b_0\ket{0}+b_1\ket{1}$ such that we obtain the qubit multiplication as follows:
\[
 {\ket{a} \tens \ket{b}} = (a_0\ket{0}+a_1\ket{1}) \tens (b_0\ket{0}+b_1\ket{1})
\]

\[
= (a_0b_0\ket{0} \tens \ket{0}+ a_0b_1 \ket{0} \tens \ket{1} + a_1b_0\ket{1} \tens \ket{0} +a_1b_1\ket{1} \tens \ket{1})
\]

This translates $(u_1,u_2,u_3,u_4) $ in $(a_0b_0,a_0b_1,a_1b_0,a_1b_1)$. We applied the following rules for the above computations:
\[
 {\lambda(\ket{x} \tens \ket{y})} = \lambda \ket{x} \tens \ket{y} =\ket{x}\tens \lambda \ket{y} 
\]

\[
 \ket{x} \tens (\ket{y}+\ket{z})  = \ket{x} \tens \ket{y}+ ~\ket{x}\tens \ket{z} 
\]

%\ket{\alpha}\ket{\beta_1}\ket{a}\ket{b}\ket{}\ket{|w_1}
Those rules can be applied to any combination of $nD$ and $mD$ spaces with
the following computation bases $V=(\ket{\alpha_1},\ket{\alpha_2},...,\ket{\alpha_n>}$ and $W=(\ket{\beta_1},\ket{\beta_2},...,\ket{\beta_m})$. So if we have qubit $\ket{a}=\sum^n_{i=1} a_i\ket{\alpha_i}$ and $|b>=\sum^m_{j=1} b_j\ket{\beta_j}$ then we have $\ket{a}\tens \ket{b}=\sum^n_{i=1} \sum^m_{j=1}a_ib_j \ket{\alpha_i} \tens \ket{\beta_j}$

When we tensor a qubit $\ket{a}$ n-times with itself we can note it as $\ket{a}^{\tens n}$, written n-times as $\ket{a}^{\tens n}=\ket{a}\tens\ket{a}\tens ...\ket{a}\tens$.

Now, we introduce the notion of linear operators. If we have a linear operator A defined for the $|a>$-space and the operator B for the $\ket{b}$-space, we can have the operator tensor product $A\tens B$ that can be applied to the tensor states $\ket{a}\tens\ket{b}$ using the distribution rule $A\tens B(|a>\tens|b>)=A\ket{a} \tens B\ket{b}$.  A tensor operator A tensored with itself n-times will be formulated as $A^{\tens n}$. The operator tensor product $A\tens B$ has the distributive properties of conjugation, transposition and Hermitian conjugation as follows: 
\[
 (A\tens B)^*=A^* \tens B^*
\]
\[
 (A\tens B)^T=A^T \tens B^T
\]
\[
 (A\tens B)^+=A^+ \tens B^+
\]

The next question is how to derive the matrix of the tensor operator $A\tens B $ ? To this purpose, we compute the Kronecker product as follows.  Let us assume that A is represented by a $n x m$ matrix with coefficients $A_{ij}$ with i=1,2,...,n and j=1,2,...,m. By definition,

\[
A\tens B=\begin{bmatrix}
A_{11}B & ...& A_{1m}B\\
...\\
A_{n1}B & ...& A_{nm}B
\end{bmatrix} 
\]

Here are some illustrative examples for this Kronecker product.
%\ket{\alpha}\ket{\beta_1}\ket{a}\ket{b}
\begin{itemize}
    \item For qubits $|a>$ and $|b>$ in bases $V=(|\alpha_1,|\alpha_2)$ and $W=(|\beta_1,|\beta_2,|\beta_3)$. The single column matrices are $A_{i1}=a_i$ and $B_{i1}=b_i$, for which we obtain the following. %page 329
    \[
\ket{a}\tens \ket{b}=\begin{bmatrix}
a_{1}\ket{b} \\
a_{2}\ket{b}
\end{bmatrix} 
\]
\begin{equation} \label{eq1}
%\begin{split}
  \begin{bmatrix}
a_{1} & \begin{bmatrix}
b_1\\
b_2\\
b_3
\end{bmatrix}\\
a_{2} & \begin{bmatrix}
b_1\\
b_2\\
b_3
\end{bmatrix}
\end{bmatrix}= \\ 
\begin{bmatrix}
a_{1} & b_1\\
a_1&b_2\\
a_1&b_3\\
a_{2} & b_1\\
a_2&b_2\\
a_2&b_3

\end{bmatrix} 
%\end{split}
\end{equation}

Pauli gates are single quantum gates and in general we distinguish between the X, Y and Z gates which can be combined with each other or other quantum gates.
If we consider the tensor product $X\tens Y$ of the two Pauli matrices X, Y, we obtain the following:

%\begin{itemize}

 \begin{equation} \label{eq2}
%\begin{split}
X \tens Y %&=&
  \begin{bmatrix}
0 * Y & 1 * Y \\
1 * Y & 0 * Y\\
\end{bmatrix} %&=&  
\begin{bmatrix}
0 & Y\\
Y&0\\
\end{bmatrix} %&=&
\begin{bmatrix}
0 & 0&0&-i\\
0 & 0&i&0\\
0 & -i&0&0\\
i & 0&0&0\\
\end{bmatrix}
%\end{split}
\end{equation}
%\end{itemize}
%\ket{a}\ket{}
    \item $H^{\tens n}$ is the n-tensored operator provided  where H is the Hadamard gate. The Hadamard gate creates a superposition between two or more qubits where each qubit has the same amplitude. The assumed computational base is $V^n=(\ket{0},\ket{1})^n = \ket{a}$ where $\ket{a}$ symbolises any of the n-qubits base elements generated by the n-tensor product $\ket{a}=\ket{v_1}\tens \ket{v_2}\tens ...\tens \ket{v_n}$ where $v_i=0$ or $1$.  A general property can be formulated which is 
    
 \[
H^{\tens n}\ket{a}=\frac{1}{\sqrt{2^n}}{ \sum_{V^n}(-1)^{a*b}\ket{b} }
\]   
\end{itemize}
%\ket{}
\begin{figure}[hbt]
\centering
\includegraphics[width=0.5\textwidth]{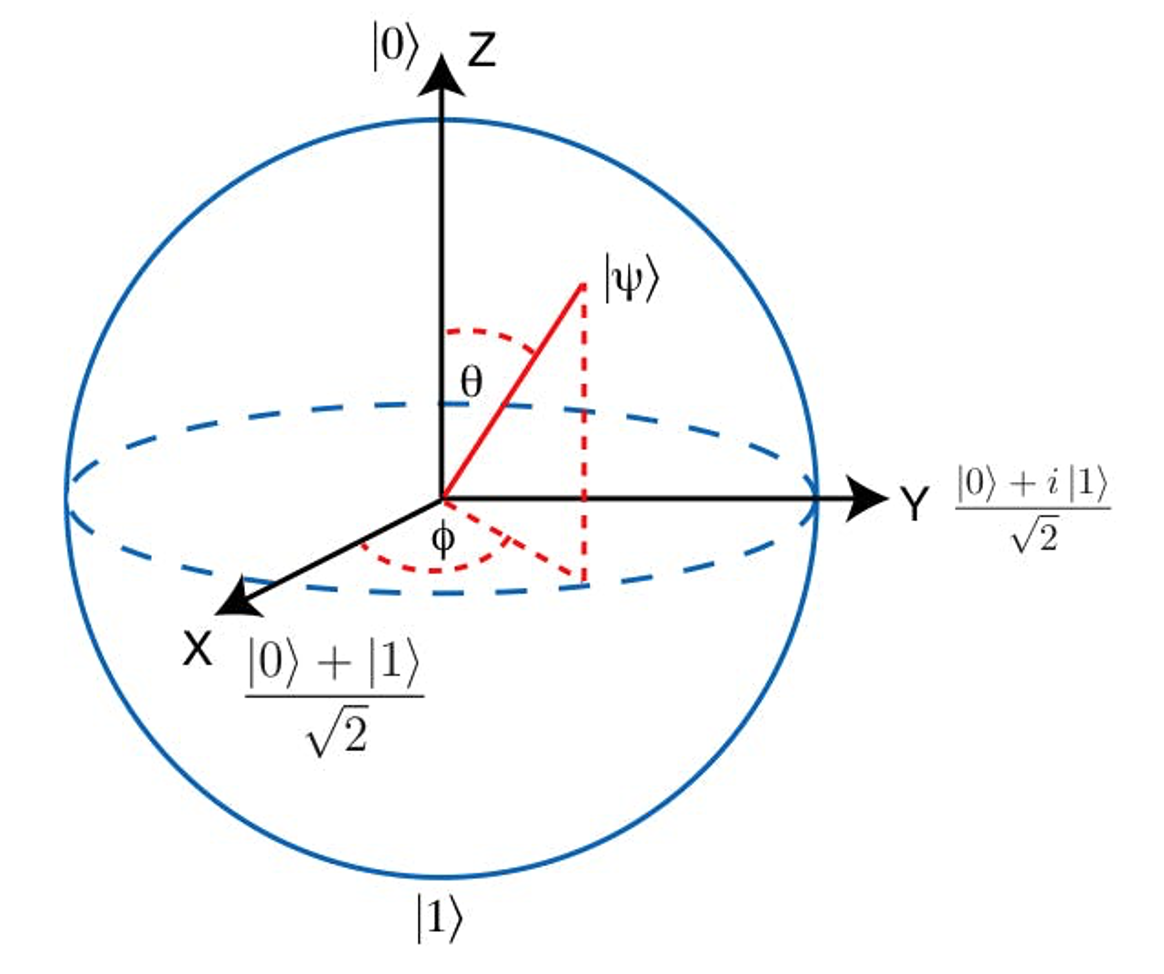}
\caption{The Bloch Sphere for the Quantum Gates}
\label{fig:bloch}
\end{figure}

where $\ket{b}=\ket{w_1}\tens \ket{w_2}\tens ...\ket{w_n} $ is any base element of $V^n$ and $a*b$ is a scalar :
 \[
a*b=v_1w_1+v_2w_2+...+v_nw_n=\sum_{i=1}^n v_iw_i
\]   

\subsubsection{Noncloning theorem}
 %\ket{b} \ket{}
An important property of qubits is the impossibility to clone the state of a qubit, called the noncloning theorem. So the question is whether a qubit $|\psi>$, which is in any state and running in system A can be duplicated into a second qubit $|s>$ in any pure state and running in system B ?  Theoretically, such a cloning operation would correspond to the following formula:
\[
\ket{\psi}\tens|s> \xrightarrow{} U(\ket{\psi}\tens \ket{s})=\ket{\psi}\tens\ket{\psi}\equiv\ket{\psi\psi}
\]   
where U is a unitary tensor operator.  We can assume that that an operator U exists and can be applied to any state $\ket{\psi}$ of state A.  If the qubit $\ket{\phi}$ is one of the states of A such that $\ket{\psi}\neq \ket{\phi}$. We should be able to duplicate it into state B according to 
\[
U(\ket{\phi}\tens|s>) =\ket{\phi}\tens \ket{\phi}\equiv\ket{\phi,\phi}
\]
When applying the linearity of the transformation, for any state mixture $\ket{\chi}=\lambda\ket{\psi}+\mu\ket{\phi}$ of A where $\lambda$ and $\mu$ are complex numbers such that
%\ket{}
\begin{equation} \label{eq4}
%\begin{split}
U(\ket{\chi}\tens \ket{s}) =\ket{\chi}\tens \ket{\chi}\equiv\ket{\chi,\chi} \\
%\end{split}
\end{equation}
If we develop the left-hand side of equation \ref{eq4}, we obtain the following
%\[
%U(|\chi>\tens|s> )=U(\lambda|\psi> + \mu  |\phi>)\tens |x>
%\]

\begin{equation} \label{eq5}
%\begin{split}
U(\ket{\chi}\tens \ket{s} )=U(\lambda \ket{\psi} + \mu \ket{\phi} )\tens \ket{x} 
=\lambda \ket{\psi}\tens\ket{\psi}+\mu\ket{\phi}\tens\ket{\phi} \equiv\lambda \ket{\psi}\ket{\psi,\psi}+\mu\ket{\phi,\phi}
%\end{split}
\end{equation}
Having developed the left-hand side of equation \ref{eq4}, we now have to develop the right-hand side of that same equation. 
%\ket{}
%\begin{equation} \label{eq6}
\[
\ket{\chi}\tens \ket{\chi} =(\lambda\ket{\psi} + \mu \ket{\phi} )\tens (\lambda \ket{\psi} + \mu \ket{\phi})
\]
\[
=\lambda^2\ket{\psi} \tens \ket{\psi}+ \mu \lambda (\ket{\psi}\tens \ket{\phi} +\ket{\phi}\tens \ket{\psi} )+\lambda^2\ket{\phi}\ket{\phi}
\]
\begin{equation} \label{eq6}
\equiv \lambda^2|\psi,\psi>+\mu\lambda|\psi \phi>+\mu\lambda|\phi \psi>+\mu^2|\phi \phi>
\end{equation}
If we merge the results in equations \ref{eq5} and \ref{eq6}, we obtain the following expression
\begin{equation} \label{eq7}
\lambda(\lambda-1)|\psi \psi>+\mu\lambda|\psi \phi>+\mu\lambda|\phi \psi>+\mu(\mu-1)|\phi \phi>\equiv 0
\end{equation}

If we assume that $|\psi>$ and $|\phi>$ are pure states, that implies that equation \ref{eq7} assumes that $\mu\lambda=0$ as the qubits are orthogonal to each other. This leads to $|\chi>=|\psi>$ or $|\chi>=|\phi>$.  The meaning of this observation is that if an operator U exists that can clone two pure states $|\psi>,|\phi>$, the operator U cannot clone any of their mixtures 
$|\chi>=\lambda|\psi>+\mu|\phi>$. This is a very restrictive conclusion.  

Concluding this section and as stated in the beginning of this section, the no-cloning theorem is normally stated and proven for pure states. Wikipedia states that the no-cloning theorem shows that it is impossible to create an identical copy of an arbitrary unknown quantum state. \footnote{https://en.wikipedia.org/wiki/No-cloning_theorem.} It was generalized to mixed states and is called the no-broadcast theorem.

It was shown also in \cite{desurvire2009} that one conclusion stays valid, which we quote here. "There always exists a unitary operator U capable of cloning a pure state $|\psi$ or any pair of pure states $|\psi>$ and $|\phi>$... It is not possible to clone quantum states in the general case, known as the noncloning theorem."

\subsubsection{Quantum logic for a small quantum circuit}
With this example, we simply show how the supercomputers can be used to make an exact computation of the quantum algorithm, based on the value of the amplitudes rather than any classical probability. The amplitudes can become negative as well as positive and in the last step, we translate the amplitude in a classical probability which allows the developer to assess the accuracy of the qubit.
\begin{figure}[hbt]
\centering
\includegraphics[width=0.7\textwidth]{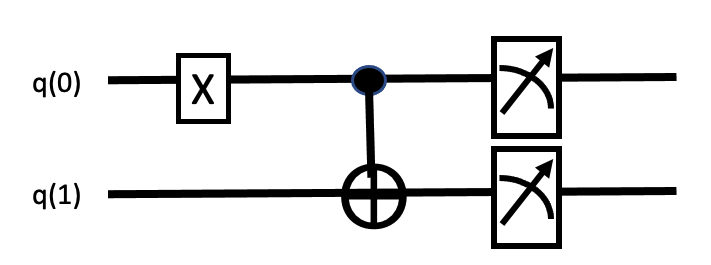}
\caption{The OpenQL Small Quantum Circuit}
\label{fig:qmicrodna}
\end{figure}

At the top of Figure~\ref{fig:qmicrodna}, a small circuit is shown that has 2 qubits in entanglement after the two quantum gates are applied. The first gate is the Hadamard gate on $q(0)$ and following is the CNOT-gate where $q(0)$ is the control qubit. The circuit closes with a double measurement of qubits $q(0)$ and $q(1)$.  We briefly describe how that circuit is translated in a mathematical form and finally in the cQASM version that we shown on the right part of the figure.
\begin{itemize}
     \item Quantum Logic and cQASM version: The quantum circuit shown at the top of the Figure can be translated in the cQASM version.  The last kernel listed there is the measurement of the two qubits so we will have the two qubits combined, leading to 4 qubit states with each a separate amplitude. That overview is shown in the last column in the Figure.
   % \begin{lstlisting}
%        version 1.0
%        qubits 2
%        .min_kernel
%            prep_z q[0]
%            prep_z q[1]
%            display
%            h q[0]
%            display
%            cnot q[0],q[1]
%            display

%        .kernel_measure
%            measure q[0]
%            measure q[1]
%            display
%    \end{lstlisting}
    \item Tensor Mathematics and  Hadamard gate: the formula for the Hadamard gate has two versions, one for the $|0>$ and one for the $|1>$.
    \[     H q(0) = \frac{|0>+|1>}{\sqrt(2)} \] or \[ H q(1) = \frac{|0>-|1>}{\sqrt(2)}\]
    
    The Hadamard gate is only performed on the $q[0]$ qubit so the state of that qubit is given in the above formula.
    \item CNOT-gate: the control-NOT-gate looks at the value of the black rounded qubit to determine if the X-gate needs to be applied on the second one.  The control qubit is $q[0]$ and the target qubit is $q[1]$. Given the two Hadamard intermediate results, that leads to the following mathematical steps.
    \begin{itemize}
        \item State $q(0)=|0>$ and $q(1)=|0>$ or $q(1)=|1>$ so the X-gate is not executed for $q(1)$ which stays in its value. So if $q(1)=|0>$ it stays $|0>$ and if it is $|1>$ that is also what it stays.
        \item State $q(0)=|1>$ and $q(1)=|0>$ or $q(1)=|1>$ so the X-gate is  executed for $q(1)$ which changes their value so if $|q(1)=|1>$ the X-gate makes it $|0>$ and if $q[1]=|1>$ then it becomes $|0>$.
        
    \end{itemize}
    The result of the short circuit is the superposition of the two qubits, resulting in  the amplitudes of 0.5 that are connected to the combination of the two qubits $q(0)$ and $q(1)$.  The exact value of the $\alpha_i$-amplitudes depends on the initial value of the amplitudes.
   So the final result of this very short quantum circuit is as follows:
   \begin{itemize}

       \item The amplitude value is 0.5 for each component of the superpositioned qubit. Computing the total probability for the qubit is  $|0.5|^2 +|0.5|^2 +|0.5|^2 +|0.5|^2 =0.25+0.25+0.25+0.25=1$ .
       \item if the first qubit is $ q(0)=|0>$ then the superpositioned qubit is $0.5|00>+0.5|01>+0.5|10>+0.5|11>)$

       \item if the first qubit is $q(1)=|1>$ then the superpositioned qubit is $0.5|00>-0.5|01>+0.5|10>-0.5|11>)$
   \end{itemize}
       
\end{itemize}

\subsubsection{Parallelising a quantum circuit}

The next step to reason about is how we can execute a quantum application on a classical hardware computer. So we split up this section in a short discussion about classical parallelisation and then we explain what the quantum logic requires.

\begin{figure}[htb] 
    \centering
    \captionsetup{justification=centering}
    \subfigure[Amdahl's law]
    { 
        \centering
        \includegraphics[width=.37\textwidth]{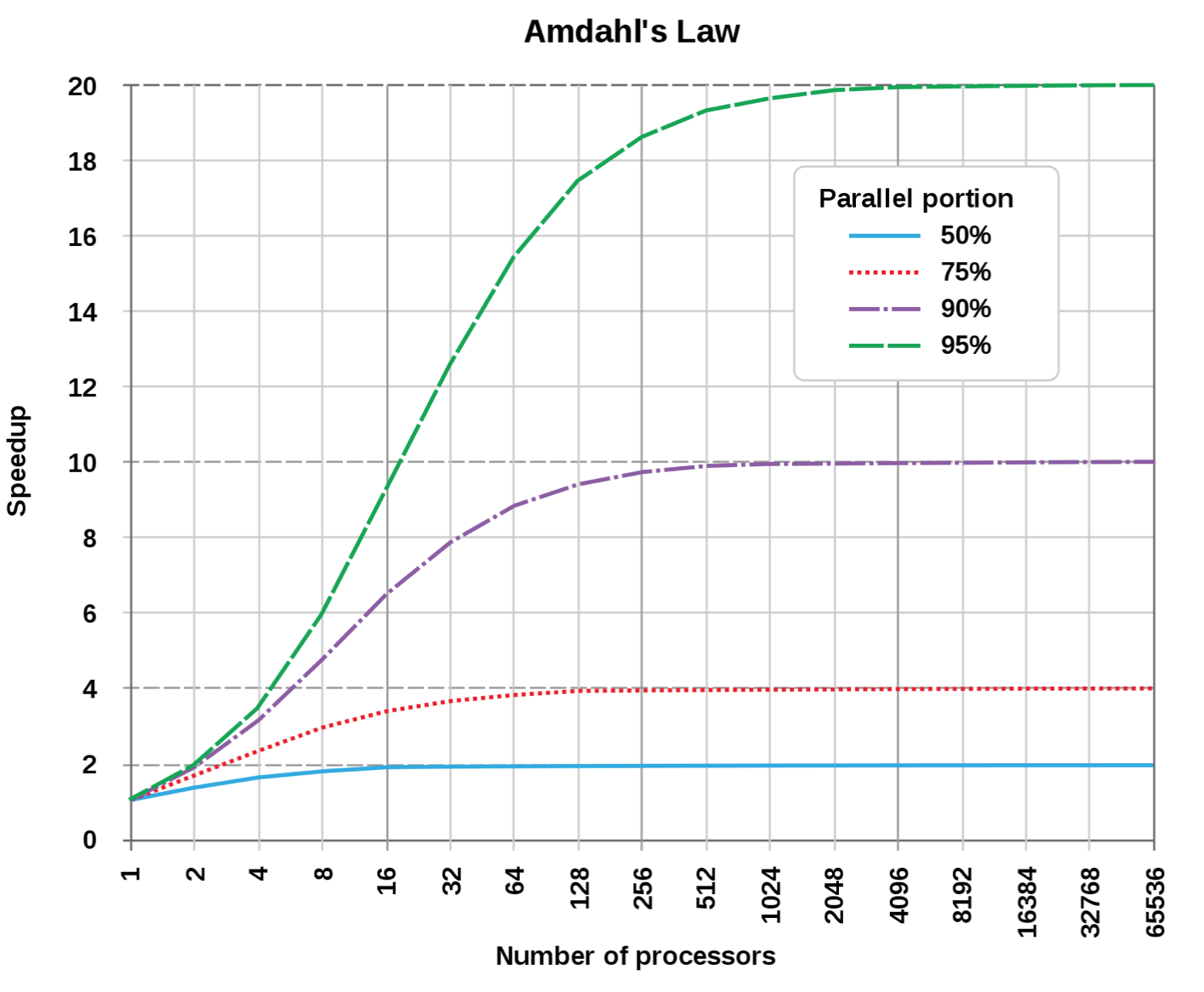}
        \label{fig:amdahl} 
    } 
    \subfigure[Gustafson-Barsis law]
    { 
        \centering
        \includegraphics[width=.42\textwidth]{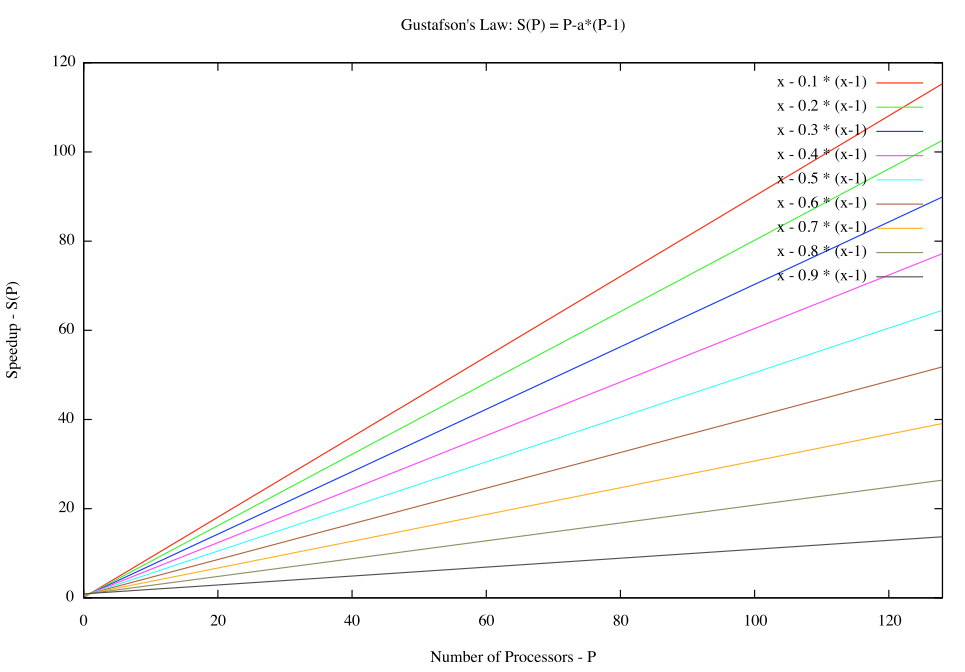}
        \label{fig:gustafson}
    }
    \caption{The Amdahl and Gustafson-Barsis law} 
    \label{fig:parallel}
\end{figure}
\begin{itemize}
    \item \textbf{Amdahl and Gustafson-Barsis laws :}
There are basically two ways in which we can use the computer hardware in a parallel way to increase the overall computer power.  The simplest way is  what we do on any modern computer. Today, we can buy and use co-processors that execute part of the application in a faster way. As shown in Figure \ref{fig:parallel}, there are two important laws in the parallelisation, namely Amdahl and Gustafson-Barsis law.  The first law that is used in parallelisation is Amdahl's law. It shows what kind of speedup one can expect when a smaller or larger part of the application can be parallelised. One sees that if 50\% of the application is parallelised, the maximum speedup is 2.  If we can parallelise up to 95\% of the code  the speedup can reach up to 20 times faster than the pure serial execution. A second observation from that figure is that the number of processors applied for a parallel execution has a very limited impact. The smaller the parallel part of the application, the smaller the number of processors is that it has a positive impact. That speedup goes from 16 processors in parallel for only 50\% of a parallelised application up to 4096 processors for the 95\% parallelisation. This immediately puts a major constraint on the number of processors that we can use for any parallel execution.
The second graph that is shown is Gustafson-Barsis law. There the logic is similar to Amdahl's law but there is one difference.  There is no horizontal part of the graph as the number of available processors will increase but rather a linear graph that keeps on rising with the number of processors that can be used. There is a percentage of the application that is not parallelised  which has a negative impact on the overall speedup. 

\item \textbf{Classical parallelisation :} After having explained Amdahl and Gustafson-Barsis laws, we can now go into detail on each of them. Amdahl's equation that can be used is 
\[S_{latency}(s)= \frac{1}{(1-p)+\frac{p}{s}} \] where 
\begin{itemize}
    \item \textbf{ $S_{latency}$} is the theoretical speedup of the whole application.
    \item \textbf{$s$} is the speedup of the part of the task that will execute in parallel on the hardware.
    \item \textbf{$p$} is the part of the execution time where there is no parallelisation.
\end{itemize}

If we take as an example  that about 30\% of the total execution time will be the subject of the speedup, \textbf{$p$} will be 0.3.\footnote{https://en.wikipedia.org/wiki/Parallel_computing} Parallelising the code implies that this part of the application will run twice as fast, giving the value 2 to \textbf{$s$}.  Filling those numbers in the equation, we obtain \[S_{latency}(s)= \frac{1}{(1-0.3)+\frac{0.3}{2}} =1.18\]
If we assume that the parallelisation can be increased by accelerating multiple parts of the application in the following way: $p_1 = 0.11, p_2 = 0.18, p_3 = 0.23,$ and $p_4 = 0.48$. The acceleration of the different components can be assumed as follows: $s_1$ no speedup so that value is 1, $s_2=5$, $s_3=20$ times faster and the last component will accelerate 1.6 times so $s_4=1.6$. This leads to the following computation.
\[S_{latency}(s)= \frac{1}{\frac{p_1}{s_1}+\frac{p_2}{s_2}+\frac{p_3}{s_3}+\frac{p_4}{s_4}}= \frac{1}{\frac{0.11}{1}+\frac{0.18}{5}+\frac{0.23}{20}+\frac{0.48}{1.6}} =2.19\]

This example shows that even when certain parts are substantially accelerated,  like the second and the third part, the overall acceleration is much lower because the first does not accelerate and the fourth part accelerates 1.6 times.

An additional extension on the parallelisation effort is given by the computer scientist Gustafson and Barsis, called the Gustafson-Barsis law. \footnote{https://en.wikipedia.org/wiki/Gustafson\%27s_law}  The main idea is that the speedup $S$ is realised by P processors for an application where not all components can run in parallel. The equation is   \[S_{G-latency}(s)= P +s (P-1)\]. The variable s represents the percentage of the application which is not parallelised. 
The main logic behind the Gustafson-Barsis law is that programmers will always use the available hardware resources as the technology will evolve. That means that as more resources are available, the acceleration of the application will always be bigger. In Amdahl's law logic, the size of the application is fixed and will not really evolve when the technology improves.

\item \textbf{Quantum parallelisation :}
We can be relatively short about the quantum parallelisation. There is a parallel execution when  quantum circuits run on a physical quantum  device.  There is an implicit way of parallel executing  the circuit, which also defines that we need multiple runs of the same circuit on the same quantum chip to have an overview of all possible outcomes of the circuit. We know that the measurement of a physical qubit chip is after multiple runs of the  circuit and the solution with the highest frequency of measurements is the correct output of the quantum circuit.  As explained in the EdX lesson by Vazirani from Berkeley University,  there are many unknown things in quantum computing.  Just like in the example shown in this section, there is a mathematical way to describe what those gates are and what their effect is. Taking again the Hadamard gate as an example, it is clear that when one applies a Hadamard gate on the $|0>$ qubit, it results in the double qubit state.  The two qubits that exist are the $|0>$ and the $|1>$ qubit with the amplitude ${\frac{1}{ \sqrt(2)}}$. From then on, any additional quantum single or multiple qubit gates, will be split around the direction either $|0>$ or the $|1>$ qubit.  In the quantum computing world, there is no fundamental understanding of this property and behaviour but that is how a quantum computation is done. 

\begin{figure}[hbt]
\centering
\includegraphics[width=0.7\textwidth]{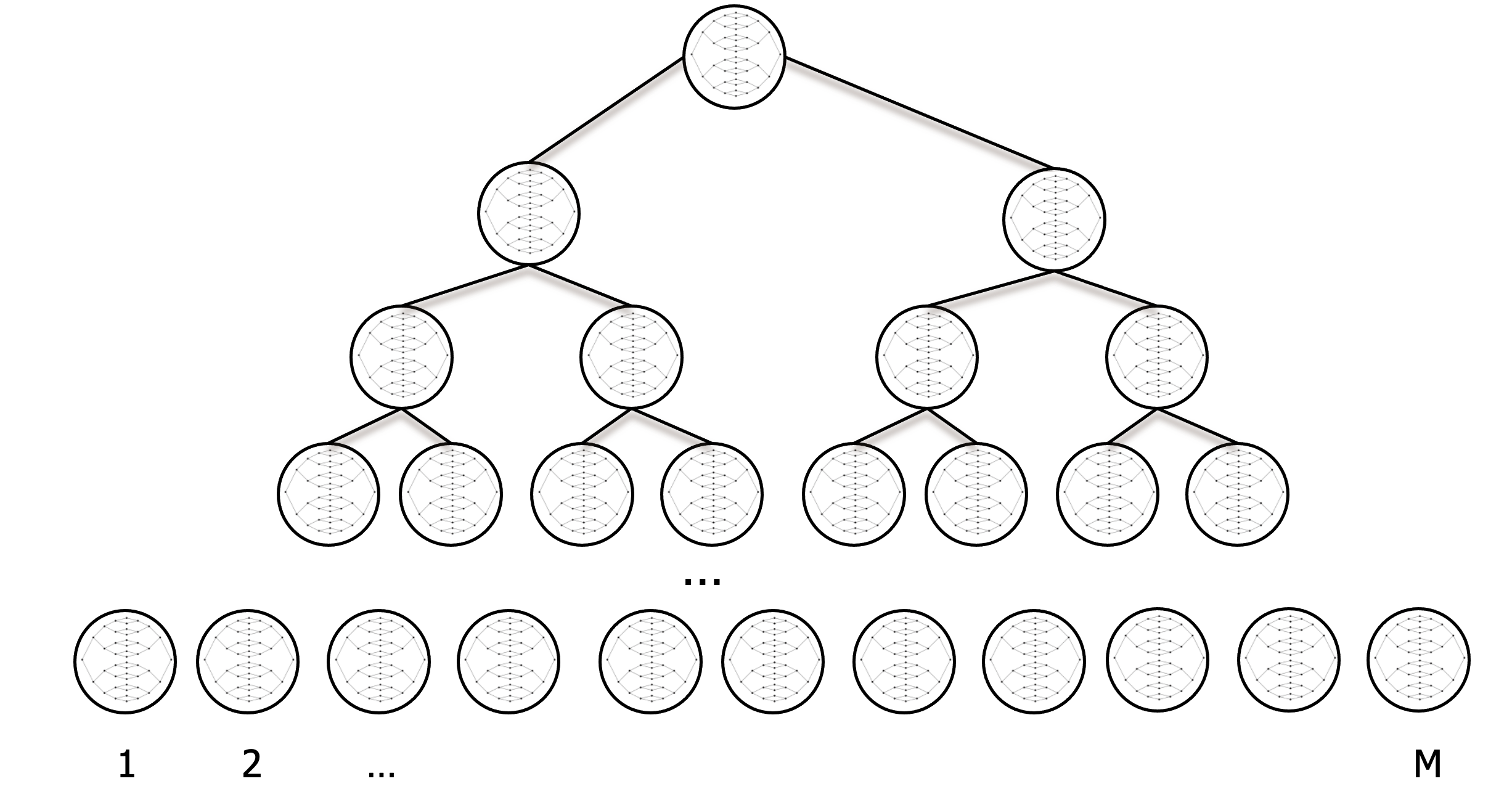}
\caption{Classical and quantum parallelisation}
\label{fig:qparallel}
\end{figure}

\item \textbf{Classical quantum parallelisation}
 which is just our way of describing what we need to do with any quantum algorithm when we execute it on a classical computer.   When running it on a classical computer we do not have this implicit parallelism but we need to have an explicit parallel version of the circuit that is executed on the classical computer.
The final step is to focus on the parallelisation of the quantum circuit. As we showed in the previous section, the parallelisation is implicit in the execution on any quantum device. However, as we do not have any good number of good qubits, the use of these properties are still very far away in the future. As we presented and will present in the paper, one of the useful alternatives is to use very powerful computers to run the quantum algorithms.  As we made clear up to now, there is no reason why we should not run the quantum algorithm in a parallel way on the available computers.  When we look at Figure~\ref{fig:qparallel}, we see a graphical presentation of how a quantum algorithm can also be parallelised where the lowest level can contain up to M different outcomes, representing all the possible outcomes of the parallel algorithm. A positive effect also on the classical programming and executing it on very parallel hardware.

As explained above, parallelising software is very difficult topic and there is no compiler that will automatically parallelise any software code in a scalable and correct way.  The main implication therefore is that when we look at quantum logic, we need to do the effort ourselves to have a parallel version. The graphical presentation in Figure~\ref{fig:qparallel} shows that every node in the graph in this case is a quantum graphical presentation of the gates that need to be executed. It is not a representative graph but at least it shows in whatever branch of the quantum logic, there will be a repetition of the quantum gates but each time on a different set of qubits. It is enough to look at the exercise that we presented in this section, that even for a very small circuit, the end results are quite different.  One of the challenges in the classic-quantum parallelisation is the effort to specify all the different branches in the parallel tree that represents the application. However, even in classical programming this stays a very big problem which is not easy to solve.  So any research done on these aspects, will also have a positive effect on classical software running on very powerful supercomputers.

\end{itemize}

\subsection{Quantum Software Testing}
Quantum programs involve high-dimensional quantum search spaces. Even in
classical programs, for example, the overall potential inputs to test a 
program are often infinite, and optimisation techniques are frequently 
used to find the best test inputs~\cite{harman2012search}. 

Such complexity is significantly higher in quantum programs, owing to quantum
principles. Thus, to optimise quantum software testing and debugging
techniques (qV\&V, short for quantum verification and validation), the research
community needs novel quantum search and optimisation
techniques that can efficiently search in quantum search spaces. At this point, 
works on qV\&V are uncommon yet~\cite{miranskyy2019testing,gomesaoff}, let alone 
those on the 
optimisation of qV\&V techniques. Most works are about quantum-inspired
meta-heuristics~\cite{karmakar2017use}, which run on classical computers, 
and they are thus limited 
to the computational power of classical computers. This line of work is 
however in its infancy, and breakthroughs are needed to defining quantum 
search operators embedded in, for example, Ising
models~\cite{moessner2001ising,chakrabarti2008quantum}, 
data representations of tentative solutions in qubits, and
mutation/crossover/neighbour operations as mathematical functions. All this needs 
to be done in the context of perfect,  realistic or real qubits so that we are 
able to adapt to current technologies or the requirements of the end-user.

Faults found in quantum programs with qV\&V solutions must be located, isolated, 
and patched; therefore, debugging is needed. To our knowledge, there are currently 
no (even limited) debugging techniques or development environments to end users. 
Hence, debugging is difficult and cumbersome. The development of effective debugging
solutions must overcome the following challenges: (1) difficulty examining values of
quantum variables in superposition; (2) even when observations or simulations are
available, quantum states are generally high dimensional and difficult to interpret,
limiting their usefulness in debugging; and (3) no evidence or guidelines exist for
where and what to check when debugging quantum programs. 

In order to have debugging tool-sets widely adopted, one needs to address the
following challenges~\cite{huang2019statistical}: (i) developed novel ways to 
debug, such as the combination of
simulation and real quantum computation to infer to find the distribution of states
inside a quantum program using statistical assertions; (ii) better ways to show
expected and real behaviour, and (iii) leveraging higher-level constructs to
effectively debug quantum programs. Additionally, the debugging techniques will
include novel quantum fault isolation and patching techniques at the appropriate level of abstraction. Again, as for testing, this needs to be done in the context 
of perfect,  realistic or real qubits so that we are able to adapt to current 
technologies or the requirements of the end-user.

 \subsection{Quantum Operating System}
 Many companies have launched a quantum operating system to the market. The goal is not to bring a complete overview as many things still need to evolve. We just focus on one example as it may have interesting features to look at.  One of the first quantum operating systems was Deltaflow.OS, developed  by Riverlane. They assume the hardware is FPGA based and has quantum functionality. The OS they developed allows to connect different FPGA nodes which can then be used to simulate the quantum processor. The emulation of the quantum hardware can then be used to run any kind of quantum application on the hardware.
     \begin{figure}[H]%[hbt]
\centering
\includegraphics[width=13cm, height=8cm]{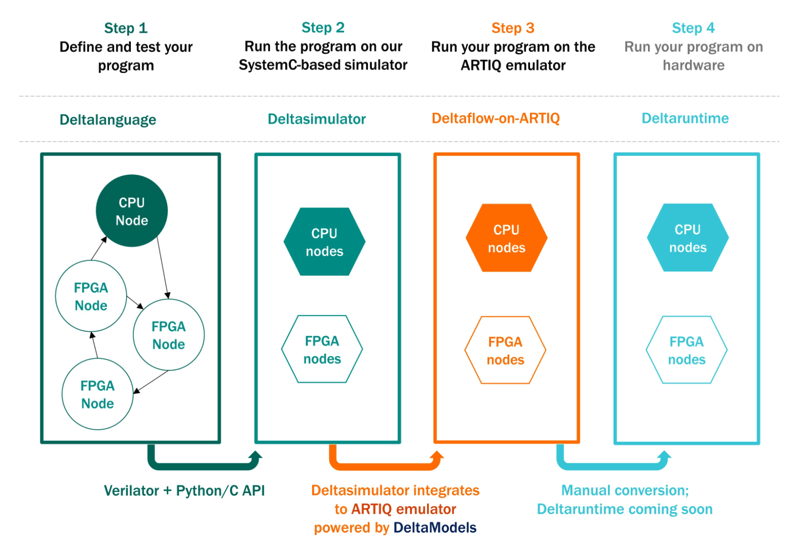} %, angle=90
\caption{\label{fig:qosArtiq}Deltaflow.OS operating system (www.riverlane.com/products/)}
\end{figure}

The goal of the Deltaflow.OS  is to outperform any qubit hardware that can be developed. It clearly focuses on the emulation program but it shows there is potential for improving the OS in the next couple of years.  Similar to heterogeneous architectures, the main feature of Deltaflow.OS  is that all stack layers accessible, including CPU, FPGAs and qubits.  It is not clear how they qubits are developed but the fact that one can combine heterogeneous components is very powerful.
In that way, operations  increase the overall performance compared.

\subsection{Quantum Micro-Architecture}
\begin{figure}[bt]%[hbt]
\centering
\includegraphics[width=15cm, height=10cm]{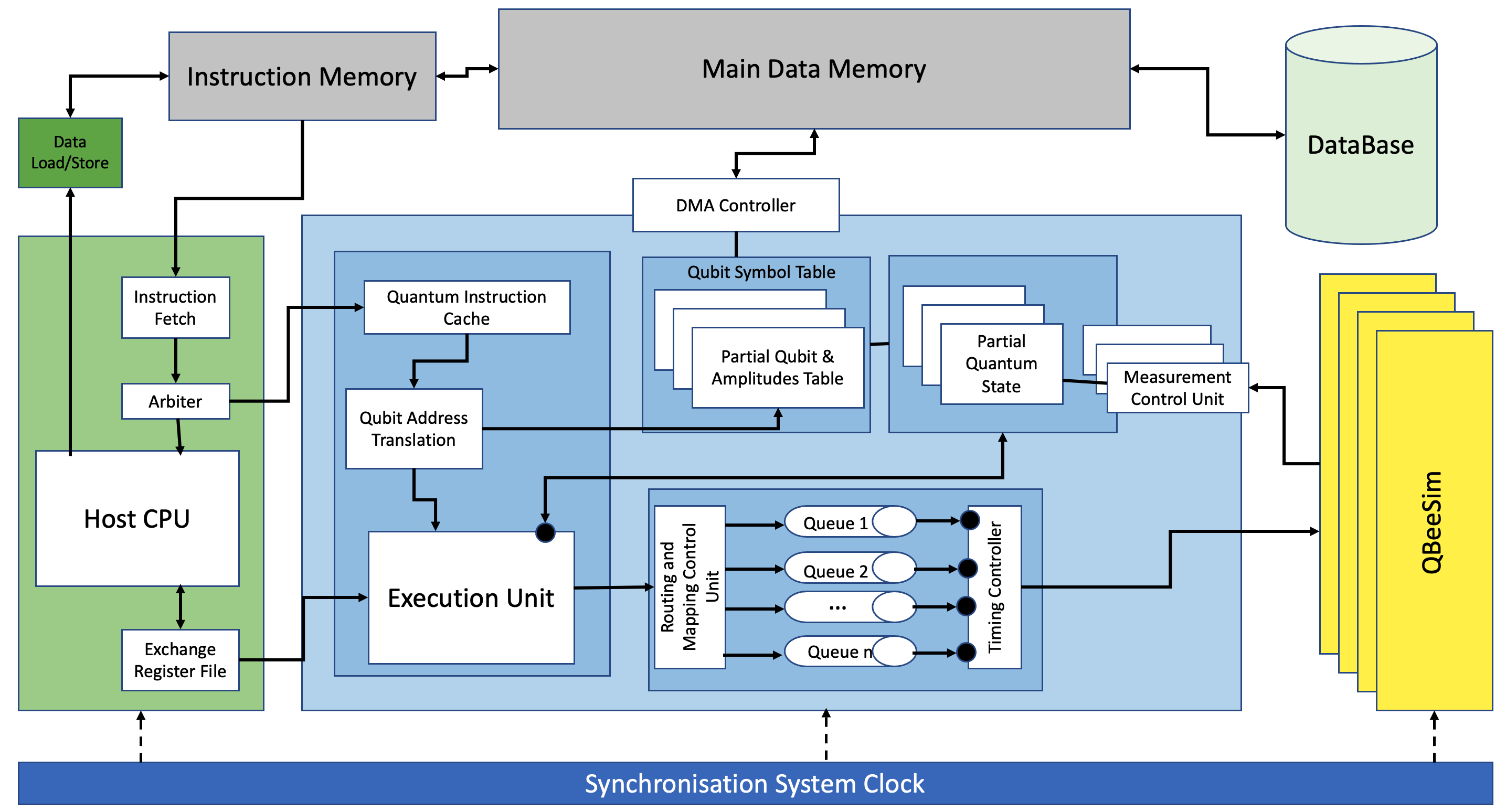} %, angle=90
\caption{\label{fig:microarchitecture}An generic Example of a Quantum Accelerator Micro-Architecture}
\end{figure}

Even though it sounds late in the overall logic of this paper but in this part we are describing the micro-architecture. It is important to realise that any future quantum computer will always have a classical and therefore a digital micro-architecture and a quantum component that will behave according to the analogue logic.  Even if, at some point in the future, we will develop a real, experimental quantum accelerator, any application that we execute on a modern computer with a quantum accelerator, will have classical logic that drives the overall application which will also have certain tasks that can be delegated to the quantum accelerator. That implies that any quantum accelerator, as it can be defined right now, will always consist of digital hardware and control blocks that will send quantum gate instructions to a quantum chip capable of executing them. As shown in Figure~\ref{fig:microarchitecture} and as just explained, the largest block in blue represent the entire quantum accelerator processor that consists of the digital, control quantum processor and the quantum chip that executes the quantum instructions. What is important to understand regarding the execution of the quantum logic,  there currently is a quantum simulator that execute that part of the quantum logic. In our case, the simulator is called the \QX-simulator. We will explain later in this paper that  the execution of any large application with quantum parts will be executed on modern supercomputers.  That way, we will emulate and simulate the execution of the quantum logic on a classical computer.

Any application that will be executed on a modern computer will have several parts that need to be executed on the main processor but there will be parts that are assigned and executed on a particular co-processor.  That implies that the application can be written in several languages and that different parts interact with each other by using the same main memory and apply the dedicated circuits on the co-processor.  To this purpose, any kind of modern processor has a particular architecture capable of executing any sequence of the instructions, irrespective in what original language they were written.
This also holds for the quantum processor, which also has a series of instructions that it can execute, some of which are classical logic and others are the quantum instructions that will be executed on the quantum chip.
So the quantum accelerator will consist of two components: the classical and digital micro-architecture part that has a classical processor to execute part of the accelerator logic and the quantum chip that contains the qubits that need to be executed in an analogue way.

Essential to any kind of computational device is the presence of one or multiple computer architectures that are responsible for executing the instructions that are delegated to the co-processor.
The architecture of a machine connects the physical hardware to the applications that can run (on that hardware) and dictates how instructions are executed. 
This is also true for the case of a quantum accelerator.
For the quantum algorithms to be understood by the quantum accelerator, a low level representation of the quantum instructions is required that the classical control hardware of the quantum chip can understand.
This is known as the Quantum Instruction Set Architecture (QISA).
The content of the QISA can be modified for each accelerator logic that needs to be implemented.
% Extensions to the compiler may therefore be needed but the micro-architecture will need hardware components that will execute the instructions that are sent to it.
% We want to be very precise in how the instructions are formulated and executed.
One example of a micro-architecture is given in Figure~\ref{fig:microarchitecture}.
For any micro-architecture, there are a number of properties that we have to estimate, such as the appropriate instruction-length, pipeline depth (for parallel quantum gates) and targeting multiple control channels per single instruction.  Based on these principles, the basic blocks are constructed, such as timing control unit and the microcode instruction set of the overall micro-architecture.  We describe the components of the micro-architecture according to the execution of the different instructions.  All instructions are expressed in cQASM.
\begin{itemize}
    \item \textbf{I-Instruction Memory:} In the main memory of the classical computer, the instructions are stored and  expressed, either in a classical programming language such as Java or C++ but it can also be expressed in OpenQL and cQASM. The openQL-compiler will always output the cQASM version of the algorithm as that language can be understood and executed by the quantum hardware. The arbiter determines whether the instructions need to be sent to the classical processor or whether it should be sent to the quantum processor.
    \item \textbf{II-Quantum Instruction Cache:} The instructions are decoded by the quantum processor. This means that it is determined what qubits are used in the instruction and where they are physically located in the quantum chip. That information is stored in the Qubit Symbol table where the name of the qubit is linked to a topographical location where it resides and what the amplitude values are. The decoding also identifies what the quantum gates are that need to be executed. If there is some independent parallel execution possible, it is also identified at this stage. 
    \item \textbf{III-Execution Unit:} The actual and if possible parallel execution of the cQASM-instructions is done by the Execution Unit in the micro-architecture. At that moment, the cQASM code has been changed in view of the physical addresses listed in the Qubit Symbol table.  The quantum code contains also the different logical operators such as the branching and loop instructions. Whenever results from the classical processor are needed, they can be loaded in the Execution unit, coming from the Exchange Register File of the classical processor.
    %There is a Pauli Arbiter that looks at the instructions that need to be executed and that are based on the Pauli gates. There is no need to execute the Pauli gates on the quantum chip. The qubits as well as the operation that will be applied on the Qubit is saved in the Pauli Frame Unit. Whenever a non-Pauli gate needs to be executed on a particular qubit, that information will be first collected from the qubit table in the Pauli-Frame-Unit and will then be sent to the quantum chip.
    
    \item \textbf{IV-Routing and Mapping Control Unit:} Once we reach the execution unit of the micro-architecture, there is the need to route the two-qubit instructions as well as the qubit states to locations close to each other and at the same time respecting all the timing requirements that are needed. This step is needed to deal with larger number of qubits that are needed for the quantum logic and we still assume that the physical distance should be as short as possible imposing the need for qubits to route to a location such that the two qubits for which a quantum gate needs to be performed, are close to each other.  The result of this step is a reformulation of the different cQASM-instructions such that the routing steps needed each qubit are inserted. This will be further explained in part \ref{mapping}. 
    \item \textbf{V-Queues:} The modified cQASM-instructions are now put in different queues with a particular time-stamp.  The timing is not that important when we execute on the \QX-simulator but understanding the overall timing conditions needed for quantum algorithms is important.  Just like in classical computers, that language that allows hardware components to interact and talk to other hardware parts. In this step, the cQASM is translated in a series of low-level micro-code instructions that will be executed by the different hardware blocks in the Wave-Control-Unit. That unit consists of a series of queues where the micro-code versions are inserted with a time stamp that defines the exact moment at which they need to be executed.
    \item \textbf{VI-CMOS and \QX:} The component that we neglect for now is the cryoCMOS component that allows the CMOS-components to be operational at very low temperatures such as 4-10 Kelvin.
    \item \textbf{VII-\QX:} A major component of our quantum accelerator is the \QX simulator that is capable of executing the cQASM-code. The \QX simulator receives the cQASM instructions and also has a local memory where all the qubits are stored that will be manipulated in the quantum instructions.  %This will be further explained in part \ref{qx}.
    \item \textbf{VIII-Measurement:} When one wants to know what the result of the quantum algorithm is, the only way to find that out is looking at the Qubit Symbol Table. All the amplitudes are listed there and the measurement simply means that you update the qubit symbol table with the update version of the last run of the quantum algorithm.  
    \item \textbf{IX-Qubit Symbol Table:} Every result that is represented by a vector or matrix will be stored in the Qubit Symbol Table. The content of that table is a reflection of whatever quantum memory the quantum device will have in the future but it is too early to specify that. In the classical quantum accelerator that information is stored in the classical memory where the qubit name, type and composition is stored. The composition is consisting mostly of the different amplitudes that represent the result of the computation up to that moment. At the end of the circuit, the amplitudes of the active qubits represent the quantum result.
    \item \textbf{X-DMA controller:} whenever qubit information needs to be written to the classical memory, the DMA controller is responsible for transferring that. The DMA will also be used when previously computed classical or quantum results need to be loaded to the quantum accelerator.

\end{itemize}

\subsection{Routing and Mapping of quantum circuits}
\label{mapping}

An important feature is the routing of the computation logic on the different qubits on the quantum device or simulator where constraints such as \NN needs to be respected.  Quantum Computing is a good example of in-memory computing. This implies that we have to investigate the effect of mapping of the qubits on the target topology. A challenge in modern computers is to route the data and the logic to the right processor, coming from very large and highly distributed memories. Quantum Accelerators face a similar challenge but it is different as the operation is executed on in-memory locations, implying that the logic is ported to and applied directly on the qubit. Even though the \QX simulator does not impose restrictions such as the \NNN it is more logical to assume that this limitation also applies in simulation. This is why this topic is one of the concepts of the research and we have to investigate what the influence is of the \NN on the overall performance of the software algorithm.

Mapping of quantum circuits is considered in two different contexts: the first is when applied on small real quantum processors and the second one targets a simulation engine that addresses larger number of qubits. Depending on the target technology, we can either take into account large number of qubits or stay at a small scale and closer to the experimental state-of-the-art.  But as explained earlier in this paper, we reason in terms of perfect qubits. That implies that we abstract away from any quantum technology that will become accepted as the dominant way to make physical qubits. However, we still need to reason about moving qubits around, irrespective of what technology is used to implement them.  The example that we briefly present, called Surface Code, will appear again in the quantum computing world but it takes time until the quality has substantially improved.\cite{preskill2018}
Surface Code is a very popular logical qubit mechanism but one needs to have up to 49 physical qubits in a specific mapping to have one logical qubit.  When working with simulated processors, we can assume that the qubits are physical, logical or perfect. So the assumption that the qubits are perfect does not alleviate the nearest-neighbour constraint for those qubits. Placing highly interacting qubits close to each other could help to reduce the communication overhead and then the circuit latency.  A similar challenge needs to be researched for perfect qubits where there can also be \NN-requirement. One can choose to abstract that \NN away but it needs to be addressed at some point.
Finally, not all qubits can be placed in the necessary adjacent positions. Therefore, some of them will have to be moved or routed for which the compiler will insert a MOVE-operation for the run-time routing logic. When targeting a real or simulated quantum processor, the mapping of circuits is an important topic as described in~\cite{lin2015paqcs,dousti12min}. The circuit description of the algorithms does not usually consider a physical location of the qubits and assumes that any kind of interaction between qubits is possible. However, even perfect qubits need to be placed on a specific  qubit layout such that interactions between qubits can be performed.
    \begin{figure}[t]
    \centering
    \includegraphics[width=10cm, height=10cm]{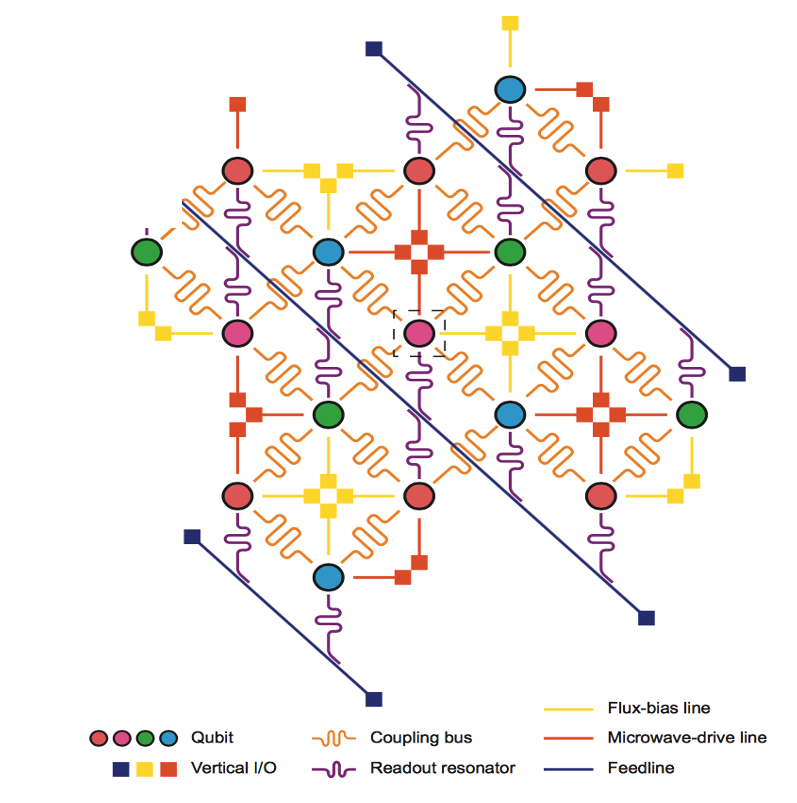}
    \caption{Surface Code Logical Qubit \cite{versluis2017scalable}}
    \label{SC}
    \end{figure}

When the algorithmic behaviour and content is not yet defined, which is the case in most of the situations, it is important to be able to use perfect qubits that are more reliable and predictable than the experimental ones, as that have no decoherence and execute reliably the quantum gates of the quantum circuit.

\subsubsection{Mapping and Routing Challenges}
A quantum algorithm can be described as as quantum circuit that consists of a set of qubits and gates (single-qubit and multi-qubit gates) operating on them.  Such a circuit is hardware agnostic and assumes, for instance, that any kind of interaction between qubits is possible. In other words, it assumes that a two-qubit gate can be performed between any pair of qubits (full qubit connectivity). However, there are some constraints coming from the quantum chip such as the nearest-neighbour interactions that must be taken into account.  The \textbf{mapping} process adapts a quantum circuit to the hosting device constraints.
Any quantum circuit will require a mapping step before being run in any quantum chip. This mapping is also called the topology step which defines the way the (perfect) qubits are placed on the surface.
Note that there are also some other constraints coming from the control electronics that limits the parallelism of the quantum operations.

After the mapping, the number of operations in the algorithm and/or the circuit depth (also called circuit latency) will increase, affecting the reliability of the algorithm (higher number of operations or higher circuit depth, results in higher error rates). Therefore, it is key to have an effective mapping that produces the minimum overhead, specially when no quantum error correction (QEC) is used. %Note that noise sets a limit of the maximum size of a computation without QEC.

%Although needed for any quantum chip layout, this topic is crucial for chips with qubits amount greater than 100 units, based on any QEC code or quantum technology.

The mapping procedure consists of the following steps. \textbf{i) Scheduling} the quantum operations in the algorithm, \textbf{ ii)} assigning an (possible optimal) \textbf{initial placement} to the qubits in the chip, and iii) \textbf{routing} the qubits that need to interact when performing a two-qubit gate.

\begin{itemize}
    \item Scheduling of quantum operations: the quantum operations are scheduled based on the gate dependencies and the chip/control electronics constraints.
    
    \item Initial placement: virtual qubits (qubits in the circuit) need to be mapped to  a physical location; that is, they have to be mapped into the qubits of the chip. One of the most common hardware constraints is the limited connectivity between qubits that is usually reduced to nearest-neighbour interactions. This means, that if non-neighbouring qubits need to interact- i.e. perform a two-qubits gate- they have to be \textbf{moved} to adjacent positions. A possible way to reduce this movement of qubits is by placing heavily interacting qubits next to each other.
    
    \item Routing of qubits: as we just mentioned qubits (quantum states) need to be moved or routed during computation due to the nearest-neighbour constraint. Different paths will need to be computed in a way that the communication overhead (extra movement operations and/or increased circuit depth) because of routing is minimum.  
    
\end{itemize}

\subsubsection {Metrics for the mapping models}
Most of the mapping models developed so far use as a cost function  for the routing algorithm  and as an output metric (after the mapping) the circuits' depth growth (or increased latency) or the amount of added gates to the circuit (gates inserted for moving qubits around). Of course, these parameters should be as small as possible. However, these metrics are not complete and do not tell anything about how the mapping affects the algorithm's reliability. Or in other words, is the algorithm still producing `good' results after mapping? Note that, as we already mentioned, the higher number of operations and/or the higher circuit depth, the higher the probability of errors.

As the mapping process increases the probability of having an error in the quantum circuit, we think that more convenient metrics would be \textbf{quantum fidelity} and \textbf{probability of success}. These metrics will tell us if an algorithm can still be successfully run after the mapping. Right now, we can obtain the probability of success and quantum fidelity by running several simulations of the same algorithm (around 1000) on a quantum computing simulator, called the \QX simulator). The idea is to investigate what are the most convenient metric(s) to be minimised by the routing algorithm and that will be also used as a metric to determine how good the mapping process is. 
Summarising, we will need to define more accurate mapping metrics and investigate how to implement them in our routing model. The metrics we will use to profile our algorithm and to evaluate how it is affected by the mapping process are described later in this paper.

\subsection{\QX}
An important and very powerful tool is the quantum simulator, called \QX , that simulates quantum circuits.
The memory requirements of simulating a quantum circuit scale exponentially with the number of qubits.
The primary contributor of this large memory requirement is the matrix of the unitary quantum gate that needs to be stored in memory.
By using a novel approach that removes the need to store the quantum gate in memory, \QX is able to simulate more qubits.
\QX converts matrices into mapping functions.
These functions map between the basis states in the computational basis, $\ket{0}$ and $\ket{1}$.
These mappings can have scaling coefficients.
Furthermore every quantum gate can be split into one type of mapping, either a one-to-one mapping or a one-to-two mapping.
Gates that perform one-to-one mappings do not create a superposition of states, whereas gates that perform a one-to-two mapping to create superposition.
An example of a gate that has one to one mapping is the $X$ gate, as it maps $\ket{0}$ to $\ket{1}$ and vice versa as shown in equation \ref{eq:xMapping}.
\begin{equation}
    \label{eq:xMapping}
    X\ket{0} \rightarrow \ket{1}
\end{equation}
An example of a gate that maps one state to two is the Hadamard gate, $H$, as shown in equation \ref{eq:hadamardMapping}.
\begin{equation}
    \label{eq:hadamardMapping}
    H\ket{0} \rightarrow \frac{1}{\sqrt{2}}\ket{0} +  \frac{1}{\sqrt{2}}\ket{1}
\end{equation}
The use of these mappings functions removes the need to store the entire quantum gate matrix.
This section will be split into the 2 parts of \QX, the allocation of memory and the operation of the quantum gates.

\subsubsection{Memory}
\QX stores the amplitudes of the qubits.
The amplitudes are complex numbers.
Therefore, \QX, stores the amplitudes as 2 \texttt{tuples}, where the first entry is the real part of amplitude and the second entry is the imaginary part.
For an $n$ qubit state there are a possible $N = 2^n$ states that can be represented, hence the number of 2 \texttt{tuples} is equal to $N$.
The total size of a qubit state vector is then given by
\begin{equation}
\label{eq:vectorMem}
    s_v = 2\cdot s_T \cdot N = 2\cdot s_T \cdot 2^n,
\end{equation}
where $s_T$ is the size of the data type that is being used to store each part of the \texttt{tuple}, which is scaled by 2 due to each complex number have 2 parts. 
\QX creates 2 arrays, one for the amplitudes of the qubits that are inputs to the gate and one for the amplitudes of the qubits that are the outputs of the gate.
Therefore the total memory requirement for \QX  is the double of $s_v$,
\begin{equation}
    m_T = 2\cdot 2\cdot s_T \cdot 2^n = s_T \cdot 2^{n+2}
\end{equation}
\begin{table}[!htp]
    \centering
    \renewcommand{\arraystretch}{1.3}
    \begin{tabular}{|c|c|c|c|c|c|c|c|c|c|c|c|c|}
    \hline
         Number of qubits ($n$)&2&3&4&5&6&7&8&9&10&11&12&13\\
         \hline
         Memory (KiB)&0.125&0.25&0.5&1&2&4&8&16&32&64&128&256\\
         \hline
         \hline
         Number of qubits ($n$)&14&15&16&17&18&19&20&21&22&23&24&25\\
         \hline
         Memory (MiB)&0.5&1&2&4&8&16&32&64&128&256&512&1024\\
         \hline
    \end{tabular}
    \caption{\centering Memory requirement for different number of qubits}
    \label{tab:memBench}
\end{table}

As an example, when storing each amplitude as a complex \texttt{double}, $s_T$ = 8 bytes, the memory requirement for \QX  is shown table \ref{tab:memBench}. 
To store the matrix of the quantum gate, a larger amount of memory will be required.
This can be calculated in the same way as $s_v$.
The matrix for the quantum gate has $2^{2n}$ 2 \texttt{tuples} (each \texttt{tuple} is a complex element in the matrix).
The memory requirement for this matrix, $s_m$, is
\begin{equation}
\label{eq:matrixMem}
    s_m = 2\cdot s_T \cdot N^2 = 2\cdot s_T \cdot 2^{2n}.
\end{equation}
The space requirement of both the qubit state vector and the quantum gate matrix is exponential, assuming a constant $s_T$. 
For the qubit state vector it is of $\mathcal{O}(2^n)$ and for the unitary matrix it is of $\mathcal{O}(2^{2n})$. 
The memory requirement for the quantum gate matrix grows much quicker than that of the qubit state vector.

\subsubsection{Gate operation}
Each of the quantum gates is implemented as a classical function, that moves around amplitudes and performs arithmetic operations on them if needed.
These classical functions are mappings between the states in the computational basis.
They are also able to map between states that are in superposition.
This implementation of the quantum gate removes the need to store a matrix in memory, thereby reducing the memory requirements of \QX  as shown in the previous section.
Each gate iterates through the $2^N$ states and performs its operations on each of these states.
Therefore each gate is called $2^N$ times every time it is applied.
%\par
\begin{figure}[hbt]
\centering
\includegraphics[width=0.7\textwidth]{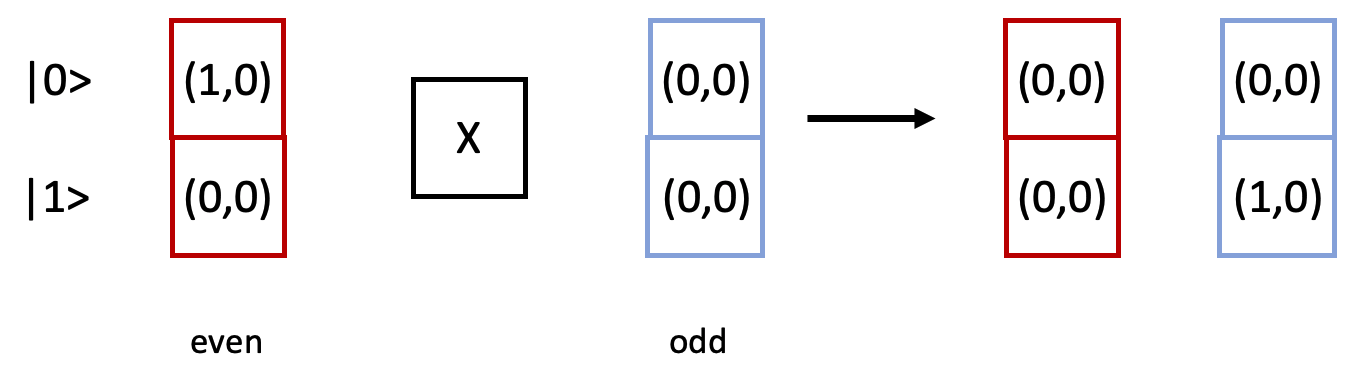} %, angle=90
\caption{The $X$-gate application}
\label{fig:gateApplication1}
\end{figure}
An example of a gate application ($X$-gate) is shown in figure \ref{fig:gateApplication1}.
There are 2 array of 2 \texttt{tuples} as discussed earlier-one for the input to the gate and one for the output.
For visual purposes the 2 arrays are marked with colours.
The role of each array switches after every gate application, i.e. the input array is toggled to the output array and vice versa.
This parity system is used so that the next gate can carry on from the output of the previous gate, without having to copy data to a fixed input array that hold the amplitudes for the input qubit vector state.
After every gate, the parity is toggled, so that \QX  can keep track of which array stores the amplitudes of the input qubit vector state.
The red blocks correspond to the array that has \textit{even} parity and the blue blocks correspond to the array that has \textit{odd} parity.
The X gate maps the amplitude of the state $\ket{0}$ to the state $\ket{1}$.

\begin{figure}[hbt]
\centering
\includegraphics[width=0.7\textwidth]{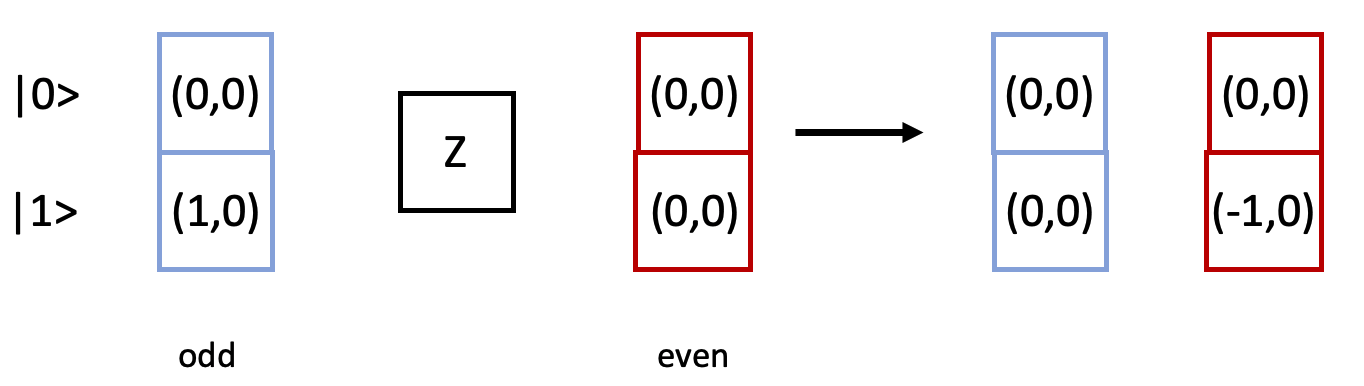} %, angle=90
\caption{The $Z$-gate application}
\label{fig:gateApplication2}
\end{figure}

After the gate is applied the \textit{even} array is reset to all 0s.
It also toggles the parity, which switches the input and output arrays.
Applying another gate, a $Z$-gate for example, is shown in figure \ref{fig:gateApplication2}.
The role of the arrays is switched as discussed earlier.

\begin{figure}[h]%[hbt]
\centering
\includegraphics[width=0.5\textwidth]{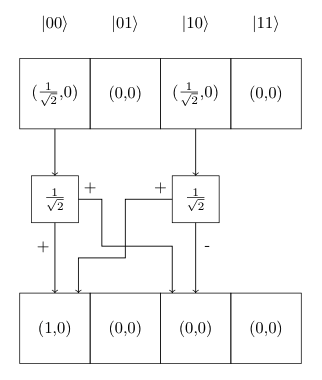} %, angle=90
\caption{The $H$-gate application}
\label{fig:hGate}
\end{figure}

The input is the \textit{odd} array this time, which was the output of the previous gate, $X$.
The output is the \textit{even} array which was reset to 0.
If a third gate were to be applied it would look like figure \ref{fig:gateApplication1}, where the input would be the \textit{even} array and the output the \textit{odd} array.
An example using the Hadamard gate will provide the final piece to understand how these quantum gates are used as mappings.
Consider the 2 qubit state $\ket{\psi}$,
\begin{equation}
    \ket{\psi} = \frac{1}{\sqrt{2}}(\ket{00} + \ket{10}).
\end{equation}
Applying the $H$-gate changes this state to
\begin{equation}
    H\ket{\psi} = \ket{00}.
\end{equation}
\QX  iterates over the $2^2 = 4$ states and performs this $H$-gate on the first qubit of every state.
If a state has no amplitude it skips that state.
The application of the Hadamard gate is shown in figure \ref{fig:hGate}.

\subsubsection{Performance}
The system specifications for the benchmarks that are performed on are given in table \ref{tab:sysSpecs}.

\begin{table}[!htp]
    \centering
    \renewcommand{\arraystretch}{1.3}
    \begin{tabular}{|c|c|}
        \hline
         Processor speed&2.6 GHz\\
         \hline
         Number of cores&4\\ 
         \hline
         L1 cache&32 KB\footnotemark\\
         \hline
         L2 cache/core&256 KB\\
         \hline
         L3 cache&6 MB\\
         \hline
         RAM & 16GB @ 1600 MHz\\
         \hline
    \end{tabular}
    \caption{\centering System specifications}
    \label{tab:sysSpecs}
    \footnotemark[1]{\small same for instruction cache and data cache}
\end{table}
\par
The number of qubits, the circuit depth and the exact gates being used are the 3 main factors that determine the execution time of a complete circuit.
For a classical simulator, the circuit depth will refer to the total number of gates, since the entire program runs sequentially.
The first item to benchmark is the execution times of each gate.
Each gate is implemented classically and to measure the execution time for each gate, 1 qubit is used for single qubit gates, 2 qubits for 2-qubit gates and 3 qubits for 3-qubit gates.
This is done to highlight the difference between the gate times and more importantly the reasons for these differences.
Table \ref{tab:singleQGateTimes} shows the gate times for the basic gates implemented in the quantum simulator.
\begin{table}[!htp]
    \centering
    \renewcommand{\arraystretch}{1.3}
    \begin{tabular}{|c|c|c|c|c|c|c|c|c|c|c|}
    \hline
         Gate&$X$&$Y$&$Z$&$H$&$R_x$&$R_y$&$R_z$&CNOT&CPHASE&Toffoli\\
         \hline
         T (ns)&298.4&422.2&308.0&464.6&490.4&448.0&521.0&363.2&690.6&503.2\\
         \hline
    \end{tabular}
    \caption{\centering Single qubit gate times}
    \label{tab:singleQGateTimes}
\end{table}
\begin{figure}[!htp]
    \centering
    \includegraphics[width=0.6\textwidth]{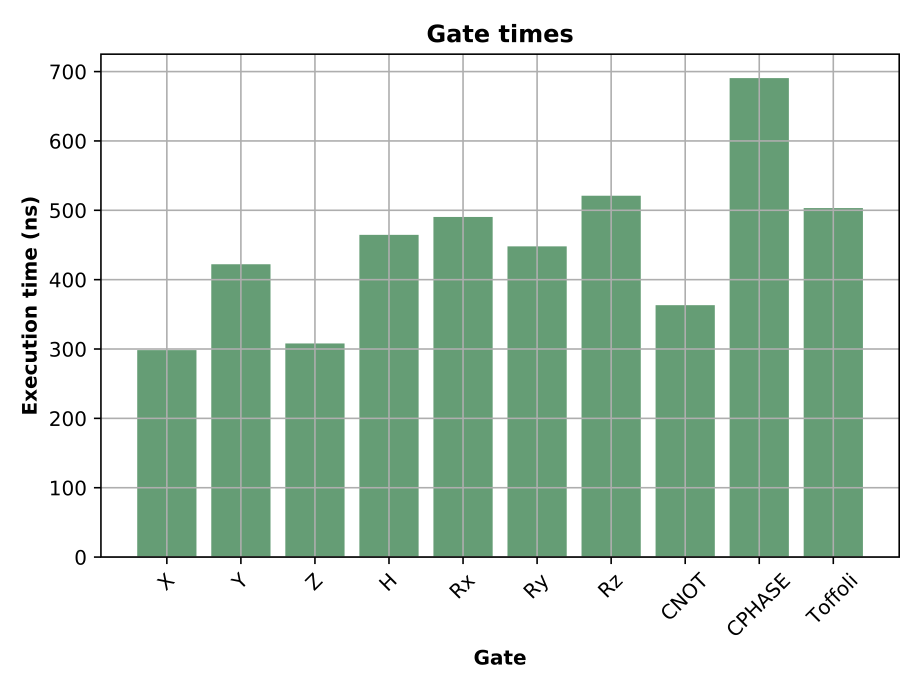}
    \caption{\centering Quantum Gate times}
    \label{fig:gateTimes}
\end{figure}
\par
The results are also shown in figure \ref{fig:gateTimes}.
The y-axis shows the execution time in nanoseconds.
The execution time for each gate is affected by 3 factors:
\begin{enumerate}
    \item The state of the qubit
    \item The number of operations performed by each gate
    \item The number of qubits
\end{enumerate}
Each of the gates are only performed on states that have a non-zero amplitude.
Therefore the execution time when performing an $X$-gate on the state $\ket{0}$ is lower than if it were performed on the state $\ket{+} = \frac{1}{\sqrt{2}}(\ket{0}+\ket{1})$.
The input state for the single qubit gates is $\ket{0}$.
The fastest gate is the $X$ gate.
This is because in the computational basis, it simply moves a value from one address in the 1D array stored in memory to another address.
The $Y$ and $Z$ gates have to check the qubit state before before applying their operations.
This makes them slower than the $X$ gate.
Furthermore both of them also need to multiply a scalar constant to the amplitude.
The $Z$ gate need only apply the phase -1, but the $Y$ gate applies a phase of $\pm i$ depending on the state of the qubit.
This makes $Y$ slower than $Z$.
The Hadamard gate is not only applying a scalar multiple of $\frac{1}{\sqrt{2}}$ to the current state, but it is also "creating" another state based on the input state.
Recall that the action of the Hadamard gate is $\ket{0} \rightarrow \frac{1}{\sqrt{2}}(\ket{0}+\ket{1})$.
It is multiplying a constant of $\frac{1}{\sqrt{2}}$ to $\ket{0}$ and "creating" a new state $\frac{1}{\sqrt{2}}\ket{1}$.
The rotations gates taken longer as they all need to compute sines and cosines of the rotation angle.
$Rz$ takes the longest as it has to multiply both the sine and cosine of the rotation angle to the states.
$Rx$ takes longer than $Ry$ as it also has has a complex term in the scalar multiple for the states.
The CNOT gate operates on 2 qubits, and does exactly what the $X$ gate does except with an additional checked on the control qubit.
The CPHASE gate has to perform a scalar multiplication by -1 based on the target qubit and that makes it slower than the CNOT gate.
The Toffolli gate acts on 3 qubits but is considerably faster than both the $Rz$ gate and the CPHASE gate as it is simply moving data and not performing and computations.
\par
\begin{figure}[!ht]
    \centering
    \includegraphics[width = 0.7\textwidth]{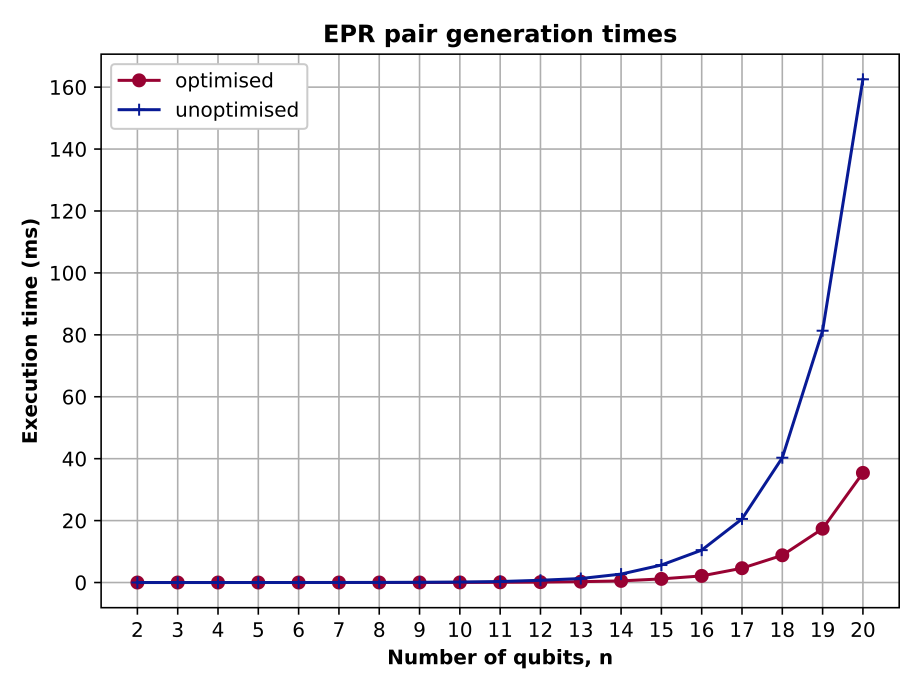}
    \caption{\centering Generating EPR pairs as the number of qubits scale}
    \label{fig:qubitsExp}
\end{figure}
The second factor that determines the execution time of a circuit is the number of qubits.
The experiment is carried out by using the same number of gates and while increasing the number of qubits.
\begin{figure}[!htp]
    \centering
    \includegraphics[width=0.7\textwidth]{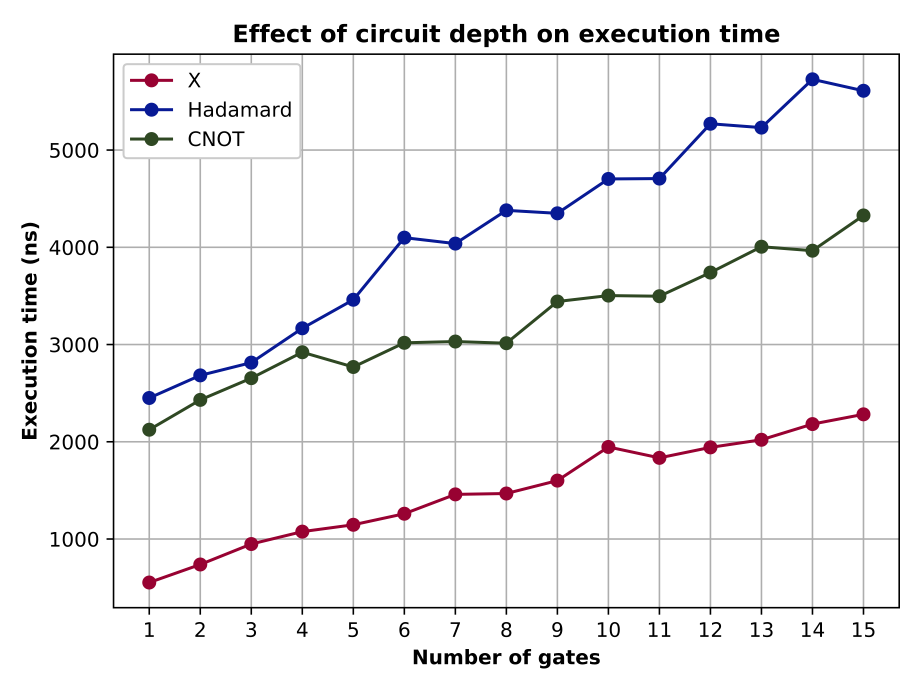}
    \caption{\centering Effect of number of gates on the execution time}
    \label{fig:depthBench}
\end{figure}
The goal of this experiment is to investigate whether gates acting on larger arrays result in longer execution times.
This experiment uses a simple EPR pair generation circuit.
It uses an unoptimised version of the simulator and an optimised version.
The optimisation is the skipping of states that have 0 amplitude, hereinafter referred to as zero-state skipping.
Every gate skips states that have no amplitude, as there is no reason to waste computation time on these states.
The results of this experiment are shown in figure \ref{fig:qubitsExp}.
The optimisation is clearly effective, especially as it has almost no overhead.
Another result to note is that the execution time of the circuit increases with the number of qubits.
This is because there are more states to iterate over.
The amplitude checking optimisation will skip 0 amplitude states, but the gate will still have to iterate over all $2^n$ states and perform this check for each of those states.
\par
The last factor that determines the execution time of a circuit is the depth of the circuit, i.e. how many gates are being used.
Intuitively it is expected that the more gates a circuit uses, the longer its execution time.

The experiment is carried out by adding an extra gate every time and measuring the execution time of the circuit.
As can be seen in figure \ref{fig:depthBench}, the execution time increases as the number of gates applied does. This relationship however, is linear, compared to the previous experiment where the effect of the number of qubits on the execution time was explored and the trend was exponential.
This experiment is carried out on 2 qubits (to test two qubit gates as well) and the timing is measured after memory allocation has been performed, so only the the execution time of the gates is taken into account.
It is carried out on 3 different gates to show that the relationship is indeed linear.

%\newpage
\section{Two Quantum Accelerators}
We now present two possible quantum applications which could be implemented on a physical quantum accelerator but which are now tested on a classical supercomputer.  The applications are Genome Sequencing and a first content description of Quantum Economics.
\subsection{Quantum  Genomics Sequencing} \label{3stack}

In this part, we present and briefly describe an example of the quantum genomics application.  
We explain with a small example case the different execution steps in accelerating genomics using quantum logic.  
Figure~\ref{fig:microarchitecture} represents the generic micro-architecture for the Genomics accelerator that we are currently building out of it. 
% This accelerator is just one example of a quantum accelerator one can make. 
Other quantum accelerator applications could, for instance, focus on traffic management in big cities, cyber security, quantum artificial intelligence etc.
In this paper, we first introduce the quantum logic for genomics and then translate that in hardware blocks that compose the micro-architecture.

\subsubsection{Quantum Genome Sequence Reconstruction on Perfect Qubits}

Genome sequencing involves taking fragments of the DNA (called short reads) of an individual organism, from the sequencing machines and stitching them together to reconstruct the original genome of the individual. 
Reconstruction can either be carried out by aligning these reads to an already available reference genome, or via de novo assembly.
For human beings the reference genome has around 3 billion base-pairs.  
The base-pairs in the DNA sequence consist of the 4 nucleotide bases, cytosine [C], guanine [G], adenine [A] or thymine [T]. 

%When introducing the quantum concepts as described in this paper, this requires the algorithmic primitive of searching an unstructured database or graph-based combinatorial optimisation respectively for the two types of reconstruction. 
As we introduced in the quantum concepts described in this paper, this problem requires the algorithmic primitive of searching an unstructured database and graph-based combinatorial optimisation respectively for the two types of reconstruction. 
Translating such quantum kernels to an efficient implementation on a quantum accelerator requires in-depth tuning of both an architecture-aware quantum algorithm and the underlying micro-architecture.  
In current state of the art genome sequence reconstruction research, the focus is on performance optimisation by extending the existing algorithms with heuristics such that it becomes tractable to determine the genetic profile of the specific application.
Medical diagnosis in the near-future will be based on the genetic profile of every individual.

\subsubsection*{Ab Initio Quantum Genome Sequence Reconstruction}

\begin{figure}[hbt]
\centering
\includegraphics[width=0.85\textwidth]{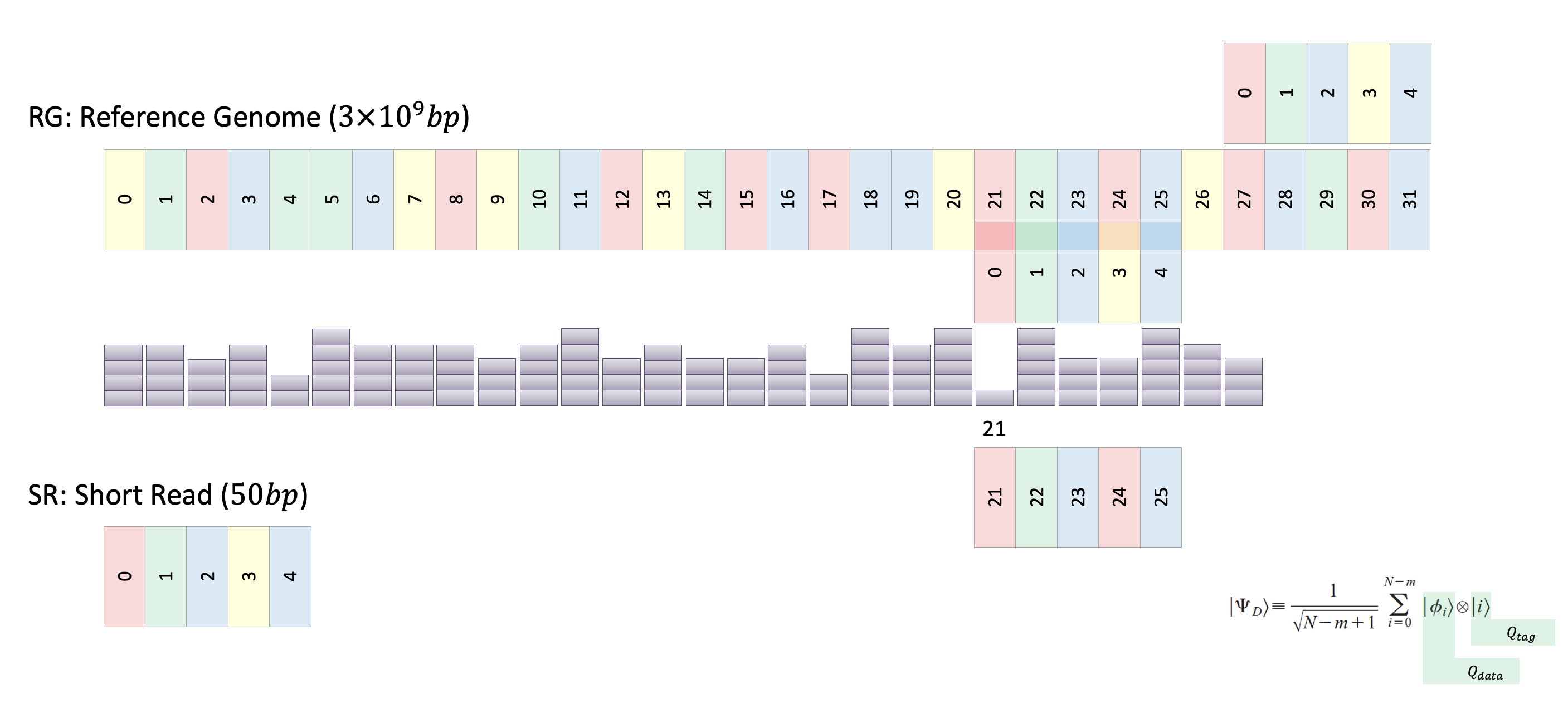}
\caption{Quantum Genomics Sequencing Illustration}
\label{fig:qgshow}
\end{figure}

Ab Initio genome sequencing is the method that is still very intensively used and is based on a reference genome to which individual read-outs will be matched. Figure~\ref{fig:qgshow} shows a high-level view of a reference genome to which the different read-outs of the measured genetic profile have to be matched.%\footnote{\textbf{TODO: finding how the number of qubits grows or shrinks when we execute an algorithm}} 
For the example, we used a reference genome which has 32 base-pairs which, in total, is made up of 4 colours representing the 4 DNA alphabet consisting of C,A,G,T.
Then there is the read-outs (produced for instance by an Illumina machine) from the organism we want to identify and characterise. 
The alignment algorithm aims to find the position (e.g. 21 in this example) where the read matches best in the reference.
As there are read errors from the sequencing machine, the match might not be exact (e.g. we find only 4 out of 5 bases matching in this example, at position 21, but that is the best possible).
The technical details of the development of this algorithm, called QiBAM, can be found in \cite{sarkar2019algorithm}.

\begin{figure}[hbt]
\centering
\includegraphics[width=0.5\textwidth]{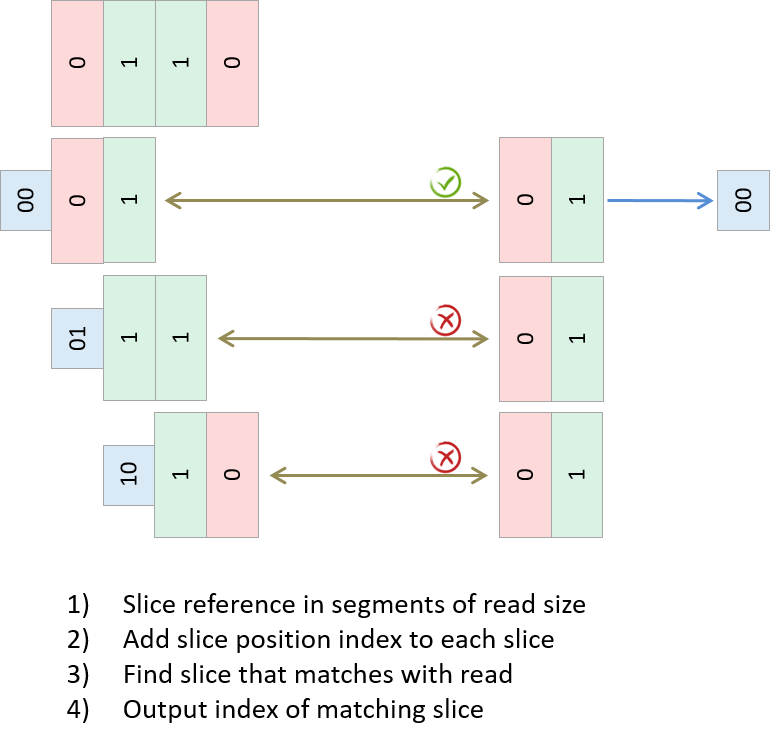}
\caption{Read-alignment algorithm outline}
\label{fig:qgsex02}
\end{figure}

To provide a logical view of how the algorithm work, we will assume a much shorter genetic profile and a very short read-outs of the genomes by the DNA-machine.
Let us assume here a reference genome of length 4 which in this case only consists of 2 characters (binary, rather than the original 4 as real genetics profiles).
We have a short read-out of length 2.
This is shown in Figure~\ref{fig:qgsex02}.
%\begin{figure}[hbt]
%\centering
%\includegraphics[width=0.5\textwidth]{fig/align01.png}
%\caption{QGS example reference and read data}
%\label{fig:qgsex01}
%\end{figure}

We need to figure out where it fits in the reference genome.
While it is easy to see that the read-out can be matched on the first two positions of the reference genome, with the colours red and green, we develop the scalable algorithm here, as shown in Figure~\ref{fig:qgsex02}.  
The reference is sliced in segments.
The segment size is same as the read size, i.e. 2.
Each slice can be indexed with the position where it belongs in the reference.
We need to find the slice that matches with read and output the index of the matching slice.

% As was stated before in this paper, the majority of the quantum gates that are going to be executed are related to the X- and Z-dimensions of the Bloch sphere. That is why the qubits are initialised in the  0-values and the last  the superpositioned qubit has amplitudes without a complex number. In total, we are combining up to 5 qubits so that in total we have 32 possibilities.

\begin{figure}[hbt]
\centering
\includegraphics[width=0.8\textwidth]{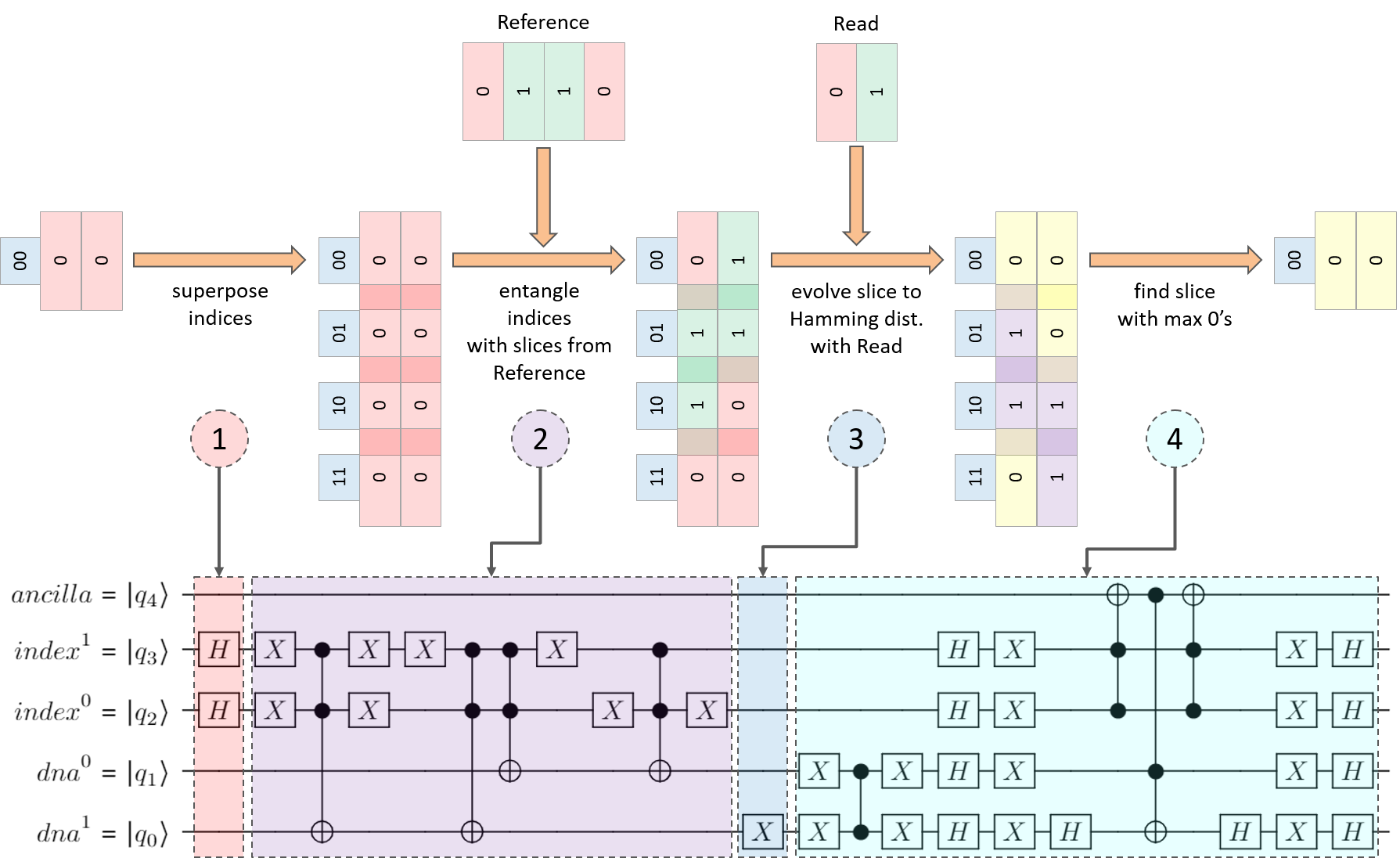}
\caption{Quantum read-alignment algorithm outline}
\label{fig:qgsex03}
\end{figure}

There are 4 steps in the quantum algorithm as shown in Figure~\ref{fig:qgsex03}.
The first step is to superpose the indices in the reference.
This is done by the Hadamard gate on the index qubits.
In the next step, each index is entangled with the data from one slice from the reference.
This is done using multi-controlled NOT gates.
The control qubits is based on the index, while the target depends on what data needs to be stored for each slice.
In the third step, the data part of this quantum database in evolved to the Hamming distance with respect to the read.
This is done using X-gates on the position where the read bits are 1.
The last step is essentially a quantum search algorithm.
The oracle always marks the all-0 state as that is the entry with the minimum Hamming distance.
This state's amplitude is amplified by the Grover iterations~\cite{grover1997quantum}.

The algorithm is based on the Grover's search which allows searching an unstructured database in a quantum superposition giving a quadratic speedup in query complexity (with respect to a linear search).
It was modified to take into account 3 features, (i) read errors mean we might not find an exact match in the reference, (ii) there can be multiple locations where the read might match approximately, (iii) the location of the matches needs to be sampled as the output.

\begin{figure}[hbt]
\centering
\includegraphics[width=0.8\textwidth]{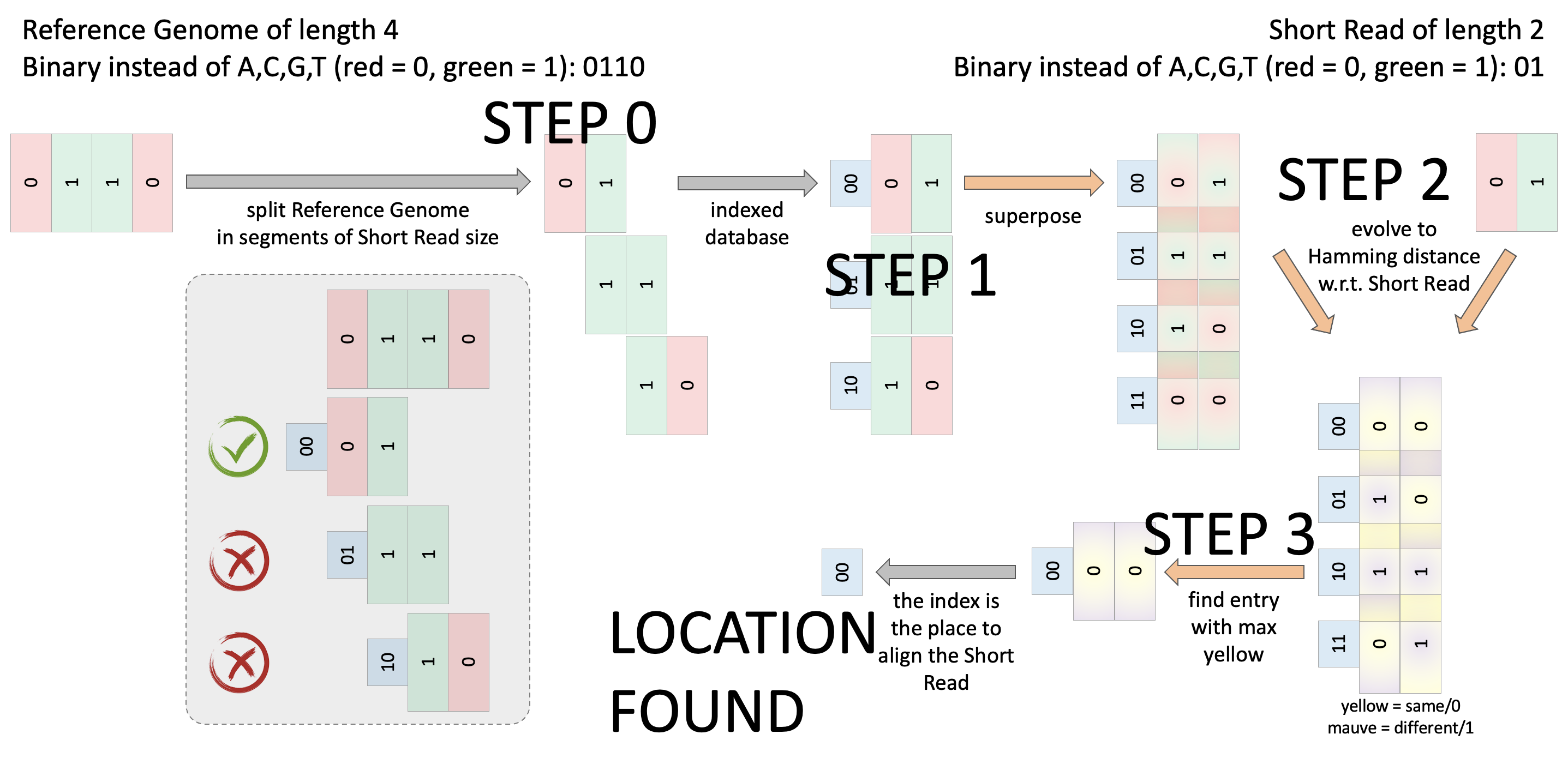}
\caption{Final results of the Quantum Genomics Sequencing}
\label{fig:qgsex10}
\end{figure}

These quantum gates in these 4 steps can be coded in quantum assembly language as shown in Figure~\ref{fig:qgsex10}. 
The entire state vector at the end of each step can be queried from the simulator.
The initial state is the all zero ground state.
In the final state vector we find that the amplitude corresponding to the $\ket{00000}$ state is $0.625+0.0i$.

%\begin{figure}[hbt]
%\centering
%\includegraphics[width=0.8\textwidth]{fig/align04.png}
%\caption{Quantum read-alignment algorithm cQASM code and state evolution}
%\label{fig:qgsex04}
%\end{figure}

The final state is shown as a plot in Figure~\ref{fig:qgsex10} of the real amplitude and the squared probability.
The highest probability state represents the position of best match in the reference.
Due to the reference database and index, being entangled, the closest-match index can be estimated.
%\textbf{Aritra, maybe you can say something about the number of steps in the algorithm and the wall clock time it takes to get the result ?}

% \begin{figure}[hbt]
% \centering
% \includegraphics[width=0.5\textwidth]{fig/QGS_02.png}
% \caption{cQASM code for Quantum Genomics Sequencing}
% \label{fig:qgsex02}
% \end{figure}

We have obtained initial results from combining domain-specific modification on the Grover's search and quantum associative memory~\cite{ventura} approaches. 
This new alignment algorithm, described and analysed in~\cite{AritraMSc}, has been tested on the \QX simulator platform. 
The reference DNA is sliced and stored as indexed entries in a superposed quantum database giving exponential increase in capacity. 
The designed algorithm~\cite{sarkar2019algorithm} considers inherent read errors in the sequence, incorporating the requirement for approximate optimal matching.
A quantum search on the database amplifies the measurement probability of the nearest match to the query and thereby of the corresponding index.

Current explorations involves designing optimisation algorithm for genomics applications using near-term Quantum Machine Learning~(QML) primitives like the Quantum Approximate Optimisation Algorithm~(QAOA).

\subsubsection*{De Novo Quantum Genome Sequence Reconstruction}

\begin{figure}[hbt]
\centering
\includegraphics[width=0.5\textwidth]{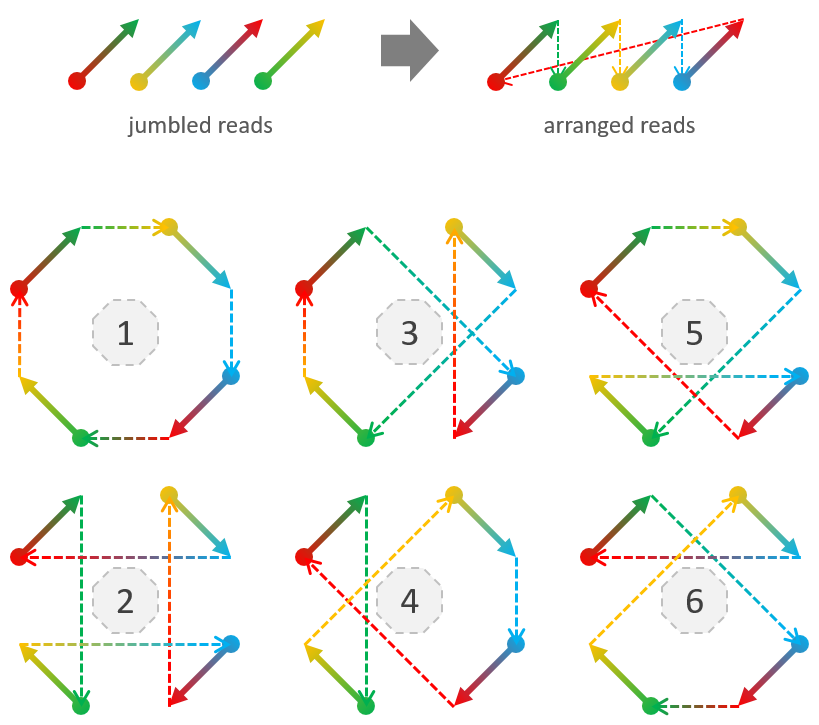}
\caption{De novo sequencing example}
\label{fig:qgdenovo}
\end{figure}

Another type of reconstruction involves directly stitching the reads based on pair-wise overlap.
This takes a set of jumbled reads as shown in Figure~\ref{fig:qgdenovo} and arranges them in order of how the ends of each align.
The number of possible ways to arrange grows exponentially (in this case 6), and is related to the Traveling Salesman Problem (TSP).
We find that for this case, Option 4 is the best, as it always stitches ends with the same colour.

We designed a quantum approach~\cite{sarkar2020quaser} for this problem using the QAOA algorithm.

\newpage

\subsubsection{The Micro-Architecture for Quantum Genome Sequencing}

\begin{figure}[hbt!]
\centering
\includegraphics[width=0.8\textwidth]{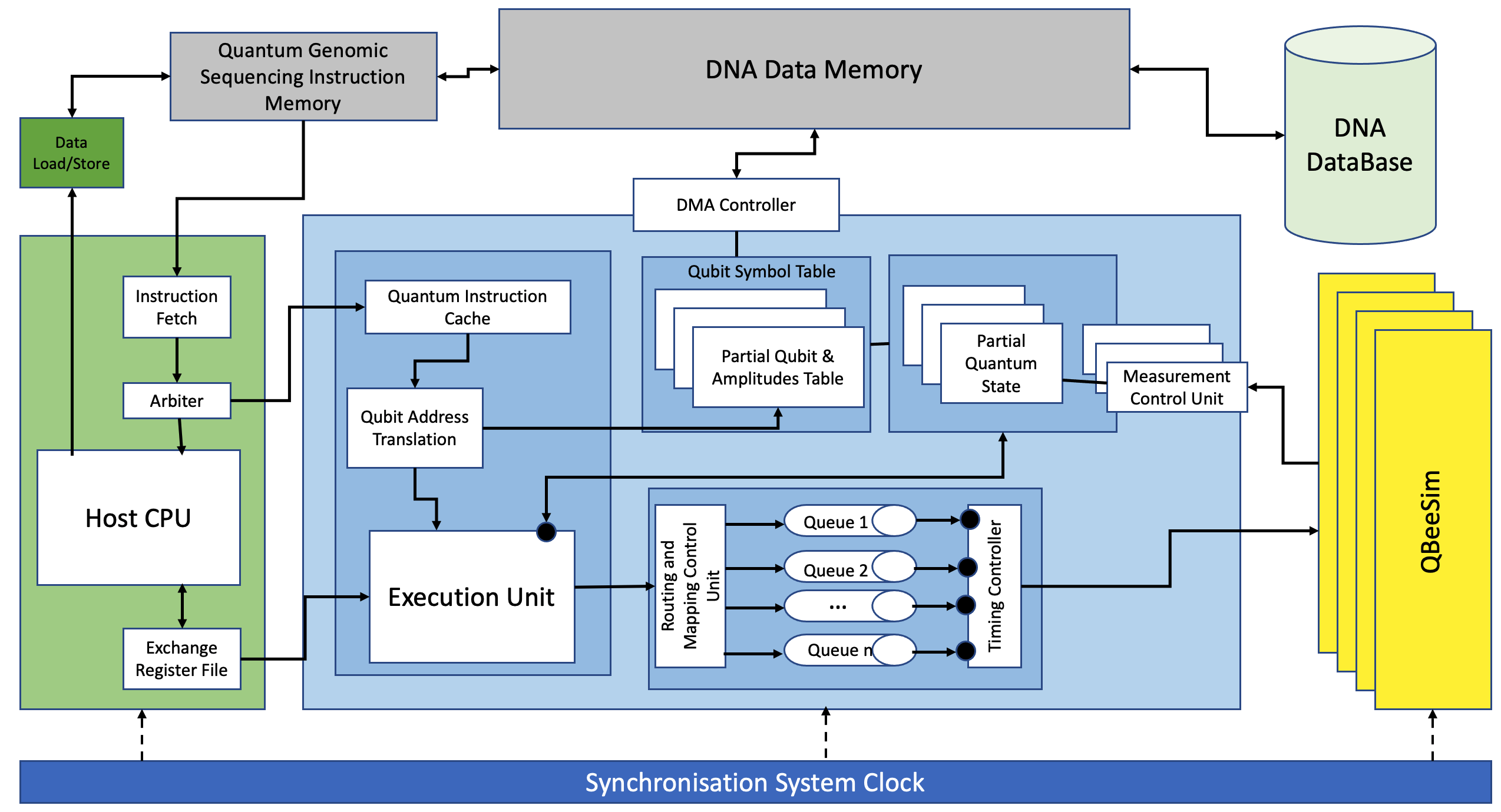} %, angle=90
\caption{\label{fig:FullArchitectureGen} Quantum Accelerator Micro-Architecture for Genomics}
\end{figure}

As already mentioned, the proposed  quantum accelerator will not be a standalone machine, but rather a quantum co-processor that will be part of a heterogeneous system in which classical processors are connected to the quantum accelerator.
Each processor will have its own instruction set.
A first tentative view of the quantum genome sequencing micro-architecture is shown in Figure~\ref{fig:FullArchitectureGen},  which very closely resembles the proposed generic micro-architecture earlier in the paper.

%Connected to that external memory is the internal, quantum memory of the accelerator where all the short reads will have to be placed and used for the sequencing of all the reads in the context of the reference genome.  All the short reads need to be mapped on the genome reference such that any of the later steps can lead to meaningful interpretations of the investigated DNA. That is why there is an indication of the in-memory computing logic as the short reads will need to be stored digitally and locally before being processed by the quantum co-processor.  We do assume that the queues in the Timing Control Unit will still be used but this needs to be experimentally investigated.

%The main additions will be new components such as the local quantum DNA memory and the \qgs-DMA controller, that will control the transfer of the base pairs to the accelerator. 

There is need for run-time support to coordinate the activities of the different micro-architectural components and, as discussed, be responsible for the run-time routing of qubit states for two-qubit gates. 
In the quantum accelerator, the executed instructions generally flow through modules from left to right. 
The pink block on the right of the figure represents the \QX simulation platform or an implementation of a quantum chip on which the test-runs of the quantum genome sequencing algorithms will be performed. 
The rest of the large (blue) block represents the micro-architecture. 
The DNA data-sets is to be retrieved from an external classical database and transported to a local memory in the quantum accelerator. 
The size of the local memory will depend on the capabilities of the \QX simulator platform and how that information is encoded. 
This research is based on the large-scale micro-architecture simulation platform that we have already developed.
Using the \QX simulator platform makes it possible to rapidly develop hardware prototypes and verify their behaviour and performance before a FPGA implementation is started. 
The set of queues will be relevant for feeding the DNA information to the qubit chip and for defining how the quantum gates are applied.  

\subsubsection{Nearest-Neighbour Operations for QGS}

In a specific qubit plane topology, qubits will have to move around so that two-qubit gates can be applied on adjacent qubits.
It is a prevailing idea that quantum compilers generate technology-dependent instructions~\cite{svore2006layered, abhari2012scaffold, haner2016software}. 
However, not all technology-dependent information can be determined at compile time, because some information is only available at run-time due to hardware limitations, for instance qubits that need to be re-calibrated. 

For testing the functionality of the algorithm, we use artificial DNA sequences that preserve the statistical and entropic complexity of the base pairs in biological genomes; yet in a reduced size so that they can be efficiently simulated in a classical architecture with qubit limitations.
This implies understanding which run-time and thus routing support will be necessary to make sure that the quantum accelerator always has enough data to process and that they are in adjacent positions when necessary.

From an algorithmic perspective, near-term quantum optimisation algorithms employ the variational principle, where a shallow parameterised quantum circuit is iterated multiple times while the parameters are optimised by a classical optimiser in the Host-CPU.
This model of Hybrid Quantum-Classical (HQC) algorithms requires fast feedback between the quantum accelerator and the  real-time circuit/instruction generator (i.e. the compiler and the micro-architecture).
Since most quantum algorithms expect a statistical central tendency over multiple measurements, the expected probability of the solution state can be calculated inside the quantum accelerator itself, aggregating the measurements over multiple runs.
%\begin{figure}[hbt]
%\centering
%\includegraphics[width=\textwidth]{fig/QGS_04.png}
%\caption{Step 4 - Quantum GenomicsSequencing}
%\label{fig:qgsex04}
%\end{figure}

\subsection{Quantum  Economics}

A second potential candidate for which one could consider building a quantum accelerator is for quantum economics and finance. Let us give a small example of how financial decisions are made to manage the portfolios of individuals or families.   Without going in to much detail, it always starts with a conversation between the financial professional and the customer to understand their financial needs for the next coming years. It is not wise or intelligent to invest in different financial products if one has relatively short term financial needs.  Such needs could be buying a house, the marriage of one of the children or the birth of child. These are important moments that need to have a spot in the long term planning of a family. What is needed is to understand what the risk aversion is one has when buying or selling financial products. The goal of any financial market is always long term planning of financial resources, so a horizon between 5 to 10 years is what one should have. So if one is very sensitive to financial risk, that immediately translates in a series of financial products one should not have, such as options or very volatile shares of companies.  There is also the vision that one should have in any case a spread of the risks such that one does not have the risk of losing a large amount when one company goes bankrupt or anything. There are many parameters playing around in financial management such as the international economic context, industrial areas and some personal requirements.

The advantage of the application of Quantum Computing to Economics
and Finance is twofold, with the most obvious being the computational aspect.
The promises of the acceleration of the computing times using quantum
algorithms for solving, for example, optimisation problems. Those
are problems that can take advantage of the new algorithms in its
classical formulation, at least after being written/converted to quantum algorithms.

The other less obvious side is that Quantum Economics and Finance brings the quantum formalism to describe the Economics and Financial phenomena involved like the risk profiles presented above. It is interesting and in many senses counter-intuitive
that this applies so well to Economics and Financial problems.

\paragraph{Formalism advantage}

One example of applying the quantum formalism, applied to Economics and Finance,
is Quantum Decision Theory (QDT), see~\cite{fabre2016}, a recently
developed theory of decision making. This framework theory formalises
the concept of uncertainty and other effects that are particularly
manifest in cognitive processes, which makes it well suited for the
study of decision making.  The main advantage of this type of model is that not only they provide a more synthetic and better description of reality as well as that they are are native to quantum computing.

In the case of Finance the use of the quantum formalism can be easily understood
as soon as we recognise that some of the equations used in traditional models can be interpreted as wave functions. As an example we have that with an appropriated change of variable the Black-Scholes becomes a diffusion equation, just like the Schrödinger equation from Quantum Physics, see~\cite{ivancevic2020adaptative}.

The decision making process is a corner stone of Economics theory.
In modern Macroeconomics this concept is enclosed in the utility function 
where economic agents comply, by design, with 4 axioms and are thus described as rational, see~\cite{orrell2020}.
Yet since the 1970's that behavioural psychologists and economists have
shown that the expected utility theory does not capture a variety
of cognitive phenomena. The use of a quantum variant of the utility function allows the model to incorporate risk aversion in its own utility function when different choices are considered.

Full entanglement is another feature that quantum formalism allows.
The quantum formalism can be used to model entanglement at the cognitive, social and financial level. All those levels are important to better represent the reality, that is to model the individual, the relation between individuals and the system as whole.

A practical, but of fundamental importance, example of this is the 2008 subprime crisis, a striking example of full entanglement that the classic models struggle to describe. In the subprime crisis there were several levels of financial products that were leveraged on a very shaky ground, and what later was dubbed the toxic loans. The use of quantum models allows these connections to be built from the ground up in a very natural way and not to be lost when computing/evaluating the system as whole.

\paragraph{Computational advantage}

There are recent results like~\cite{phillipson2020portfolio} where quantum annealing is used to optimise portfolios. There are classes of problems like the portfolio
optimisation that can be expressed using techniques like Genetic Algorithms
that lend themselves to the use of Quantum Algorithms. The problems take benefit from the computational advantages of the quantum computation in terms of exploring multiple branches at once. This is particularly evident for problems that can be solved using parallel computing.

Another example of this computational advantage is the application of deep quantum neural networks to Finance in~\cite{sakuma2020application}. The quantum neural networks are then used in machine learning models.  In addition we can also read the findings in~\cite{fontanela2019quantum} where the original partial differential equation satisfied by the option price is transformed to a Schrödinger equation in imaginary time and the equation is then solved numerically using a hybrid algorithm, where both classical and quantum computing are used. There are two remarks that can be done here, the first is that in the context of quantum algorithms the complex numbers are a natural component and thus change of variable to complex time is in line with the other arithmetic. The second remark that should made is that classical and quantum computing are not mutually exclusive and can be both used in synergy.

In terms of research objectives, we could envision modelling financial markets where professional traders buy and sell different kinds of financial products, such as shares, bonds and options, based on the risk profile of their customers.  The formulated quantum financial model will be formulated using the concepts as described here, exploiting and using the phenomena such as superposition and full-entanglement to model financial portfolio decisions. 

\section{Quantum Simulation on Supercomputers}
%\subsection{Use supercomputers to prove results of Q algorithms and performance of Q computing}

In this section, we describe how we will be testing, debugging and executing the quantum applications that we are investigating.
We describe the results of the experiment we recently did to show how quantum algorithms can be correctly executed on supercomputer infrastructures. There are principally two kinds of exercises we executed. One was a selection of algorithms that came out of the quantum genome sequencing work, expressed in OpenQL and cQASM and one was written in C++, including a new quantum simulator that executes the quantum circuit written down in C++.

\subsection{From Errors on Quantum Chips to Supercomputers}

\begin{figure}[hbt]
\centering
\includegraphics[width=\textwidth]{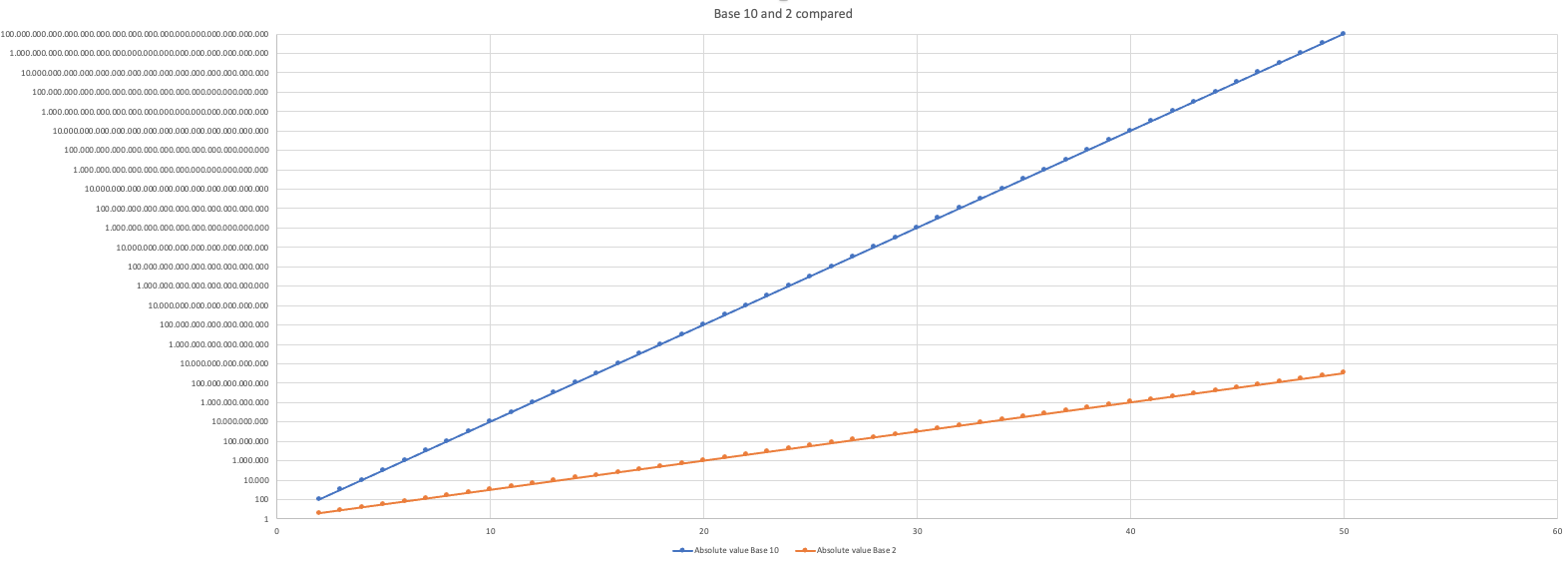}
\caption{\label{fig:errors}Error rates in Base 2 and 10 for any of the available quantum technologies}
\end{figure}

It is generally known that the qubits we are currently capable of producing have a very bad quality. Figure~\ref{fig:errors} shows that the current qubits, irrespective in what technology they are made, have error rates which are substantially worse than any of the modern CMOS-transistor that we can already produce during several decades. The left graph of that figure shows that the current qubits have errors between $10^{-2/-3}$ where in the CMOS-world we are used to having errors every $10^{-15/-16}$. The right table translates that graph in numbers such that it is clear that we currently are very far away from any of the CMOS-quality that we can have in classical hardware.  On top of that, we also have to emphasise that it is not known now in what core technology the quantum bit will be produced. That period will most likely be around 10 years or longer. There are still a lot of technologies competing with each other and it is not known at all which one will turn out to be the best one.  Looking at the estimates of the number of qubits, those estimates  for various applications vary a lot. Number of qubits go from millions to billions  so there is no agreement in the physics world on that either.

Given this incredible uncertainty, one could assume that no other field or discipline can enter the quantum computing field as there is no way to test in any reasonable way what the quantum computing result will be. That is why we are proposing an alternative testing platform where we can test and validate the way any quantum algorithm is computing a result.  As stated before, we will be using the most powerful supercomputers such that the most important feature of quantum simulation is the amount of memory that can be used.

\begin{figure}[hbt]
\centering
\includegraphics[width=0.8\textwidth]{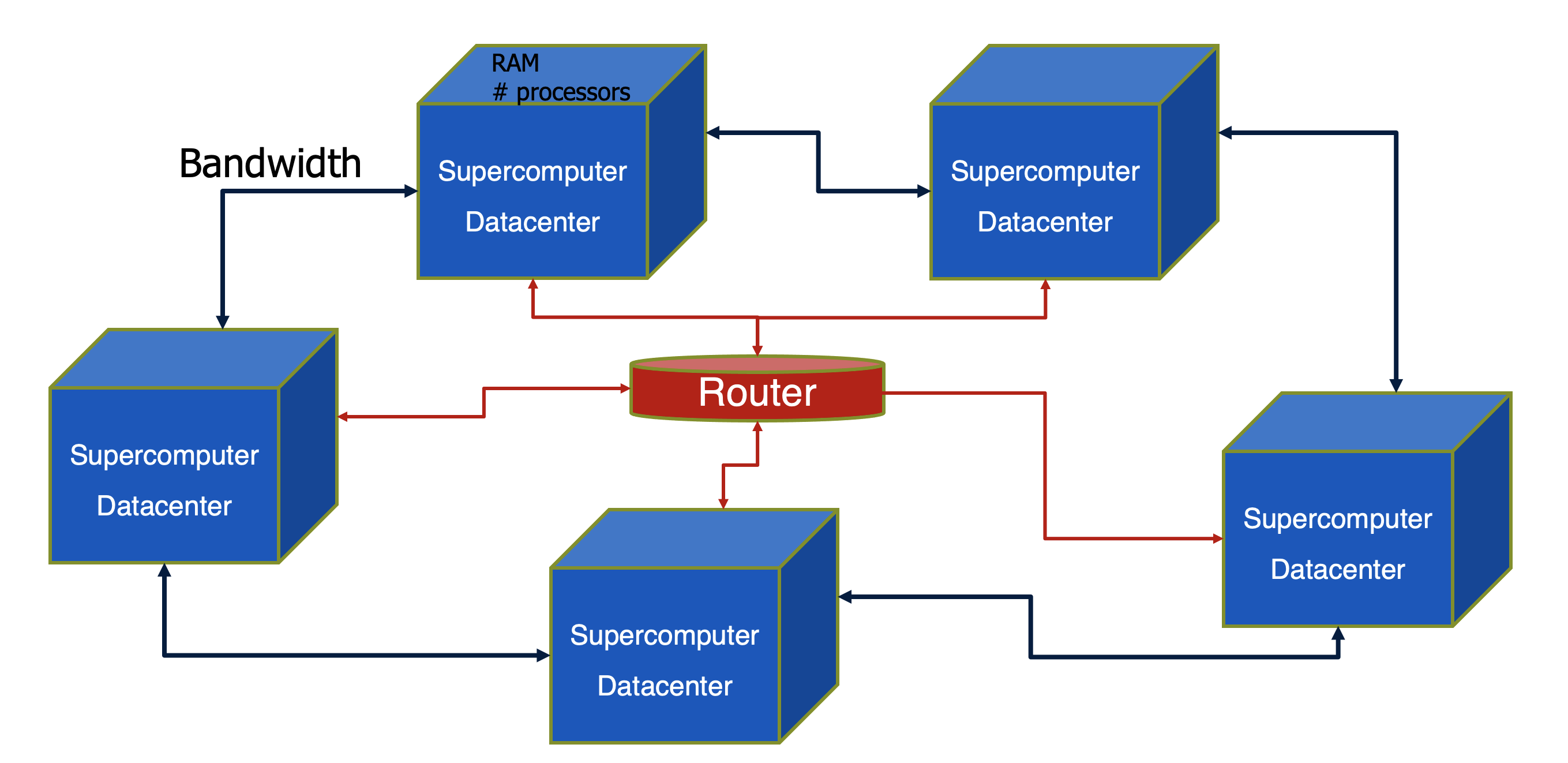}
\caption{\label{fig:supercomputer}Connected supercomputers}
\end{figure}

The way that the computation of the amount of memory is done is directly related to how many qubits are combined with each other.  We should not forget that each qubit consists of a basis state consisting either of $(|0>,|1>)$ or $(|+>,|->)$. Each basis component has an amplitude that represents the way that the qubits are the most likely to be selected at a readout of the qubit states.   However, in general we are used to apply 10 as the base number. Those figures and numbers are shown in Figure~\ref{fig:supercomputer}.

 Figure~\ref{fig:errors} gives the numbers expressed in base 2 and that goes up to 50 qubits that are fully entangled, both in terms of the basis states as well as the amplitudes.  The amount of memory therefore always be computed based on a simple equation: either $ 2^N$ for the basis states or $4^N$ for the amplitudes. We have to realise that for every amplitude, we have two numbers. The first is a real number and the second is an imaginary number combined with a real number, which explains the amount of bytes that we will need for the memory.

 \subsection{Supercomputers used for Testing}
Without going into needless details for the supercomputer platforms, it is clear that the number of such very powerful computer is pretty high. As declared multiple times, the amount of memory that we need to use is very large. This amount is very important for us to understand what the ultimate computer architecture could be for any quantum accelerator.  As we know the number of amplitudes is twice a large as the number of qubits that we will be using, this is clearly a limiting element. 

That is why we decided to use  an Ubuntu Virtual Box on top of the service provider to give us a stable and scalable platform such that we can experiment with the results of the hardware and software choices we make and for which we need to execute on all the layers the quantum application.  We intend to have two versions available. The first one is the version we would put in the public domain such that other researchers can bootstrap their own research in the quantum field without having to learn a large number of topics before having any useful result.  The second version is the one which is undergoing some changes, based on the research decisions that we will need to make.  Once tested by the colleagues in a very wide context, that version will then also be put in the public domain.  The virtual box will also contain an encrypted database where any execution on all the layers of the full-stack will also store intermediate results such that researchers can also study what the operational.
Running all our experiments on the Ubuntu virtual box makes us in principle more independent of any hardware constraint that might influence our work. As explained in the next section, we will store that information in a database that can be used to extract scientific and technical results of any full stack implementation.  We will know at any point in time during the execution what the hardware constraints are that we are using.

\subsection{From Errors to a Metrics Database}

\begin{figure}[hbt]
\centering
\includegraphics[width=0.8\textwidth]{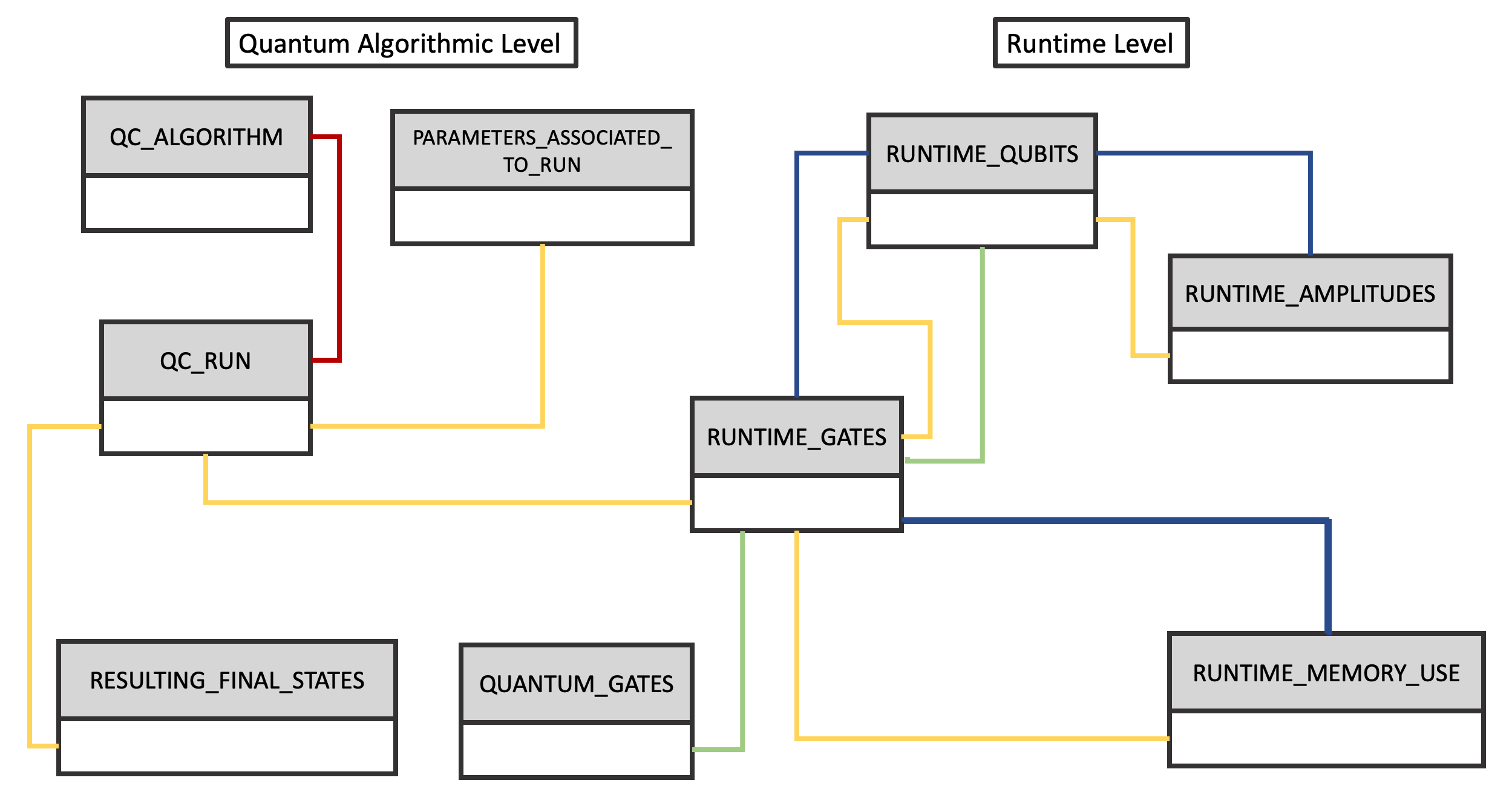}
\caption{\label{fig:DBQ}Quantum Metrics Database}
\end{figure}

A key part of any research initiative is to discover what the effect is of certain changes to the overall performance of the quantum application we are running. To this purpose we have developed a database that collects all the errors that occur when we are executing an algorithm.  Figure~\ref{fig:DBQ} gives a very high-level view of the database that we have developed. The details of the content of the tables are not specified as we are currently running tests on it.  The goal is to experiment with different computer infrastructures to run the quantum instructions on them.
There are two sets of tables in the database. There is the left part that contains all the tables describing the algorithms that we are running.  A quantum application will need to be executed multiple times to understand what information is generated, which is why we have the \textbf{QCRUN} table, where certain parameters may change and where we obtain in the table \textbf{ResultingFinalStates} a final result of the computation.  These tables drive the  use of the quantum logic gates that need to be applied on the qubits. That is the moment where we reach the part of the application that will be executed and which is a composition of the different quantum gates that are used.  The quantum gates table will feed the basis for any quantum circuit that we are making.  The \textbf{ Runtime tables} combine the qubits, their amplitudes and the quantum gates that are applied to the qubits. We emphasize again that quantum computing is one way of implementing in-memory computing. That implies that the execution logic is brought to the qubit and applied on them rather than moving the qubits around to a quantum processor. What is important to underline and understand is that even when we execute the quantum algorithm on a classical hardware computer, and we need to bring the qubit information to the processors, the basic principle of in-memory computing stays valid.  If ever we can use an operational quantum processor with real qubits, it is only a small step in the compiler approach to generate the cQASM that can be executed such that the micro-architecture can execute the quantum circuit on the quantum processor, implying to bring the gates to the qubit than the other way around. In the transition to that period,  we need to keep track of how much of the available memory we are using.  In this context, it is equally important for each run of the algorithm, that we understand what the number of qubits is that is used. That number of qubits can grown and shrink at any point in the algorithm.

\begin{itemize}

    \item Number of gates: This is clearly a main metric that describes how many quantum operations need to be done on the qubits. Again, the quantum physics people need to make this number very large to understand how stable the qubits are but for quantum application developers, the number will be more around several hundreds of gate operations.  It is important to realise that many of these gates can be executed in parallel if there is no direct connection and data-dependency between them. So, one can have 10.000 gates in a serial way or execute them in parallel on, for instance, 1000 parallel paths.
    \item Circuit depth (or latency): irrespective whether the number of gates is very large or executed in parallel, an important metric is how long each quantum gate will take to execute.  The execution time can be expressed in wall-clock time or the number of CPU cycles and can be applied to either a physical quantum device or a classical computer. In essence, it can be computed by the sum of all quantum gates in one single path of the parallel execution and which will take the longest time to execute. And this time determines the circuit depth or latency of the quantum circuit which is then the overall execution time of the quantum circuit.
    %\item Quantum fidelity: 
    %\item Probability of success
    \item \textcolor{black}{Quantum Volume (QV): A metric that was introduced by IBM researchers is the quantum volume.\footnote{https://medium.com/qiskit/what-is-quantum-volume-anyway-a4dff801c36f}} It is inspired by the well-known Moore´s law for classical transistors but which are clearly different from the qubits used in quantum computing. The focus is here clearly on physical qubits with a limited coherence time. It measures how well a quantum computer can run a circuit consisting of random two-qubit gates acting in parallel on a subset of the device´s qubits, irrespective of the kind of quantum technology used to make the qubits. This metric allows the Qiskit compiler to recompile the quantum circuit such that its execution complies to the constraints of the quantum chip involved.What is interesting is that the first step assumes that the quantum circuit is run on a classical computer. This is to gather the \textbf{heavy output generation} for the problem you are investigating. After computing the heavy output values, the circuit is executed multiple times on the quantum processor and measure for each run the circuit´s depth and width. The Quantum Volume algorithm stops at the highest depth and getting results bigger than two-thirds, with a confidence interval of 97.725\%. The confidence interval represents the probability of a random interval containing a (unknown) deterministic true parameter. The Quantum Volume is then computed by computing it as follows: $QV=2^D$ where QV is the Quantum Volume and D is the depth of the circuit. The metric is interesting to compare the classical computer to the quantum computer by computing the time it takes to execute a quantum circuit with a particular number of qubits. There are still a lot of shortcomings to the QV metric but it has to evolve as the technology is also evolving.\cite{Jurcevic_2021} \cite{Cross_2019}
    
\end{itemize}
As we explained in this paper, the biggest challenge is to have of course the parallel version of the quantum algorithm and to understand how much physical memory we need on the large computer systems.  It is also important to understand how many qubits, in superposition or fully entangled are needed at any point in time during the execution of the circuit. That is what we need for any improvement at any layers, going from the quantum application up to the micro-architecture control of the quantum simulator. 
\section{Hardware and Software Long-Term Vision}

% Software

There are different ways of building a computer and the way it is currently done is to combine multiple heterogeneous multi-core processors
There are several models of quantum computation.
The theoretical models, like the quantum circuit model, adiabatic quantum computing, measurement-based (cluster state) quantum computation and topological quantum computing are equivalent to each other within polynomial time reduction.  
One of the most popular and by far the most extensively developed is the circuit model for gate-based quantum computation. % paradigm.
This is the conceptual generalisation of Boolean logic gates (e.g. AND, OR, NOT, NAND, etc.) used for classical computation.
The gate set for the quantum counterpart allows a richer diversity of states on the complex vector space (Hilbert space) formed by qubit registers.
The quantum gates, by their unitary property, preserves the 2-norm of the amplitude of the states thereby undergoing a deterministic transformation of the probability distribution over bit strings.
The power of quantum computation stems from this exponential state space evolving in superposition while interacting by interference of the amplitudes.

Most of the quantum computers which are made today are based on superconducting qubits but in the past there have been attempts on ion traps and semiconducting qubits are becoming very popular.  
We are just starting to reach the 50 qubit mark in processors but are way below the required coherence.
The big system is shown in Figure~\ref{fig:layers}, where we include both the quantum annealer and the quantum gate accelerator. %, even though the real  algorithm for the quantum annealer is quite generic tuning of biases and couplings.
The same holds for the micro-architecture, for which, the components need to be developed.
% But the idea is that the quantum annealer is a closed box and should be relatively easy to be used in any kind of testing environment.

\textbf{Full connectivity:}  An important limitation that is not yet solved in any scalable way is the connectivity between the qubits, as for two-qubit gates the qubits need to be close to each other.
It means that there is direct connectivity only in the neighbourhood of each qubits.
This has important implications on the initial mapping of the qubits on the topology and especially the routing of the qubits to a location close to the other.
% There is not a current high-level way of analysing and describing them so everything still needs to be explored. 
Evidently, the kind of logical qubit one uses is very important. 
That is also an open issue currently brought to light by Preskill's paper~\cite{preskill2018} stating that surface codes are too expensive.
It suggests to move to small codes where much less qubits are needed to create a logical qubit.  
That is also why we have introduced the notion of a perfect qubit such that some of the complexities and problems can be abstracted away for the application developer.
%\begin{figure}[bt]
%\centering
%\includegraphics[width=0.8\textwidth]{fig/layers.png}
%\caption{Full Stack Execution}
%\label{fig:layers}
%\end{figure}

Figure~\ref{fig:layers} shows our long-term schema of what a quantum computer can look like in the two directions that are currently being explored, quantum gates-based and the quantum annealing approach. 
To give an overview of what is available on the market is very difficult as there are no commercially available computer systems that can be used in any reasonable way. 
% Just like in most part of this paper, we are naming these devices quantum accelerators.  
The market can be split in two parts: companies that are building a quantum-gate based computer and ones that are focusing much more on optimisation problems that can be solved with quantum annealing.

Gate-based quantum algorithms are designed such that the solution states interfere constructively while the non-solutions interfere destructively, biasing the final probability distribution in favour of reading out the solution(s).  However, the error rates are still around $10^{-2/-3}$ and need to be substantially improved.

\begin{itemize}[nolistsep,noitemsep]
    \item \textbf{IBM} : they offer a quantum processor up to 65 qubits and the goal is to arrive at 1121 qubits in 2023. The qubits have all the normal error behaviour but they can be programmed. They also have started and implementing any micro-architectural control of the physical level. Qiskit also supports perfect qubits as a meaningful extension.
    \item \textbf{Intel} : is looking at both semi- and superconducting qubits but are in essence more interested in the semi-conducting qubit processor. The essence is fixing a lot on the qubit production, partly supported by a solid micro-architecture.
    \item \textbf{Microsoft} : has some preference for the majorana-based approach but they still have to make the first qubit based on that quasi-particle. They are very active in the software development.
    \item \textbf{Alibaba} : is a strong player in this field and they have a quantum lab that focuses on a range of activities going from the development of a quantum processor, quantum-classical algorithms up to simulation of quantum physics.
    \item \textbf{Google} : also one of the leaders in superconducting qubits (John Martinis's team)
    % One of the leaders in superconducting qubits is John Martinis who was hired by Google a couple of years ago. He is one of the leaders world-wide in superconducting quantum computing
    \item \textbf{Rigetti} : is a start-up %in California and 
    focusing on the superconducting quantum processor. They advance well but there is not yet any applicable processor in the market even though there is a processor that can be used for some testing purposes.
    
    \item \textbf{Xanadu} : the team focuses on continuous variable quantum computing based on photonics of squeezed light.
\end{itemize}

\section{Towards In-Memory Computing}

In-memory computing is becoming increasingly important as a new computer architecture. 
Rather than moving the huge amounts of data around to the logic, it is much more meaningful to move the logic around and keep the data as local as possible without moving it around, using, for instance, innovative technology such as memristors. 
Memristors were theoretically defined already several decades ago by Leon Chua, but recently the semiconductor manufacturers are seriously investigating their production.
The key idea of a memristor is that it can be used to store data but also to make calculations. 
This is why memristors are an ideal candidate for making an in-memory architecture.
The concept of in-memory computing is described in a paper where the concept is illustrated using memristor based devices~\cite{hamdioui2015}. 
The main advantage of memristors is that they can be used both to store information and to work on it. 
So an intelligent merging of logic with data storage is the key of an in-memory architecture.  
It is a completely new way of designing algorithms and computing systems and it is far from evident what the design rules are that are needed to fully exploit the in-memory computing potential. 
    % The data movement is an unsolved issue in modern day supercomputers and is becoming increasingly important with the huge amounts of data that need to be processed and stored.

The link with quantum computing is very straight: the quantum logic is directly applied on the qubits and the qubits do not need to be transported to any Quantum Arithmetic and Logical Unit (ALU) before being  processed.   
In quantum computing, the routing of qubit states is therefore also a very important problem. 
The qubits need to be put on the quantum chip in a way that the movement of qubit states is as minimal as possible. 
Also what routing protocols will be used for any quantum chip is a big open area of research in quantum computer engineering. 
Currently, in any of the semiconducting or superconducting quantum implementations the interaction between qubits has a nearest-neighbour constraint. 
That induces the need for deciding where to map and how to route the qubits used in the algorithm on the quantum chip. 
This qubit routing is an important and illustrative example of what in-memory quantum computing actually means.
When adopting an in-memory computing architecture, a crucial challenge is to decide on the placement of the data that needs to be processed and to have a programming language and compiler, such that the appropriate logic can placed close to the data. 
Any kind of algorithm will have data that the algorithm is changing to get a result and it is quite unlikely that there is no dependency between any of those data items. 
What that implies is that intermediate results will have to move around in the architecture such that it reaches the place where that result is used in the next computational step. 
Even though in-memory puts all the data in some kind of memory, those data items still have to move around such that a final result can be computed by the classical Host-CPU.  
From a quantum physics point of view, the main challenges are the coherence of the qubits, the fidelity of the operations and the overall error rate of the quantum computation, involving both the qubits as well as their operation and the involved error-corrections.  
This is already being sufficiently studied by the quantum community but there are also clearly other challenges that need to be researched as soon as possible.

One of the main problems is the error-proneness of the qubit behaviour which consumes up to 90\% of the (quantum) computer time. 
As explained, the routing and moving around of qubit states is a very important challenge.  
So any progress the physics community is making in that respect is extremely important as it will reduce substantially the pressure on the micro-architecture and the overall system design.  
In \cite{brennen2003quantum}, the authors present a quantum computer architecture that addresses the important problem of qubit state routing for nearest-neighbour two-qubit gate execution. 
They use an idea from the von-Neumann architecture of classical machines such as a quantum bus which is a refreshable entanglement resource to connect distant memory nodes. 
The overall approach is at the level of entanglement purification and qubit pairs with different fidelities.
Given that a quantum computation on qubits complies to the same overall in-memory computing logic, that particular architecture is definitely interesting for any quantum device.  
    % Quantum gates are applied on the qubits to change the state of that qubit. 
    % In principle, qubits therefore do not need to be transported somewhere else before one can apply the quantum operation.  
The challenges involved with in-memory computing are therefore the same as for quantum computing. The underlying technology are not memristors or other technology but any of the quantum technologies and require also a full-stack integration of the different layers.  
In that sense, the quantum computing research should be based very much on the basic principles of in-memory computing.

\section{Future Prospects}

\begin{figure}[htb] 
    \centering
    \captionsetup{justification=centering}
    \subfigure[Development time frame]
    { 
        \centering
        \includegraphics[width=.56\textwidth]{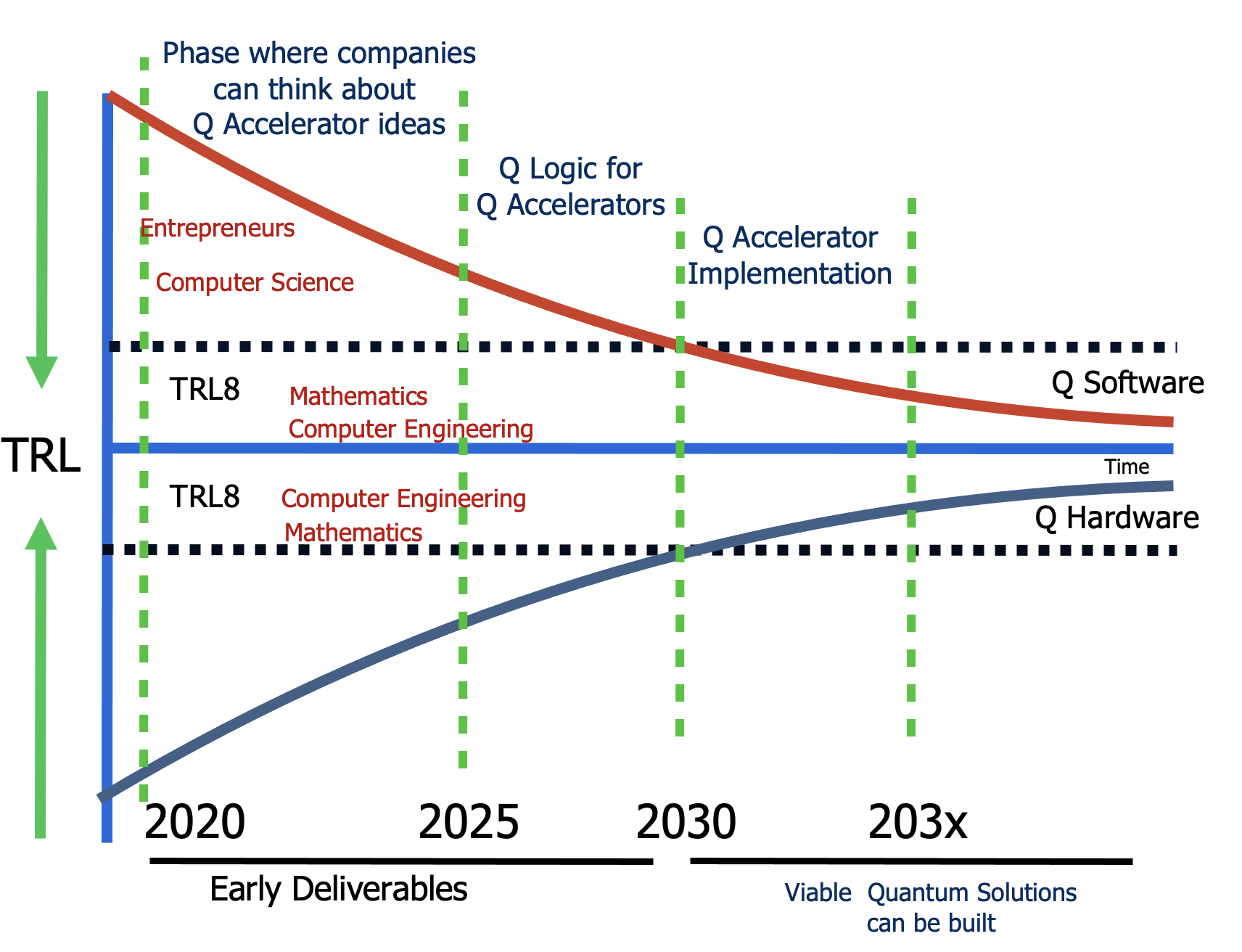}
        \label{fig:time-frame} 
    } 
    \subfigure[Structural division between perfect and realistic qubits]
    { 
        \centering
        \includegraphics[width=.36\textwidth]{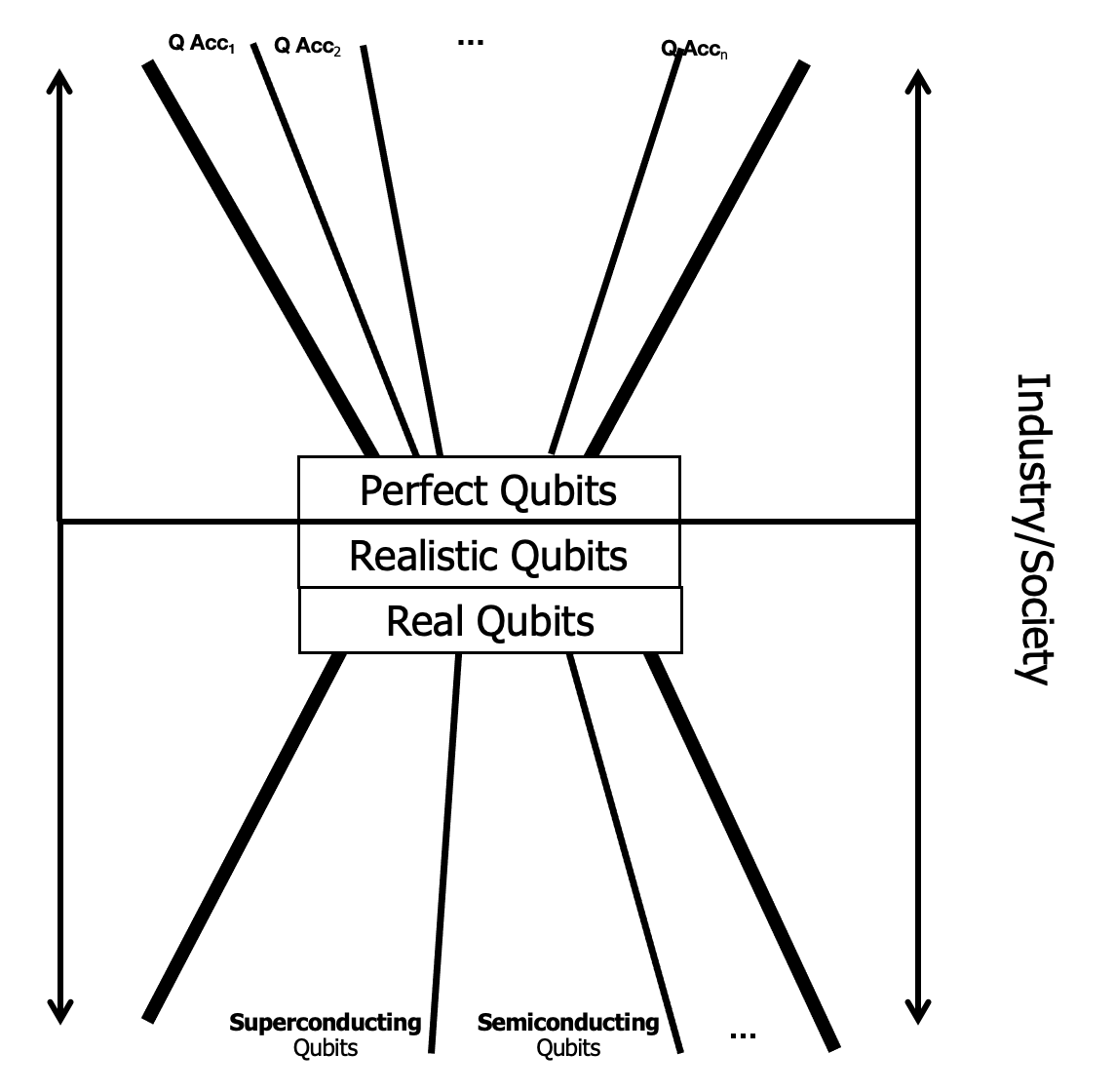}
        \label{fig:structural-division}
    }
    \caption{Quantum computer development future projections}
    \label{fig:future-projections}
\end{figure}

It is very important that companies and other organisations start investing as soon as possible in Quantum Technology.
Figure~\ref{fig:future-projections} shows a projection of when different parts of software and hardware development will be required, to create an efficient quantum computer.
The distinction is made between the use of quantum accelerators and that of manufacturing a quantum chip.  
In general, any commercial or other organisation is interested in new technology if the Technology Readiness Level~(TRL) is high enough.  
If we adopt the same levels as for classical technology, the TRL needs to have reached level 8 and that is sketched in the red and black line that are shown in Figure~\ref{fig:time-frame}.  
There are 4 vertical, green-dotted lines to illustrate 3 moments leading to the last phase where we assume there is enough software or hardware maturity that can be used for any accelerator one wants to build.  
Phase I focuses on the reflection by the organisation on the concrete need that exists and for which a quantum accelerator logic can be developed.   
Phase II resembles the team members brainstorming on the logic for the quantum accelerator. 
They will express that logic in OpenQL and develop some prototype micro-architecture and executed the logic on the \QX simulator.  
Phase III then focuses exclusively on the actual implementation and execution of the Quantum Accelerator logic, whether on an experimental quantum chip or on the \QX simulator. 
This is the moment when the top and low curves can be combined in a real quantum prototype of the accelerator.  
Figure~\ref{fig:structural-division} represents the way that the two lines of research are currently separated and which will be joined in maybe over the next decade.
The division was used in this paper where we made the distinction between the use of perfect and realistic qubits and how that determines the different layers in the full-stack.
\section{Conclusion}

Over the last couple of decades, quantum computing has been a one-dimensional research effort focusing on understanding how to make coherent qubits and how to implement the different universal quantum gate sets on any of the multiple quantum approaches.
As far as computer architectural choices were made, the community has been focused very much on the von-Neumann computer architecture and defined qubits in terms of memory and processing qubits.
However, computer engineering as a field has understood by now that this approach never scales to the size needed for handling, for  instance, the Big Data volumes that world wide are being generated and collected.
Two approaches seem to be very promising: the first comes from the accelerator community and involves the full stack integration of the different layers that are needed to build the quantum accelerator.
The use of perfect qubits in that context makes sense as the end-users of any quantum accelerator can focus their reasoning on the quantum logic of the application and verify it through some implementation of the micro-architecture and the execution of the quantum instructions on the quantum simulator.
The second option is to use the full-stack for the control of, for instance, superconducting and semiconducting qubits with a micro-code layer where we translate any kind of common QASM (cQASM) into an operational set of micro-instructions, for a meaningful adoption of existing computer technology.
It is very difficult to predict the performance improvement of a quantum computational device but that it will be much higher than any existing computational technology is clear.
It also depends on the quantum application that is being looked at and the way the qubits are manufactured.
Research is still needed for at least a decade before the full-integration effects become visible and verifiable.
% \newpage
%\input{07-Ravi.tex}

% \nocite{*}
%\input{12-AppendixMetrics.tex} 
 %\projname
%\bibliographystyle{unsrt}       %References go to end of Section III
\bibliography{refs}

\appendix

\end{document}